# Applications of Nanomagnets as Dynamical Systems

Bivas Rana[1,2,*], Amrit Kumar Mondal[3], Supriyo Bandyopadhyay[4,*] and Anjan Barman[3,*]

[1]Center for Emergent Matter Science, RIKEN, 2-1 Hirosawa, Wako 351-0198, Japan;

[2]Faculty of Physics, Adam Mickiewicz University, Poznań, Uniwersytetu Poznanskiego 2, Poznań 61-614, Poland;

[3]Department of Condensed Matter Physics and Material Sciences, S. N. Bose National Centre for Basic Sciences, Block JD, Sector III, Salt Lake, Kolkata 700 106, India;

[4]Department of Electrical and Computer Engineering, Virginia Commonwealth University, Richmond, VA, 23284, USA.

[*]Email:
abarman@bose.res.in (AB)
sbandy@vcu.edu (SB)
bivasranaiitd@gmail.com (BR)

## Abstract

When magnets are fashioned into nanoscale elements, they exhibit a wide variety of phenomena replete with rich physics and the lure of tantalizing applications. In this topical review, we discuss some of these phenomena, especially those that have come to light recently, and highlight their potential applications. We emphasize what drives a phenomenon, what undergirds the dynamics of the system that exhibits the phenomenon, how the dynamics can be manipulated, and what specific features can be harnessed for technological advances. For the sake of balance, we point out both advantages and shortcomings of nanomagnet based devices and systems predicated on the phenomena we discuss. Where possible, we chart out paths for future investigations that can shed new light on an intriguing phenomenon and/or facilitate both traditional and non-traditional applications.

1. INTRODUCTION

Magnetism has a fascinating history that can be traced back to at least the sixth century BC. Current interest in magnetism stems from its potential to offer new insights into fascinating physical phenomena and seed new applications. The ability to produce magnets of nanometer dimensions has provided an additional impetus to study "nanomagnets", which, in their own right, have carved a niche in both physics and engineering. In this topical review, we will discuss many topics related to nanomagnets and their applications. However, by its nature, no review can be truly exhaustive and hence we must omit topics, which, while they may be important and interesting, are too distant from the other topics discussed here. We do this for the sake of focus and brevity.

Over the last several decades, research in nanomagnets has explored various types of magnetic structures such as thin films (single or multilayer); dots, i.e., isolated magnetic islands; antidots, i.e., holes in magnetic thin films; nanowires; nanoparticles; and magnetic two-dimensional (2D) materials, i.e., a monolayer of magnetic film. They have been harnessed for myriad applications, some of which have been commercialized. Perhaps the most important application of nanomagnets today is in digital data storage. Magnetic data storage industry was initially concerned with magnetic tapes and floppy discs. Because of poor performance in terms of speed, storage capacity, reliability, etc. they have now been replaced by magnetic hard disc drives (HDD) which are widely used in computers with a storage density of about few Terabytes/inch$^2$. Some of the major advantages of HDD include cheaper cost per digital bit for storage, faster access and shorter retrieval times compared to other storage devices. However, they still are much more sluggish in accessing data than non-magnetic solid-state drives (SSD). A later development in this field is magnetoresistive random access memory (MRAM), which is now used as the primary data storage units for computers. Magnetic memory is non-volatile, meaning that data stored in them is retained for a long time (more than a decade) after power to the unit is switched off. In contrast, electronic volatile memory, such as static RAM (SRAM), dynamic RAM (DRAM) require refresh cycles to retain stored data while non-volatile flash RAM foregoes the refresh cycles but relies on storing electrical charge which increases power consumption and reduces endurance.

2. MAGNETIC TUNNEL JUNCTION (MTJ) AS A MAGNETIC "SWITCH"

The "magnetic tunnel junction" (MTJ) is likely the most prolific magnetic device for digital applications and is the workhorse of many logic and memory systems. It is also the quintessential *spin-to-charge converter* which converts the spin degree of freedom of charge carriers (electrons and holes), embodied in the magnetization states of nanomagnets, into the charge degree of freedom, embodied in the electrical resistance of the device. The schematic of the MTJ is shown in figure 1. The reason it has been the centerpiece of magnetic digital circuits is that it can act as a binary switch with two resistance states – high and low – which can be utilized to encode binary bits 0 and 1 in much the same way as the two resistance states of a field effect transistor (ON and OFF) encode the binary bits 0 and 1 in modern digital electronic circuits. The difference is that the transistor is always a 3-terminal device, whereas the MTJ can be either a 2-terminal or a 3-terminal device. In the 3-termional configuration, the MTJ can even have a "gain" like the transistor, which it cannot have in the 2-terminal configuration [1].

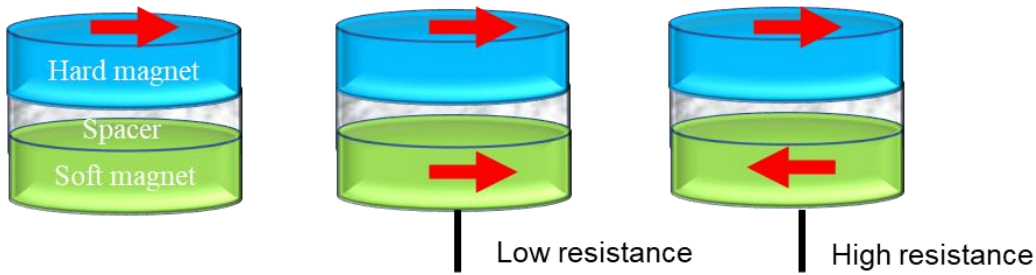

**Figure 1.** A magneto-tunneling junction, consisting of a ferromagnetic hard layer, an insulating spacer and a ferromagnetic soft layer. Electrons tunnel between the two ferromagnetic layers though the spacer and because of spin-dependent tunneling, the resistance of the device measured between the ferromagnetic layers is low when the latter have parallel magnetizations and high when they have antiparallel magnetizations. These two resistance states encode the binary bits 0 and 1.

The MTJ device has three layers – a "soft" ferromagnetic layer, an insulating spacer, and a "hard" ferromagnetic layer. Each layer is shaped like an elliptical disk and the magnetizations of the two ferromagnetic layers can point either to the left or to the right along the major axes, thereby making the magnetization orientation of either layer "bistable". These two possible orientations encode the binary bits 0 and 1. The magnetization of the last layer is made "hard" or stiff, usually by using various material combinations, so that it remains always fixed and pointing in one pre-determined direction along the major axis of the elliptical hard layer. The soft layer's magnetization can be flipped with an external agent. When the magnetizations of the hard and soft layers are mutually parallel, the electrical resistance of the MTJ (measured between the hard and the soft layers) is small because of spin-dependent tunneling of electrons from the hard layer to the soft, or vice versa. When the two magnetizations are anti-parallel, the resistance is high. Therefore, by measuring the resistance and knowing the magnetization of the hard layer (which is invariant), we can tell in which direction the magnetization of the soft layer is pointing and therefore what the stored bit is. Figure 1 illustrates this modality. Just as the high-conductance state of a field effect transistor used in electronic chips encodes one bit and the low-conductance state the other, similarly the high-conductance state of the MTJ encodes one bit and the low-conductance state the other. Thus, there is a one-to-one correspondence between a transistor switch and its magnetic counterpart, the MTJ. The difference is that the conductance state of the transistor is volatile while that of the MTJ is *non-volatile*. On the flip side, the conductance on/off ratio of a transistor may be ~$10^5$:1 at room temperature, while that of the MTJ would hardly exceed 7:1 at room temperature [2]. The low on/off ratio of the MTJ is a serious disadvantage in many digital applications (e.g. Boolean logic where it will make logic level restoration very challenging and the standby power dissipation enormous because of the high leakage), but may not be a show stopper in many non-Boolean applications (e.g. neuromorphic computing, Bayesian reasoning machines, probabilistic computing, etc.). There, the low on/off ratio is not debilitating, while the non-volatility may be a boon.

## 3. NANOMAGNETIC BOOLEAN LOGIC FOR DIGITAL COMPUTING

More than two decades ago, considerable excitement was generated around the notion of magnetic Boolean logic – fashioned out of nanomagnetic devices – because the *non-volatility* of magnetic switches portends certain advantages. Traditional logic chips built with complementary metal oxide semiconductor (CMOS) switches, or its various avatars such as tunnel field effect transistors (TFET), fin field effect transistors (FINFET), negative capacitance transistors (n-CFET), etc. have a shortcoming in that the transistor is volatile, meaning that if we turn off the power, information stored in the transistor (i.e. whether it was on or off) will be lost quickly. This is the primary reason why most computing architectures are of the *von-Neumann type* which consists of a processor, a memory and a "switch" that communicates between the processor and memory (see figure 2). The processor is made of fast but volatile elements while the memory has slow but non-volatile elements. The instruction sets are stored in the memory and are fetched to the processor via the switch when a program is executed. This is an inefficient approach since the back-and-forth communication between the processor and memory slows down the program execution (it is, in fact, responsible for the boot delay in a computer). If the processor could be made of non-volatile elements as well, then the instructions sets could have been stored in the processors in-situ which would eliminate the need for the switch and a partition between processor and memory. In fact, this is the driving force behind *non*-von-Neumann architectures and "processing in memory" (PIM) and "computing in memory" (CIM) approaches [3] which have gained traction because they can potentially speed up computation, improve reliability and reduce hardware overhead.

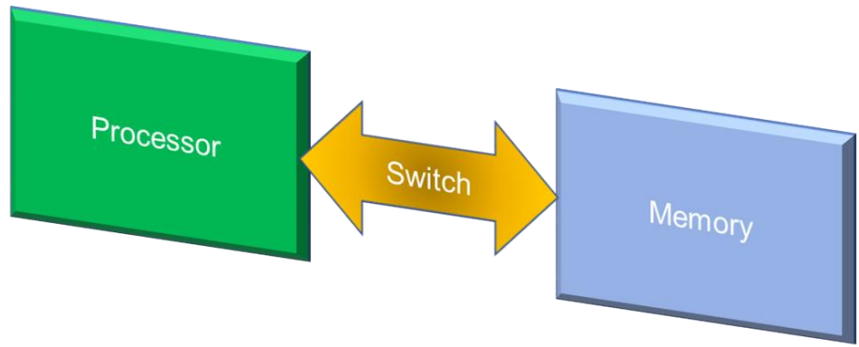

**Figure 2.** Basic von-Neumann architecture.

Clearly, transistor type volatile devices will not be ideal for non-von-Neumann architectures. This was the motivation to explore "magnetic" Boolean logic paradigms. Unfortunately, it turns out that magnetic devices are too error prone for Boolean logic. Logic has stringent requirements for reliability. This is primarily because errors in logic chips are *contagious*, unlike in memory. If the output bit of one Boolean logic gate is corrupted and that bit is fed as input to another gate, then the outputs of the latter are also corrupted. Thus, error propagates rapidly in logic. In contrast, errors in memory chips are not contagious. If one bit stored in a cell is corrupted, it does not infect any other bit. That is why error correction protocols (e.g. parity check) are relatively easy to implement in memory but very difficult to implement in logic.

There are two main genres of magnetic Boolean logic: dipole-coupled architectures, sometimes referred to as "magnetic quantum cellular automata" (a misnomer since it is classical Boolean logic and not quantum cellular automata) [4] and MTJ-based logic [5-7]. The former is too error-prone for Boolean logic as shown by many authors [8-10]. In fact, an experiment purporting to demonstrate a majority logic gate based on this paradigm reported an error-probability of 75% (only one out of four gates worked) [11]! In 1956, John von-Neumann had shown that the maximum tolerable error probability in a single majority logic gate is 0.0073 [12]. Hence, these constructs fall far short of the requirements for logic.

MTJ-based logic is more reliable than dipole-coupled logic ideas, but the lowest error probability reported for MTJ based logic at room temperature, to our knowledge, is $10^{-8}$ which falls far short of the $10^{-15}$ error probability exhibited by CMOS [13]. The lack of error-resilience is further exacerbated by defects and imperfections in nanomagnets [14], which is why it appears that magnetic devices may not be suitable for Boolean logic after all [15]. Fortunately, there are a host of non-Boolean computing applications for which magnetic devices may be eminently suitable. They include application-specific-integrated-circuits (ASICs) for image processing [16, 17], simulated annealing in energy minimization computing for solving combinatorial optimization problems [18], probabilistic computing with p-bits [19-22], computer vision [23, 24], Bayesian inference engines for belief networks and machine learning [25-28], restricted

Boltzmann machines [29], ternary content addressable memory [30] and non-volatile and reconfigurable equality bit comparators [31] for electronic locks and hardware countermeasures against cyberattacks.

## 4. NANOMAGNETIC MICROWAVE OSCILLATORS – MINIATURIZED MICROWAVE SOURCES

An intriguing analog application of nanomagnets is in microwave oscillators [32-36]. Oscillators are very important passive components of electronic circuits. In the quest for new types of energy efficient oscillators, spin torque nano-oscillators (STNOs) have emerged as a new class of nano-scale microwave oscillators with a wide range of operating frequencies and ultrahigh frequency modulation rates. Their manufacturing processes are compatible with CMOS technology. In a typical STNO, two magnetic layers (polarizer and free layer) separated by a nonmagnetic conducting or insulating layer is used, much like the MTJ. Passage of a charge current across the junction induces a spin-transfer torque on the spins in the free (or soft) layer which sets the latter in oscillation when the current density is high enough. This results in a time varying (oscillating) resistance of the device and hence a time varying voltage if the device is powered with a constant current source. The time varying voltage results in a time varying electric field at microwave frequencies that produces microwave emission.

These types of oscillators have very well-defined frequencies which can be tuned by a dc current passing through the junction. An important use of STNOs is in microwave assisted magnetic recording (MAMR). Microwave heating, generated by an STNO, is used to lower the coercivity of magnetic elements in recording media to facilitate the writing of data. Once the writing is complete, the microwave heating is turned off which raises the coercivity back up and protects the written data from being overwritten by external perturbations. There has also been some effort in devising neuromorphic computers based on coupled oscillators implemented with STNOs [37], although it has not yet become mainstream possibly because of the challenges encountered in synchronizing such oscillators.

## 5. MAGNONIC CRYSTALS FOR DIGITAL AND ANALOG APPLICATIONS

Another important application of nanomagnets is in magnonic crystals (MCs), which are basically periodic arrays of nanomagnets used for tailoring magnonic band structures, i.e., the frequency versus wavevector dispersion relations of spin waves (SWs). This is useful for developing various types of magnonic filters, attenuators, and logic devices, i.e., both digital and analog (signal processing) applications There are various ways to develop MCs. Static MCs can be created by periodically modulating the thickness, the saturation magnetization and the anisotropy of magnetic thin films, or the width and the shape of waveguides made of a magnetic film. Arrays of magnetic dots and antidots are some popular examples of static MCs. In static MCs, the magnonic band structures and band gaps can be tuned only by an external bias magnetic field. Dynamic MCs can be created on a magnetic thin film by a current-induced periodic Oersted field, laser induced periodic heating (which modifies the saturation magnetization), periodic arrays of stripe domain structures, and travelling surface acoustic wave-induced periodic strain (Doppler shift of SW frequency), etc. The magnonic bands of dynamic MCs can be reconfigured more easily than those of static MCs. Very recently, the development of dynamic MCs by voltage controlled magnetic anisotropy (VCMA) was predicted based on micromagnetic simulations [38] and it has now been demonstrated experimentally [39].

Exploration of MC in the third dimension has just commenced [40] where very complex and intriguing magnon dispersion and confinement effects are anticipated. The third dimension can control planar SW propagation, opening a new route for designing non-reciprocal SW spectra with chiral properties. Three-dimensional (3D) integrations will increase the density of elements and lead to scalable and configurable magnonic networks and waveguides for SWs. Moreover, curved surfaces in a waveguide can lead to a geometrical effective magnetic field that is proportional to the square of the ratio of the exchange length to the radius of curvature of the waveguide. Hence a stronger or multiple winding helical structure could lead to designer magnonic waveguides with desired properties [41]. It may also exhibit unconventional spin textures leading towards very complex and rich magnonic band structure.

Spin textures are nonuniform magnetic microstates, which are stable, resilient, but possess a high degree of tunability and scalability. They are very promising for energy-efficient, dynamically reconfigurable and reprogrammable components in magnonics. Various spin textures, namely, magnetic domain wall [42], vortex [43], onion [44], skyrmion [45], bubbles as well as quasi-uniform microstates such as S, C, leaf, flower states, etc. have been used to control SW propagation in magnonics [46]. Moreover, graded-index magnonics is emerging as a field laden with exciting possibilities, where engineered graded magnonic landscapes from nonuniform internal magnetic field has been proposed to create practical devices such as magnonic Luneburg lens [47].

Short wavelength magnonics is another niche area of magnonics, which promises high processing speed. Because of their isotropic character, they may allow for a 3D device architecture. This is advantageous for both high-density integration and other applications, e.g. neuronal networks. To this end, magnonic grating coupler (MGC) allows for interconversion at microwave frequencies and generation of excited magnons with wavelength $\lambda = 68$ nm with state-of-the-art microwave equipment [48]. Such microwave-to-magnon transducers consist of nanomagnets (grating elements) periodically arranged underneath a coplanar waveguide (CPW). The reciprocal lattice vectors of the nano-gratings, when added to the wave vector provided by the CPW, create short wavelength magnons following Bloch's theorem.

Antiferromagnetic magnonics offer very interesting physics and can implement ultrasmall, ultra-high frequency devices [49]. In antiferromagnets, the THz eigenfrequency stems from the very high sublattice exchange field. In addition to single antiferromagnetic structures like $MnF_2$ and $FeF_2$ possessing easy-axis anisotropy, or NiO with easy-plane anisotropy, there are AFs with more complex spin interactions, such as noncollinear antiferromagnets, for which the chirality of the spin structure results in topological effects [50]. This opens up additional pathways for influencing transport phenomena. Synthetic antiferromagnets and angular momentum compensated ferrimagnets also exhibit exotic SW dynamics. Magnon current in antiferromagnetic system can be significantly suppressed owing to the two oppositely polarized antiferromagnons resulting in counterflow of two species of magnon currents.

Magnon-based hybrid systems have evolved very rapidly during recent years owing to the rich physics associated with them, not to mention their potential application in coherent information processing [51]. They deal with strong coupling of magnonic excitations with diverse excitations for transformative applications in devices and circuits. In particular, magnon-photon, magnon-phonon and magnon-magnon couplings have attracted intense attention. The major challenge ahead is to achieve strong coupling between miniaturized magnets for on-chip integration of such devices.

The overriding goal of magnonics is to develop high-frequency and nanoscale devices and circuits. Magnonic data processing has potential advantages as discussed before. A realistic magnon computer will require magnonic logic circuits, interconnects, and magnonic memory. However, at present, a competitive replacement of all aspects of the state-of-the-art charge-based computing systems by its magnonic alternatives seems elusive. Hybrid SW—CMOS system with local SW islands embedded in the CMOS periphery has the potential, provided the signal conversion between magnonic and electric domains becomes efficient. Magnonic logic will have to satisfy all criteria for circuit design and for being integrated alongside CMOS in practical microelectronic applications. Moreover, energy-efficient scalable transducers and efficient periphery to interface between transducers and magnonic circuits with the larger CMOS segment of the chip will be required [52].

## 6. BIOMEDICAL APPLICATIONS OF NANOMAGNETS

Nanomagnets also have important applications in biomedicine such as in magnetic resonance imaging (MRI), magnetic hyperthermia and targeted drug delivery. MRI is a non-invasive imaging technology that produces three dimensional detailed anatomical images. It is often used for disease detection, diagnosis, and treatment monitoring [53]. Magnetic nanoparticles are used as contrast agents for MRI. Although, paramagnetic contrast agents have been used for a long time, superparamagnetic iron oxide nanoparticles (SPIOs) appear to be superior. Unlike paramagnetic contrast agents, SPIOs can be functionalized and their sizes can be tailored in order to adapt to various kinds of soft tissues [54]. Magnetic nanoparticles based targeted drug/cargo delivery are emerging treatment methods which are attracting the attention of many researchers for treating different types of tumors, cancers and artery diseases such as atherosclerosis [55-57]. Nanoparticles (NPs) play an important role in the controlled release of drugs in a target position. In particular, nontoxic superparamagnetic NPs with functionalized surface coatings can conjugate chemotherapeutic drugs or be used to target ligands/proteins, making them useful for targeted therapy, drug delivery and also for MRI.

## 7. FABRICATION OF NANOMAGNETS

Since all experimental research in nanomagnets must be preceded by fabrication of nanomagnetic samples, we dedicate this section to a concise review of the various synthesis techniques that have been employed to produce arrays of nanomagnets. Fabrication of relatively defect-free nanomagnets with narrow size dispersion, laid out in an ordered array over a macroscopic length scale, is always challenging. Even growth of high quality ultrathin magnetic films with smooth and defect free interfaces is difficult. Various types of thin film deposition techniques such as DC and RF magnetron sputtering, electron beam evaporation, chemical vapor deposition (CVD) [low-pressure CVD, plasma-enhanced CVD], atomic layer deposition (ALD), molecular beam epitaxy (MBE), pulsed laser deposition (PLD), etc. have been employed to produce uniform thin films down to one or few monolayers thickness. For fabricating one- or quasi-zero-dimensional nanostructures and their arrays, a number of 'bottom-up' and 'top-down' approaches have been developed including some methods that involve a combination of the two.

One of the widely used 'bottom-up' approaches for synthesizing monodisperse nanocrystals of uniform shape is solution phase colloidal chemistry [58, 59]. There are many colloidal chemistry based methods such as reduction, nonhydrolytic sol-gel, thermal decomposition processes [60], etc. Various types of reductants are used in the reduction methods to synthesize metallic nanoparticles. The nonhydrolytic sol-gel process is mainly used to synthesize metal oxide nanoparticles, whereas in thermal decomposition methods, the decomposition of organometallic compounds is performed in hot surfactant solutions to synthesize nanoparticles of various materials. Reverse micelle methods are another potential route for synthesizing various kinds of nanocrystals [61]. The self-assembly of magnetic nanocrystals (e.g. magnetite crystal chains) by biological organisms (e.g. magnetotactic bacterium) has also been demonstrated, notably by Zingsem *et al.* [62]. However, it is usually difficult to control the shape of a nanostructure in all three dimensions using the bottom-up approaches. Okuda *et al.* was able to ameliorate this challenge by combining a top-down approach like focused ion beam (FIB) milling with self-assembly [63].

In 'top-down' approaches, various types of conventional lithographic techniques such as photolithography [64], deep ultraviolet (DUV) lithography [65], holographic lithography [66], electron beam lithography (EBL) [67], X-ray lithography [68], ion beam lithography [69], nanoimprint lithography [70], etc. are used to delineate nanomagnet arrays on a film or substrate. The lithography techniques utilize an optical, electron or ion beam to expose a resist material deposited on the substrate or film to draw a pattern on it. In optical- or photo-lithography, a photoresist that has been spun onto the substrate or film, is first illuminated through a mask and the exposed regions are then dissolved away in a developer after the photoresist has been baked at elevated temperatures. This delineates a nanopattern on the substrate or film consisting of "windows" in the resist. Magnetic materials are then deposited on the patterned substrate using metal evaporation or other techniques, followed by lift-off, which leaves the magnetic materials selectively on the patterned regions (windows), thereby forming an array of nanomagnets of the size and shape defined by the original mask.

In photolithography, the feature sizes are limited by the diffraction of light and hence cannot be much smaller than the light wavelength. The advantage of photolithography is that multiple nanomagnets forming an arbitrary pattern can all be created in parallel (with a single exposure) which is very useful for commercial production because it has rapid throughput. DUV offers higher resolution than conventional optical lithography which uses visible light due to the smaller wavelength of ultraviolet light [71]. The shadow mask deposition is a very useful DUV technique for fabricating periodic arrays of binary magnetic nanostructures [72]. Holographic lithography is based on the interference of multiple laser beams with a single light exposure [66]. The major advantages of this technique over other optical techniques are it is mask-free, inherently low cost and compatible with large scale preparation of periodic nanostructures. Moreover, there are many optical parameters, such as beam intensity, polarization, and incident angles, which can be adjusted for fabricating periodic arrays of nanostructures of almost any lattice symmetry. X-ray lithography, where soft x-rays are used in lieu of visible or ultraviolet light, allows feature sizes down to sub-30 nm. Conformal or interferometric x-ray lithography is a very convenient tool for fabricating periodic arrays of nanomagnets because it is mask-less and has a high throughput [73].

EBL has resolution in sub-10 nm length scale. However, this technique has low throughput because it is a direct write technique, where patterns are exposed one at a time, serially and not in parallel. Often EBL is used for photomask fabrication and fabrication of small nanostructure arrays (containing few nanomagnets) for research purposes. Massive EBL equipment containing multiple electron beam columns have been employed for mass production of nanostructures, but they are exorbitantly expensive and hence impractical in many cases. In conventional EBL, selected areas of an e-beam resist like PMMA are exposed to an electron beam which is guided by computer software. The resist is then developed to form windows in the exposed regions. Magnetic materials are then evaporated on the patterned surface, followed by lift-off to create the desired nanomagnet pattern. Figure 3 shows a scanning electron micrograph of an array of nanomagnet fabricated in our laboratory using this method.

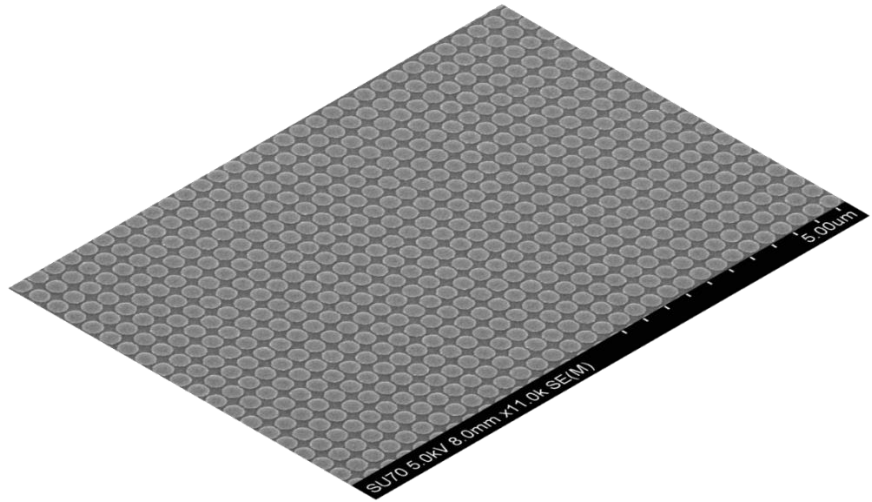

**Figure 3.** Scanning electron micrograph of an array of elliptical cobalt nanomagnets delineated on a piezoelectric substrate using EBL and e-beam evaporation of cobalt. The major axis dimension is 360 nm, the minor axis dimension is 330 nm and the thickness is 6 nm. The edge-to-edge separation between the nanomagnets along the major axis is 45 nm and along the minor axis is 65 nm. Adapted from [74] with permission from Wiley.

In FIB lithography, ion beams (e.g., Ga ions) are used to write the desired pattern directly on the resist. It, too, is a direct write technique where patterns are written serially one after the other, limiting its throughput. It is analogous to EBL, with the difference being that the magnetic lenses used to focus electron beams are replaced by electrostatic lenses owing to the heavier ion masses. The disadvantages of FIB are the slower writing speed compared to EBL and significant sample damage caused by high energy ions. However, several key advantages exist with FIB, such as highly localized doping, controlled damage (intentionally induced for device isolation), mixing, micromachining and ion-induced deposition. Scanning probe lithography (SPL) is a mask-less and direct-write approach which can get rid of the diffraction limit and thus can achieve resolutions even below 10 nm [75]. Its disadvantage is that it has a very slow throughput and hence is impractical for delineating large arrays.

Nanoimprint lithography is different from conventional lithography and has the advantage of being able to fabricate many nanomagnets simultaneously, like photolithography. At first, a pattern is created on a template by EBL consisting of raised areas like mesas. Then, a separate substrate is coated with a resist and the patterned template is pressed onto the resist forming a pattern on the substrate like in a stamping process [70]. This is a cost-effective, single-exposure technique and the feature size can be down to 5 nm or smaller.

Selective chemical etching [76], reactive ion etching, and ion milling are other useful methods to prepare periodic arrays of nanostructures on magnetic thin films with varying thickness. Two-photon lithography (TPL) is another potential method for fabricating 3D micro/nanostructures [77]. TPL is a photochemical process initiated by the femtosecond laser beam tightly focused into the volume of the photosensitive resins by a high-numerical-aperture (NA) objective. By precisely scanning the laser focus throughout the material (or by moving the resist relatively to the fixed laser spot) a 3D extended volume can be affected, within the spatial limitations of the moving stage and thus 3D micro/nanostructures can be created.

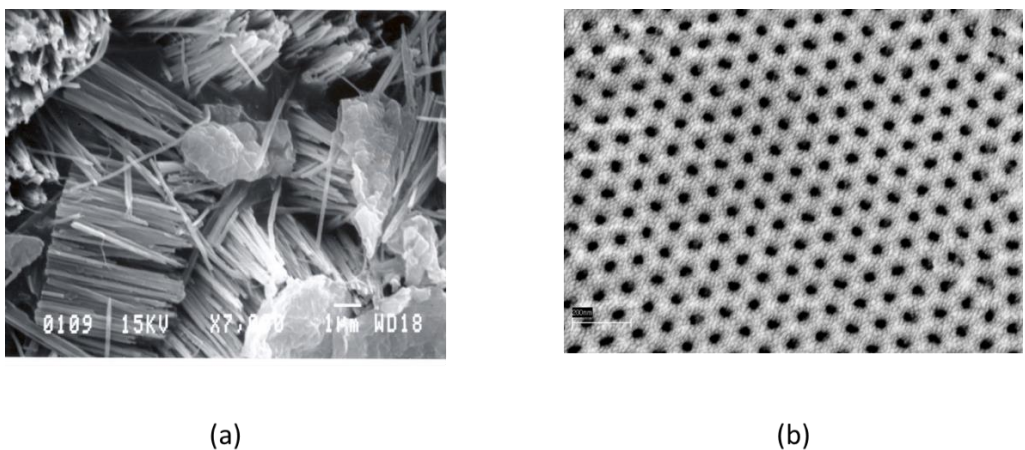

(a)          (b)

**Figure 4.** (a) Scanning electron micrograph of nanowires synthesized in our lab by electrodepositing cobalt within 50-nm diameter nanopores in an anodic alumina membrane formed by anodizing aluminium in oxalic acid. (b) Scanning electron micrograph of the anodic alumina membrane.

Other techniques for preparing ordered nanostructures are self-assembly mechanism like heterogeneous nucleation of magnetic atoms on metallic surfaces [78], seeded growth [79], and nano-template fabrication technique using templates such as di-block copolymers [80], anodized alumina membranes [81] and nuclear-track etched membranes [82]. Figure 4 shows scanning electron micrographs of cobalt nanowires of 50 nm diameter produced in our laboratory by electrodepositing cobalt within nanoporous anodic alumina membranes formed by anodizing a 99.999% pure aluminium foil in 0.3 M oxalic acid at room temperature for 15 minutes. The pore diameter in the membrane is roughly 50 nm. The cobalt is deposited from a $CoSO_4$ aqueous solution using the aluminum foil as the cathode and a silver counter-electrode. The deposited cobalt nanowires are then released from their alumina host by dissolving out the host in hot chromic/phosphoric acid at $100^0$ C. The clumped wires can be separated, if needed, by ultrasonication in water or ethanol. These magnetic nanowires exhibit high coercivities and other attractive magnetic properties [83, 84].

## 8. MAGNETIZATION REVERSAL IN THIN FILMS AND NANOMAGNETS

Most applications of nanomagnets involve rotating the magnetization orientation of one or more nanomagnets. A single-domain nanomagnet – a nanomagnet of sufficiently small size will be a single-domain nanomagnet – has a single ferromagnetic domain in which all the spins point in the same direction because of strong exchange interaction between them. That direction is the direction of its magnetization. If we rotate the magnetization, then all the spins in the single-domain nanomagnet rotate together in unison, so the magnetization acts like a giant classical spin [85]. This is called coherent rotation.

Ideally, a single-domain nanomagnet, shaped like an elliptical disk will have two stable magnetization orientations along the ellipse's major axis (referred to as the "easy axis"), either pointing to the left or to the right, as shown in figure 5.

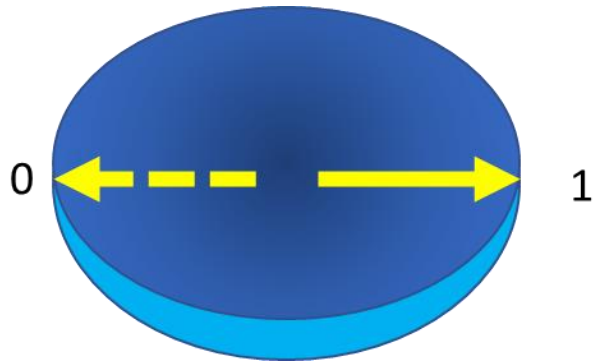

**Figure 5.** The magnetization of a single domain ferromagnet shaped like an elliptical disk will point along the major axis, either to the left or to the right when the magnet is in equilibrium. These are the only two stable orientations which are energetically degenerate and encode the binary bits 0 and 1. In digital applications (logic or memory), switching the bit from 0 to 1, or vice versa, entails flipping the magnetization, referred to as "magnetization reversal".

These two orientations encode the binary bits 0 and 1. When the nanomagnet is used in digital computation or data storage, as well as in many other applications, one would need to flip the bit and hence rotate the magnetization though an angle of $180^0$, a process known as "magnetic reversal".

There are many important parameters associated with magnetization reversal, which we discuss next.

### 8.1 Switching timescale

The switching timescale determines how quickly the magnetization of a nanomagnet can be reversed and is hence very important for the development of memory and storage devices. A nanomagnet whose dimensions are larger than the critical dimensions for single domain formation will contain multiple domains separated by domain walls. Within each domain, all the spins are aligned in the same direction, but within different domains, they are aligned in different directions. When the magnetization rotates to a particular direction, domains that are already aligned in that direction, or close to it, grow in size at the expense of other domains. This obviously involves domain wall motion. In magnetic thin films and dots, this mode of magnetization reversal occurs in nanosecond time scale. For smaller magnetic dots with dimension smaller than single domain size the magnetization reversal generally occurs through the coherent or quasi-coherent rotation of magnetization in the sub-nanosecond to few nanoseconds time scale. By solving the Landau-Lifshitz-Gilbert (LLG) equation which governs the time evolution of magnetization, Kikuchi *et al.* found in 1956 that the switching time in a thin sheet of magnetic film can be of the order of a nanosecond. However, the Gilbert damping constant that is used as a parameter in the LLG equation significantly affects the switching time. If the damping constant is either larger or smaller than the critical value for fastest switching, the reversal time increases. For larger damping, the magnetization rotates slower because the damping acts like friction, whereas for smaller damping, the magnetization oscillates before settling to its equilibrium position, resulting in longer switching time [86]. Ababei *et al.* has, however, reported anomalous damping dependence of the switching time in Fe/FePt bilayer recording media [87]. They found that the magnetization reversal time increases from sub-ns to few ns time with increasing damping when a bias magnetic field, applied to switch the magnetization to the desired direction, is set close to the coercive field, whereas the magnetization reversal time decreases below sub-ns time scale with increasing damping when the magnetic field is set much larger than the coercive field [87]. Matsuzaki *et al.* performed micromagnetic simulations to investigate the magnetization reversal behaviour and estimated the reversal time for a hard/soft magnetic composite pillar array. They found that a switching time of less than 0.33 ns can be obtained by appropriately picking the damping constant and the exchange constant between the hard and the soft magnetic layers [88]. Choi *et al.* demonstrated magnetization reversal in a 10 μm × 2 μm Py strip through domain wall rotation [89, 90]. Fast reversal was achieved by manipulating the bias magnetic field and a pulsed magnetic field. It was observed that the application of a transverse magnetic field leads to faster reversal (1.2 ns) due to domain wall motion, while the absence of it leads to slower reversal (5 ns) by domain wall nucleation. Worledge *et al.* investigated current induced magnetization switching behaviour in Ta/CoFeB/MgO based magnetic tunnel junctions and found 1 ns switching time. Interestingly, the switching time decreases with the increase of current density across the junction [91]. In similar systems, Grezes *et al.* reported a switching time of about 0.6 ns, which was found to be insensitive to the diameter of the tunnel junction in the range of 50 nm to 100 nm [92].

A switching time of 1 ns will restrict the clock speed of circuits containing magnetic elements to 1 GHz or less, which would hamstring magnetic computing of any kind. Precessional switching has now emerged as a promising route to reducing the switching time to several tens or hundreds of ps. In magnetic field induced switching, the switching speed is determined by the precession frequency of the magnetization around the bias magnetic field applied to switch the magnetization. Spin transfer torque (STT) is another method of switching a nanomagnet where a spin polarized current with the electron spins aligned in the desired direction of magnetization is passed though the nanomagnet. The electrons transfer their angular momenta to the spins in the nanomagnet, making them rotate to the desired direction and accomplishing the switching [93]. Micromagnetic simulations have shown that STT switching occurs above a threshold pulse current, and it can occur in faster than 50 ps time [94]. Femtosecond pulse laser induced all-optical switching is another potential approach to achieve switching time in the femtosecond timescale. Lu *et al.* investigated the roles of heating and helicity in ultrafast laser induced all-optical switching in TbFeCo film. They observed that the evolution of ultrafast switching occurs over different time scales that depend upon the laser heating and helicity. They found a magnetization reversal time of 460 fs [95].

**8.2 Switching field distribution in magnetic nanoparticle array**

The switching field (the threshold value for magnetization rotation to the desired direction) in magnetic thin films and magnetic dots depends upon the values of the magnetic anisotropies and the angles subtended by the field with the anisotropy axes. Depending upon the size, shape and aspect ratio (height/width), the magnetization reversal of magnetic dots may either occur through coherent rotation of magnetization or through the formation of various domain structure including vortices. This also leads to a significant dependence of the magnetization switching field on the shape, size and aspect ratio of the dots. Due to the unavoidable distribution in the shape and size of nanodots and random defects introduced during the nanofabrication, a large distribution of the switching field is generally observed for even non-interacting arrays of identical magnetic dots. However, when magnetic dots are arranged in an array very close to each other, the magnetostatic interaction among them significantly modifies their internal magnetic configuration leading to a collective reversal of magnetization. Figure 6(e) shows the switching field as a function of dot size in some magnetic nanodot arrays.

Gomez *et al.* investigated the magnetic characteristics of cobalt islands with dimensions of $0.23 \times 0.43 \times 0.02$ μm$^3$ using magnetic force microscopy (MFM) [96]. The islands were noninteracting and showed a wide variety of single and multidomain configurations due to the variation of the magnetic easy axes in the magnetic islands. A wide distribution in the switching field was observed and it caused the hysteresis loop to deviate considerably from a perfect square shape.

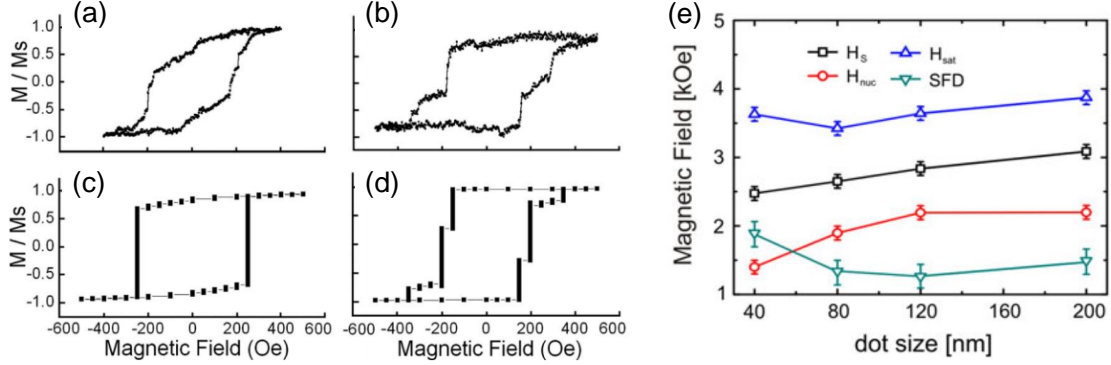

**Figure 6.** (a – b) Experimentally measured hysteresis loops of triangular shaped dot arrays. (c – d) The simulated hysteresis loops of the triangular shaped dot arrays [97]. Reproduced from [97] with permission of the American Institute of Physics. (e) Extracted values of nucleation field ($H_{nuc}$), switching field ($H_S$), saturation field ($H_{sat}$), and switching field distribution (SFD) for arrays of Co/Pt square magnetic dots [98]. Reproduced from [98] with permission of Springer.

In continuous thin films, magnetic reversal occurs in two different ways. In thin films with high concentration of structural defects, the reversal is dominated by the nucleation of the domain walls at the defect pinning sites which hinders the propagation of domain walls. This gives rise to a rounded hysteresis loop. The patterning of the film into dots does not really affect the shape of the loop. For a defect free high-quality sample, the reversal occurs through the fast propagation of domain walls which gives rise to a perfectly square hysteresis loop. After patterning the film into magnetic dots, an enhancement of the coercive field with the reduction of dot diameter is generally observed, which leads to a rounded hysteresis loop. Bartenlian *et al.* has shown that by patterning a continuous film into dots, the domain wall propagation is blocked at the edges of the dots, which leads to a wide distribution of the nucleation field inside the array [99]. Lee *et al.* showed that the switching field for 20 µm magnetic dots depends upon whether the magnetic field is applied along the easy axis or the hard axis. Interestingly, the coercive field was found to be significantly affected by the amplitude and the frequency of the applied Oersted magnetic field for switching [100]. Boukari *et al.* also showed a significant variation of the coercive field in $CoPt_3$ magnetic dots with diameter varying from 0.2 µm to 1 µm. It was also shown that the dynamic switching field in arrays of magnetic dots is strongly affected by the field sweeping rate [101]. Niu *et al.* performed magneto-optical Kerr effect (MOKE) measurement to investigate the magnetization switching mechanism in arrays of triangular shaped 30 nm FeNi dots [97]. The coercive field of the dot array was found to be significantly higher than that in an FeNi film and the hysteresis loop was also distorted from the almost perfect square hysteresis loop found for the film. Both the shape of the hysteresis loop and the coercive field for a dot array vary significantly with the direction of the applied in-plane magnetic field due to the presence of magnetic shape anisotropy (figure 6 (a-d)). Krone *et al.* also investigated magnetization reversal in arrays of rectangular magnetic dots made of Co/Pt multilayers with the diagonal varying from 200 nm to 40 nm [98]. Magnetic force microscope (MFM) studies revealed that the 200 nm sized dots are in a multi-domain state and magnetization reversal in them occurs through domain wall nucleation and propagation, while the 40 nm dots were found to be in the single-domain state. It was observed that the patterning process severely damaged the magnetic layer resulting in a decrease of the switching field and an increase in the switching field distribution (SFD) of the dot array with decreasing dot size (figure 6 (e)). By performing micromagnetic simulations, Yan *et al.* investigated magnetization reversal in Co/insulator/Fe tri-layer dots with asymmetric shape, where shape anisotropy was induced by cutting a section of the circular dot [102]. It was found that with the variation of the induced shape anisotropy, the reversal field varies significantly owing to the modification of the domain formation process during magnetization reversal. Wiele *et al.* showed that the magnetization reversal mechanism in arrays of identical magnetic dots is controlled by the competition between nearest-neighbour magnetostatic interactions among the dots and global configurational anisotropy of the whole array [103]. This leads to a distribution of the switching field among the dots and the formation of magnetic domain structures in the array during reversal. It was proposed that if one aims to obtain uniform magnetization state distributions throughout the array, then arrays with a circular global shape should be considered to suppress the effect of global configurational anisotropy. Additionally, thicker and/or wider dots should be chosen to enhance the effect of nearest-neighbour magnetostatic interactions since the latter varies as the square of the dot volume. Weekes *et al.* investigated magnetization reversal in hexagonally arranged circular Co dots with 360 nm diameter. A six-fold anisotropy in the coercive field was observed owing to the structural symmetry of the array when the bias magnetic field was oriented in-plane [104].

## 8.3 Thermally assisted magnetization switching

Thermally assisted switching (TAS) or recording exploits the temperature dependence of the anisotropy of magnetic materials to switch the magnetization and thus record data. In this method, the magnetic recording media is heated up temporarily to reduce the magnetic anisotropy and hence the switching field. The switching field is turned on to write the data and the media is then quickly cooled back to the ambient temperature to store the data. This was first demonstrated by Ruigrok *et al*, who called this "hybrid recording" [105]. A laser beam was used for heating the magnetic material. In conventional or in first generation MRAM devices, writing was performed by the application of magnetic fields generated by currents circulating in cross point lines (Figure 7(a)), which allows one to select the cell located at the intersection of the lines. However, the first generation of MRAMs had several limitations. First, the switching fields significantly increased as the element size decreased, which enhanced power consumption and limited the element size to about 100 nm. Second, the write selectivity decreased as the switching field distribution increased because of the presence of geometrical defects. Third, the long-term stability of the information deteriorated owing to smaller energy barrier to thermal perturbation in smaller elements. These problems were solved in the second generation of MRAM (mostly employing MTJs) by thermally assisting the magnetization switching of the free layer. This indeed has several advantages. First, the write power is reduced and parallel addressing of cells becomes possible. Second, the selection errors for writing are minimized as the selection is mostly temperature-driven. Most importantly, the element size can be reduced by increasing the internal energy barrier with the use of materials with higher anisotropy. Daughton *et al.* proposed writing in high resistance junctions by circulating a current in the writing lines [106]. As an alternative approach, Prejbeanu *et al.* proposed local heating by sending a current through the MTJ [107], which enabled selectivity by turning on the series transistor of the selected element during the writing procedure (Figure 7(b)). In this case, a strong temperature dependence of the writing field is achieved by the exchange coupling of storage layer with an antiferromagnetic layer in a tunnel junction. The writing procedure is accomplished by heating the junction slightly above the blocking temperature of the storage layer and subsequently cooling down under a current-generated magnetic field. This is the basis of heat assisted magnetic recording (HAMR). For electrical heating, the heating process can be made more energy-efficient by optimizing the junction area and the voltage pulse width. Prejbeanu *et al.* proposed that if two thermal barrier layers, e.g. two low thermal conductivity materials, are inserted at both ends of the MTJ layer stack, in between the junction and the electrical leads, then the heating process can be further optimized and the power consumption can be significantly improved [108]. Later, some reports were also published on thermally assisted switching in granular perpendicular media [109], perpendicularly magnetized single layers [110], tunnel junctions [111] and ferrimagnetic garnets [112]. Taniguchi and Imamura theoretically proposed STT induced magnetization switching in synthetic free layers assisted by thermal energy [113]. In a pioneering work, Pushp *et al.* demonstrated a novel method for thermally assisted switching [114]. There, a large temperature gradient was generated across an ultrathin MgO tunnel barrier in an MTJ (Figure 7(c)). It was found that a temperature difference of just a few Kelvin across an ultrathin (~1 nm) MgO tunnel barrier was able to generate giant spin currents, sufficient to significantly influence the switching of the MTJ. The thermally generated spin torque originated from the asymmetry of the tunneling conductance across the MTJ.

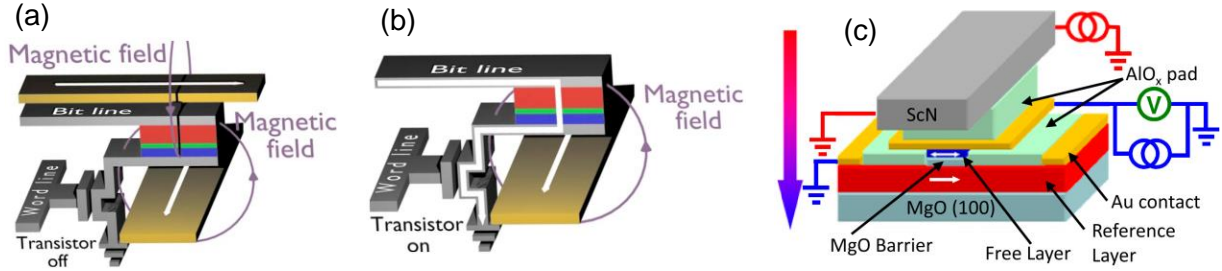

**Figure 7.** (a) Writing procedure for a conventional MRAM cell, using a cross point architecture. (b) Thermally assisted writing procedure in MRAM cell. Reproduced from [115] with permission of American Institute of Physics. (c) Schematic diagram shows various components of a device structure used for thermal spin-torque assisted magnetization switching. Reproduced from [114] with permission of United States National Academy of Sciences.

## 8.4 Microwave assisted switching of magnetization

Microwave assisted switching (MAS) is another type of energy assisted switching of magnetization similar to thermally assisted switching (TAS). In this case, switching is accomplished through magnetization precession induced by a microwave frequency magnetic field whose frequency is resonant with the natural resonance frequency of the ferromagnetic nanoelements. In MAS, large angle magnetization precession is excited so that the dynamics are in the non-linear regime unlike conventional ferromagnetic resonance (FMR) experiments. In the pioneering work of Thirion *et al.*, the authors experimentally demonstrated the efficiency of this method in a single Co nanoparticle with 20 nm diameter mounted on a micro superconducting quantum interference device ($\mu$-SQUID) [116]. The switching field was found to be significantly reduced by applying a tiny radio frequency (RF) pulse field amplitude. Following this, a number of experimental reports were published on MAS of soft [117-119] and hard magnetic [120-122] materials in the form of thin films and nanodots. It was found that the frequency and power of the microwave have major effects on the switching phenomenon. Nozaki *et al.* studied MAS of micrometer sized Co and NiFe particles. They found that the switching field of NiFe particles can be significantly decreased by MAS, whereas MAS was not effective for Co particles due to the large dispersion in crystalline anisotropy [117]. By performing LLG based calculation, Okamoto *et al.* showed that MAS takes place through very complicated precessional motion of the magnetization which is initiated with steady precession of magnetization followed by unstable motion which ends with abrupt irreversible switching [123]. By studying polarization-dependent MAS, it was found that linearly polarized-MAS (LP-MAS) is preferred over circularly polarized-MAS (CP-MAS). Although CP microwave can be absorbed by magnetization more effectively than LP microwave, the helicity of a CP microwave must match that of the magnetization precession in order to realize CP-MAS. Hence, the helicity of CP wave must be switched with the polarity of the head field, which makes CP-MAS more challenging than LP-MAS.

Granular perpendicular medium consisting of weakly exchange-coupled grains with good thermal stability is preferred for high storage densities beyond 1 Tb per square inch. Apart from in-plane magnetized media, MAS has also been demonstrated in perpendicular magnetic granular thin films [124-126]. These reports vindicate the potential of MAS for future recording technology of HDDs. In perpendicular magnetized films, a reduction of the switching field up to 75% was reported [121, 127, 128]. Okamoto *et al.* experimentally investigated MAS in a perpendicularly magnetized Co/Pt multilayer film [128]. They found that switching is initiated by nucleation of a reversed domain followed by its gradual expansion via domain wall displacement. A significant reduction of nucleation fields was found at three microwave frequencies: the low frequency Kittel mode and two unidentified higher frequency modes. Interestingly, 67% reduction of the nucleation field was observed for the highest frequency microwave frequency, which alludes to the existence of excitation modes for MAS that are more effective than the low frequency Kittel mode. However, it is quite challenging to generate large amplitude RF field required for MAS in the HDD system. Zhu *et al.* posited that the spin torque nano-oscillator (STNO) can be used to generate very localized RF magnetic fields [129, 130]. Since the structure of a STNO is very similar to that of the magnetoresistive read head of present-day HDD, it can be easily incorporated into a magnetic recording head to carry out microwave assisted magnetic recording (MAMR). Therefore, only a minimal modification of current HDD technology is required unlike in the case of some other proposed technologies such as TAS as shown in figure 8. Recently, Suto *et al.* have experimentally studied MAS of a perpendicularly magnetized nanomagnet by applying a microwave magnetic field with a time-varying frequency [131]. It was found that a larger MAS effect than that in a constant-frequency microwave field is obtained as the microwave frequency follows the nonlinear reduction of the resonance frequency to induce larger magnetization excitation.

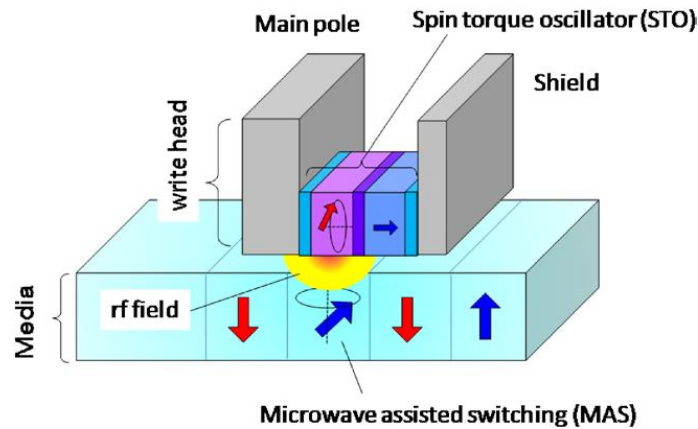

**Figure 8.** Schematic illustration of microwave assisted magnetic recording (MAMR) assembly consisting of microwave assisted switching (MAS) arrangement combined with a spin torque nano oscillator (STNO) for microwave generation. Reproduced from [132] with permission of the Institute of Physics.

## 8.5 Current controlled magnetization switching

A popular technique of switching the magnetization of a nanomagnet to any desired direction is passing a spin-polarized current through it which delivers a STT on the magnetization [133, 134]. The spins in the spin-polarized current transfer their angular momenta to the resident spins in the magnet and reorient them in the direction of spin polarization of the current. This is the basis of STT.

In figure 9, we show an MTJ to explain how the soft layer's magnetization can be switched electrically to make it either parallel or antiparallel to that of the hard layer. The spin-polarized current flows in a direction perpendicular to the heterointerfaces by tunneling through the spacer. Let us assume that the MTJ is initially in the antiparallel configuration (high resistance state) and we wish to switch it to the parallel configuration (low resistance state). If we connect the negative terminal of a battery to the hard layer and the positive terminal to the soft layer, then the hard layer will inject its majority spin electrons (spins polarized parallel to the hard layer's magnetization) into the soft layer. Of course, some minority spins will also be injected, but their population is much smaller than that of majority spins. This constitutes a spin-polarized current injected into the soft layer. The injected spins will transfer their angular momenta to the resident electrons in the soft layer and if enough spins are injected (i.e., the current exceeds a threshold value) the angular momentum transfer will align the resident spins in the direction of the hard layer's magnetization, thereby making the two magnetizations mutually parallel (figure 9(a)). The hard layer acts as the spin polarizer and generates the spin-polarized current that switches the soft layer.

If we wish to switch an MTJ from the low resistance (parallel) state to the high resistance (antiparallel) state, then we simply reverse the polarity of the battery. Now, the soft layer will inject electrons into the hard layer and generate a spin-polarized current, but only those electrons whose spins are aligned parallel to the hard layer's magnetization will be preferentially transmitted by the hard layer which acts as a spin analyzer or filter. Therefore, the soft layer will more successfully inject those spins that are parallel to the hard layer's magnetization. Those spins were originally the majority spins in the soft layer since its initial magnetization was parallel to that of the hard layer. As these spins exit the soft layer, their population is depleted so that ultimately spins that are antiparallel to the hard layer's magnetization become majority spins in the soft layer. This flips the magnetization of the soft layer, making the soft layer's magnetization antiparallel to that of the hard layer's (figure 9(b)). Thus, we can write either bit 0 or bit 1 into the memory by choosing the polarity of the battery.

This method of switching magnetization with a spin-polarized current is not particularly energy-efficient and recent estimates claim the energy dissipation to be about 100 fJ per writing event [135].

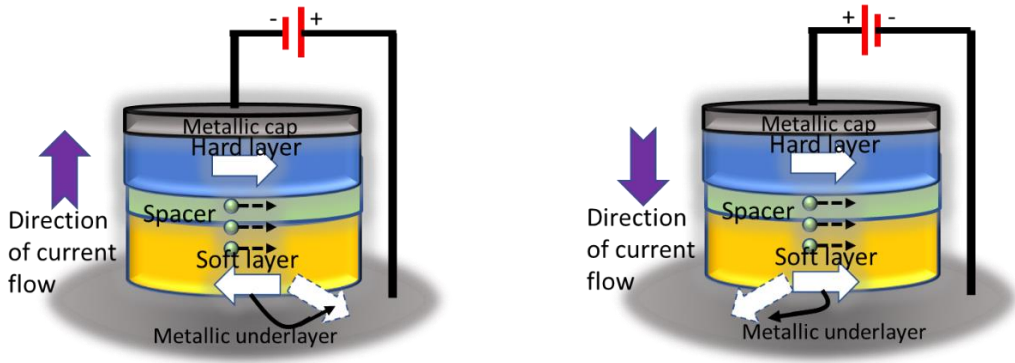

**Figure 9.** Spin transfer torque (STT) switching of a magnetic tunnel junction (MTJ). (a) antiparallel to parallel and (b) parallel to antiparallel.

A recent idea to reduce the threshold current (and hence power dissipation) needed to switch a nanomagnet with STT incorporates the phenomenon of giant spin Hall effect (GSHE) [136-138] which is elucidated in figure 10.

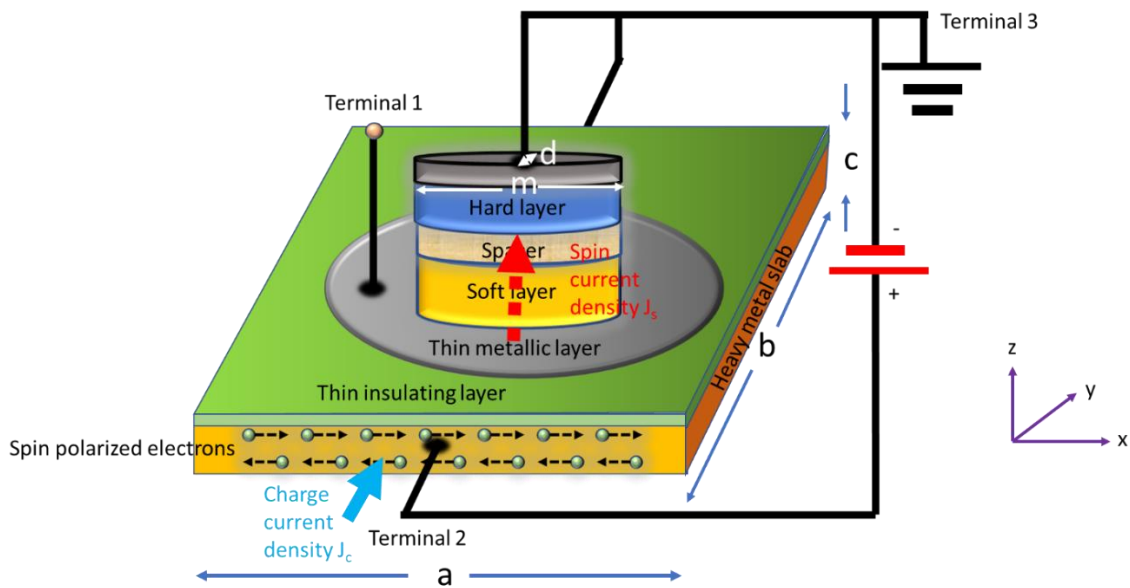

**Figure 10.** Switching with the aid of the giant spin Hall effect or spin-orbit-torque (SOT) switching. Note that a memory cell based on SOT switching or switching with the aid of the giant spin Hall effect is at least a three-terminal device, whereas a memory cell based on STT can be a two-terminal device. A three-terminal device has a much larger footprint than a two-terminal device. This has stymied the application of SOT switching in memory devices where bit density is usually the paramount consideration. The read and write terminals (or, equivalently, the read and write paths) are separate and this reduces such nuisance effects as "read-disturb" which can corrupt a bit stored in the memory cell.

Consider a two-dimensional slab of material shown in figure 10 that has strong spin–orbit interaction. It is usually a heavy metal like Pt or β-Ta. A 'charge' current of density $J_c$ (with no spin polarization) is passed through it flowing in the y-direction. The injected electrons suffer *spin-dependent* scattering as they travel through the slab (because of the spin–orbit interaction), which deflects -x-polarized spins to the bottom edge of the slab and +x-polarized spins to the top edge, causing a spin imbalance (preponderance of -x-polarized spins in the bottom surface of the slab and preponderance of +x-polarized spins in the top surface) that drives a spin current of density $J_s$ in the z-direction into the MTJ. This spin (diffusion) current flows through the MTJ and rotates the magnetization of the soft layer to the desired direction by exerting a spin torque on the resident electrons in the same manner as STT. If we wish to rotate to the opposite direction, we will simply reverse the polarity of the current which will reverse the polarization of the spin current and switch the magnetization to the opposite direction.

There may be other sources for the spin torque in the device in figure 10 such as the Rashba–Edelstein effect in the ferromagnet itself [139, 140]. In a ferromagnet with Rashba spin–orbit interaction [141], passage of a current can cause spin polarization in a particular direction and switch the magnet's magnetization to that direction. Alternately, spin–orbit interaction acts like an effective magnetic field [142] and that field can switch the magnetization of the magnet by delivering a torque. Hence, this phenomenon is sometimes referred to as spin-orbit-torque (SOT). There is some controversy about the actual mechanism for producing the torque, but there is plenty of experimental evidence that a torque is produced and that it can switch the magnet [140]. Here, we will assume, for simplicity, that the Spin Hall Effect is the source of the torque and the spin polarized current caused by the charge current is responsible for switching. The ratio of the two current densities—spin current density and charge current density—is called the 'spin Hall angle' $\theta_{SH}$:

$$\theta_{SH} = \frac{J_s}{J_c} \tag{1}$$

A more accurate expression for the spin Hall angle in this configuration is [143]

$$\theta_{SH} = \frac{J_s}{J_c}(1 - c/L_s)^{-1}, \tag{2}$$

where $c$ is the thickness of the slab (see figure 10) and $L_s$ is the spin diffusion length.

The spin Hall angle is usually quite small in most materials, but in certain materials it can be large. It is reported to be 0.15 in β-Ta [136], 0.3 in β-W [137] and 0.24 in CuBi alloys [138]. These materials are therefore said to exhibit the GSHE. The spin current $I_s$ can be used to deliver a STT on a soft magnet and rotate its magnetization. Note that the spin current does not dissipate any power since the scalar product $\vec{J_s} \cdot \vec{E}$ where $\vec{E}$ is the electric field driving the charge current and it is collinear with $J_c$ which is perpendicular to $J_s$.

Any power dissipation is due to the charge current. In our MTJ of the elliptical cross-section with major axis dimension = $m$ and minor axis dimension = $d$, the minimum power dissipation can be approximately written as

$$P_d = (I_s^{cr})^2 \left(\frac{4}{\pi \theta_{SH}}\right)^2 \rho \frac{c}{md} \qquad (3)$$

Clearly, there are two ways to make the power (and energy) dissipation small: first by using a material with large spin Hall angle, and second by using a slab with very small thickness $c$ [1, 135]. The energy dissipation can be reduced to ~1.6 fJ by using this approach, and perhaps even lower [144].

## 8.6 Voltage (or electric field) controlled magnetization switching

Current-controlled magnetization switching is very reliable (low switching error probability), but as the previous section showed, it typically consumes an exorbitant amount of energy. The energy dissipated to switch a nanomagnet with STT is on the order of 100 fJ [135], while the energy dissipation in SOT switching is about two orders of magnitude smaller [145]. Unfortunately, they are still much higher than the energy dissipated in switching a modern-day transistor, which is about 100 aJ [146]. This is obviously an unacceptable price to pay for non-volatility. On the other hand, voltage (or electric-field) induced magnetization switching is much more energy-efficient, albeit also less reliable (higher switching error probability). There is always a trade-off between energy cost and error resilience, which is also true of electronic devices like the transistor [146].

The magneto-electric (ME) effect provides a pathway for electric field controlled magnetization switching. In the case of multiferroic materials, the magnetic and electrical ordering are interlinked [147, 148]. The strong coupling between magnetic and electric polarization enables controlling magnetic properties with an electric field and vice versa. Moreover, some dilute magnetic semiconductors, such as (Ga,Mn)As and (In,Mn)As, show electric-field modulation of magnetic anisotropy, ferromagnetism and exchange interaction through electric field induced modulation of carrier density [149]. Electrically generated mechanical strain can also switch the magnetization of magnetostrictive materials through the inverse magnetostriction or Villari effect. The latter has spawned the burgeoning field of straintronics which we discuss more in the next section.

A popular modality for switching the magnetization of the soft layer of an MTJ is the so-called "voltage (i.e., electric-field) controlled magnetic anisotropy" (VCMA). VCMA is observed at the interfaces between ultrathin $3d$ transition ferromagnetic metals (e.g. Fe, CoFeB) and nonmagnetic insulators (e.g.

MgO, Al$_2$O$_3$) [150]. At ferromagnet/oxide interface, the out-of-plane 3$d$-orbitals of Fe strongly bond with out-of-plane 2$p$-orbitals of O resulting in a significant charge transfer from 3$d$-orbitals to 2$p$-orbitals. Therefore, an imbalance between the number of electrons in out-of-plane orbitals and the number of electrons in in-plane orbitals is observed, which introduces a sizeable amount of perpendicular magnetic anisotropy (PMA) through spin-orbit coupling (SOC) in a ferromagnet (FM) (figure 11(a)). When an electric field is applied at FM/oxide interface, the imbalance between the number of electrons in out-of-plane 3$d$-orbitals and in-plane 3$d$-orbitals is modified as has been shown from first principles calculations [151, 152]. This significantly influences the bonding strength between 3$d$- and 2$p$-orbitals resulting in a substantial change of the PMA (figure 11(b)). As a result, the dominant magnetic anisotropy can change from perpendicular to in-plane. When that happens, the easy axis of magnetization changes from out-of-plane to in-plane causing the equilibrium magnetization direction to rotate by $90^0$ (from out-of-plane to in-plane). If an in-plane magnetic field is present, then the magnetization rotation will not stop at $90°$, but continue to rotate further as it precesses about the magnetic field. By precisely timing the voltage pulse width such that the voltage is cut off as soon as $180^0$ precession is completed, one can flip (or reverse) the magnetization from say pointing up to pointing down as shown in figure 12. The in-plane magnetic field can be replaced by an "effective" magnetic field generated by electrically induced strain [153] for an all-electric rendition. Since the penetration depth of electric field in metal is only few angstroms, the VCMA effect is limited to ultrathin FM films, which are also more likely to have perpendicular, rather than in-plane, magnetic anisotropy.

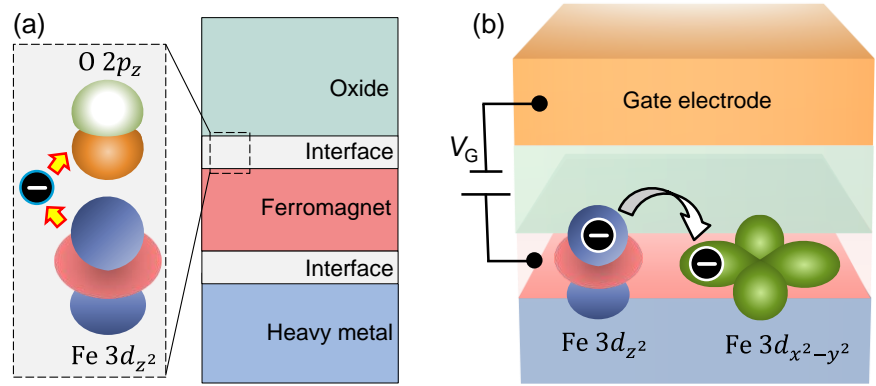

**Figure 11.** (a) Schematic diagram shows the origin of perpendicular magnetic anisotropy at FM/Oxide interface due to hybridization of out-of-plane 2$p$ orbitals of O and out-of-plane 3$d$ orbitals of a ferromagnet (FM). (b) The schematic illustration shows the mechanism of VCMA. When a dc gate voltage $V_G$ is applied across the interface, the electron density at the out-of-plane 3$d$ orbitals of the FM is modified with respect to the in-plane orbitals. This affects interfacial orbital hybridization and changes the surface anisotropy and hence the perpendicular magnetic anisotropy of the FM via spin-orbit coupling. Reproduced from [154] with permission of the Institute of Physics.

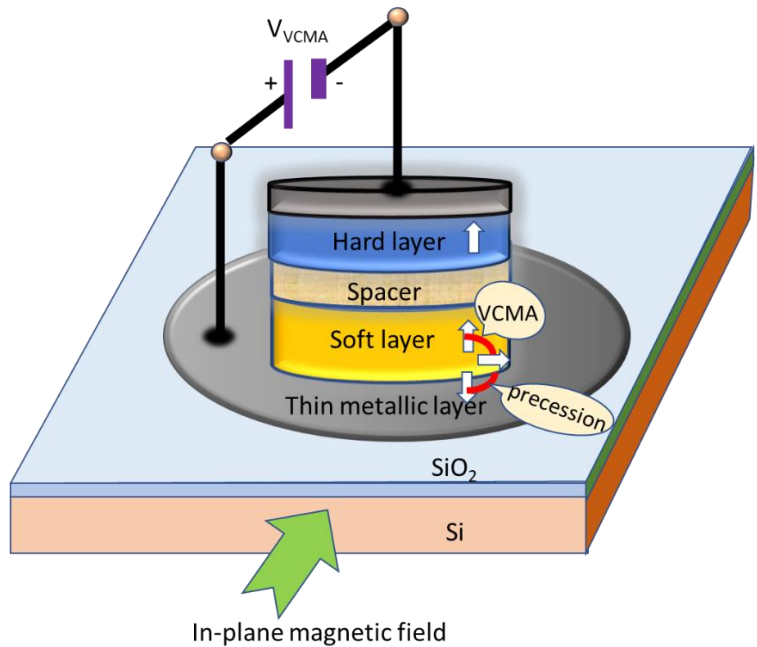

**Figure 12.** VCMA switching of the soft layer of an MTJ with PMA. The applied voltage VCMA changes the anisotropy of the soft layer from perpendicular to in-plane, thereby causing a 90° rotation. The magnetization continues to rotate because of precession around the in-plane magnetic field. If the VCMA voltage is turned off precisely after the magnetization has rotated through 180, the magnetization flips, taking the MTJ from the parallel configuration to the antiparallel configuration. The requirement of precisely timing the VCMA pulse makes this switching modality error-prone.

One of the earliest demonstrations of electric field induced magnetization switching was by Chiba *et al.* in the dilute magnetic semiconductor Mn-doped InAs [(In,Mn)As] [155, 156]. They used (In,Mn)As as the channel layer in a metal-insulator-semiconductor FET structure, which was patterned into a Hall bar for measuring anomalous Hall effect (AHE) signal. The FET was covered with a 0.9-µm-thick $SiO_2$ gate insulator overlaid with a Cr/Au metal gate electrode for applying gate voltage (figure 13(a)). In *p*-type (In,Mn)As, the holes are known to mediate the ferromagnetic interaction among Mn localized spins. The negative (positive) electric field increases (decreases) hole concentration resulting in an increment (decrement) of Curie temperature. It was observed that the coercive field is reduced significantly when electric field is varied from -1.5 MV/cm to +1.5 MV/cm (figure 13(b)). This was not all-electric manipulation of magnetization; the magnetization could not be switched with electric field alone, but the electric field could reduce the magnetic field required for switching. For switching, the magnetization of the (In,Mn)As channel layer under $E = -1.5$ MV/cm is saturated by applying a large enough positive magnetic field, and then the magnetic field is reduced through 0 mT to a small negative magnetic field of - 0.2 mT. This stage is indicated by A in the inset of figure 13(c). Subsequently, the electric field is brought down to 0 MV/cm, which triggers the magnetization switching as the applied magnetic field (-0.2 mT) is higher than the switching field at $E = 0$ MV/cm. This is the underlying principle for all types of electric-field assisted switching. The electric field-induced magnetization switching in ferromagnet-multiferroic heterostructure has been demonstrated by many groups [157, 158].

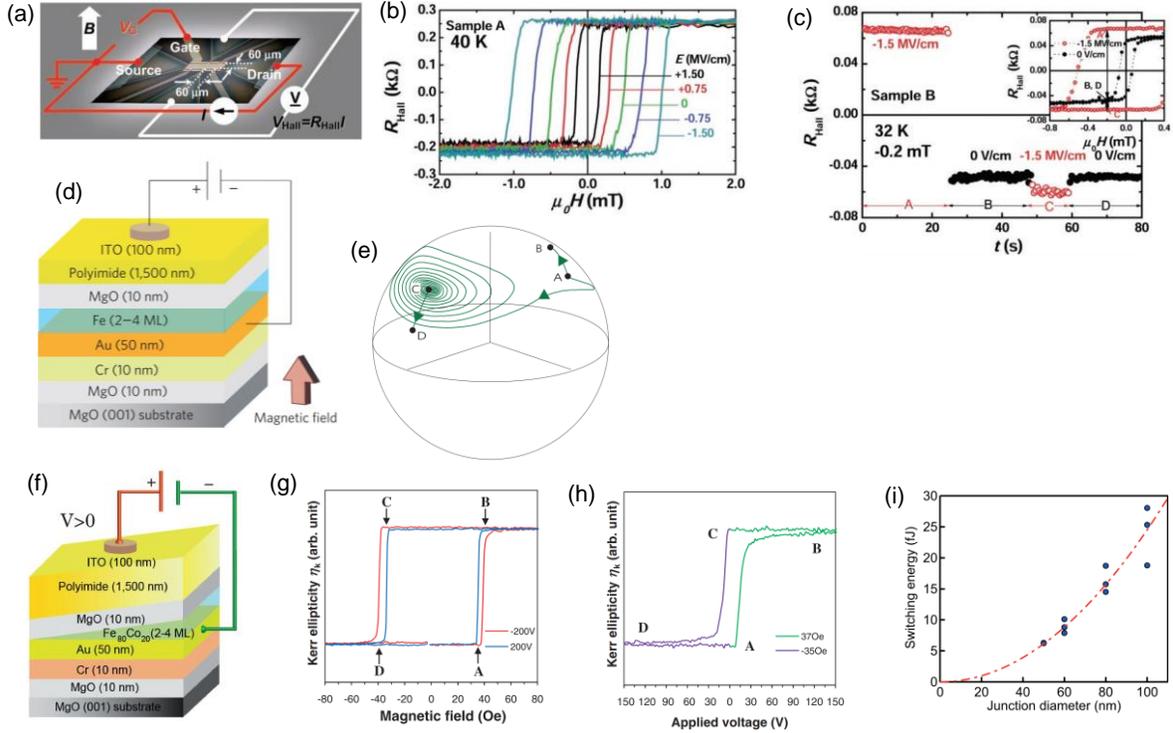

**Figure 13.** (a) Hall bar-shaped field effect transistor with a ferromagnetic semiconductor (In,Mn)As as the channel. To probe the magnetization $M$ of the channel, the Hall resistance $R_{Hall}$ is measured, which is proportional to the channel magnetization [155]. (b) Electric field dependent coercive field of the channel extracted from Hall measurements. (c) Time evolution of the Hall resistance resulting from a sequence of applied electric fields at 32 K, showing an electrically assisted magnetization reversal. Reproduced from [155] with permission of the American Association for the Advancement of Science. (d) Schematic of the sample used for a voltage-induced magnetic anisotropy change [159]. A positive voltage is defined as a positive voltage on the top electrode (i.e., ITO) with respect to the bottom electrode (i.e., Fe). (e) A macro spin model simulation of voltage-controlled magnetization switching. The green line indicates the trajectory of the spin. Reproduced from [159] with permission of the Springer Nature. (f) The sample structure used for VCMA-induced magnetization switching in a $Fe_{80}Co_{20}$ film [160]. (g) Hysteresis curves of a 0.48-nm-thick $Fe_{80}Co_{20}$ layer under an applied voltage of 200 (blue) and -200 V (red) [160]. (h) Kerr ellipticity plotted as a function of applied bias voltage. Clear transition between two magnetic states was observed under the presence of assisting magnetic fields. Reproduced from [160] with permission of the Institute of Physics. (i) Measured (dots) and calculated (dotted lines) scaling trends of the switching energy with the diameter of magnetic tunnel junction for >90%

switching probabilities. The quadratic dependence of the switching energy on the junction diameter is determined by the resistance-area product RA= 650 Ω μm$^2$ [92]. Reproduced from [92] with permissions from American Institute of Physics.

VCMA induced magnetization switching was first predicted by Maruyama *et al*. [159]. They demonstrated that the magnetic anisotropy in few atomic layers of iron can be changed significantly by an electric field. By performing macro-spin model simulation, they also suggested the principle of VCMA induced magnetization switching for a 0.48-nm Fe film (figure 13(d)). In this case, an external magnetic field is applied normal to the film plane to tilt the magnetization towards the perpendicular direction. Initially, the bias voltage is off (point A in figure 13(e)). If it is then increased with a slow rise, the perpendicular anisotropy field changes and the magnetization changes its direction to point B. However, if the rise time of the pulse is less than 1 ns, i.e., very short, a dynamic precession and switching of magnetization to another energetically stable point is achieved (point C in figure 13(e)). After switching off the voltage, the magnetization is stabilized at point D (figure 13(e)) when the system relaxes. This is VCMA induced precessional switching. The VCMA assisted switching was first experimentally demonstrated by Shiota *et al.* [160]. They showed that when negative (positive) gate voltage is applied across a 0.48-nm-thick Fe$_{80}$Co$_{20}$ film and MgO interface, the coercive field is increased (decreased) owing to the increment (decrement) of PMA (figure 13(f)). For magnetization switching, the magnetization is set at point A in figure 13(g) at zero bias voltage by applying a +37 Oe magnetic field. If a positive bias voltage is applied, then point A becomes unstable and the magnetization is switched to achieve a stable state at point B.

In another experiment, magnetization reversal was demonstrated in a CoFeB/MgO heterostructure by applying a voltage pulse whose duration was one half of the magnetization precession time [161]. The physics of this behavior was explained in figure 12. Many similar reports appeared in the literatures [162-164]. One of the major advantages of VCMA induced switching is that ultra-low energy about 6 fJ is required for switching a magnetic element of 50 nm diameter. The switching energy can be slightly reduced by increasing junction resistance [92, 165].

**8.7 Straintronic switching**

The term "straintronics" refers to the science and technology of switching the magnetization of *magnetostrictive* nanomagnets using electrically generated mechanical strain. The idea is to fashion the soft layer of an MTJ out of a magnetostrictive nanomagnet (e. g. Co, Ni, FeGa or Terfenol-D) and then place it in elastic contact with an underlying (poled) piezoelectric film. The elastic contact allows highly efficient strain transfer from the piezoelectric to the magnetostrictive nanomagnet. Such a system constitutes a "two-phase multiferroic". Application of a voltage over (not across) the piezoelectric film with electrodes delineated on the film generates biaxial strain in it. If the polarity of the voltage is such that the electric field is in the direction opposite to that of the poling, as shown in figure 14(a), then compressive stress will be generated along the major axis of the ellipse and tensile strain along the minor axis. If the voltage polarity is reversed, then the signs of the strains will reverse as well. In the event the magnetostriction of the nanomagnet is positive (FeGa, Terfenol-D), the former scenario will cause the magnetization to rotate away from the major axis (or the so-called easy axis) towards the minor axis (or the hard axis) and the latter scenario will not cause any rotation. On the other hand, if the magnetostriction is negative (Co, Ni), then the latter scenario will make the magnetization rotate towards the minor axis, while the former scenario will not cause rotation.

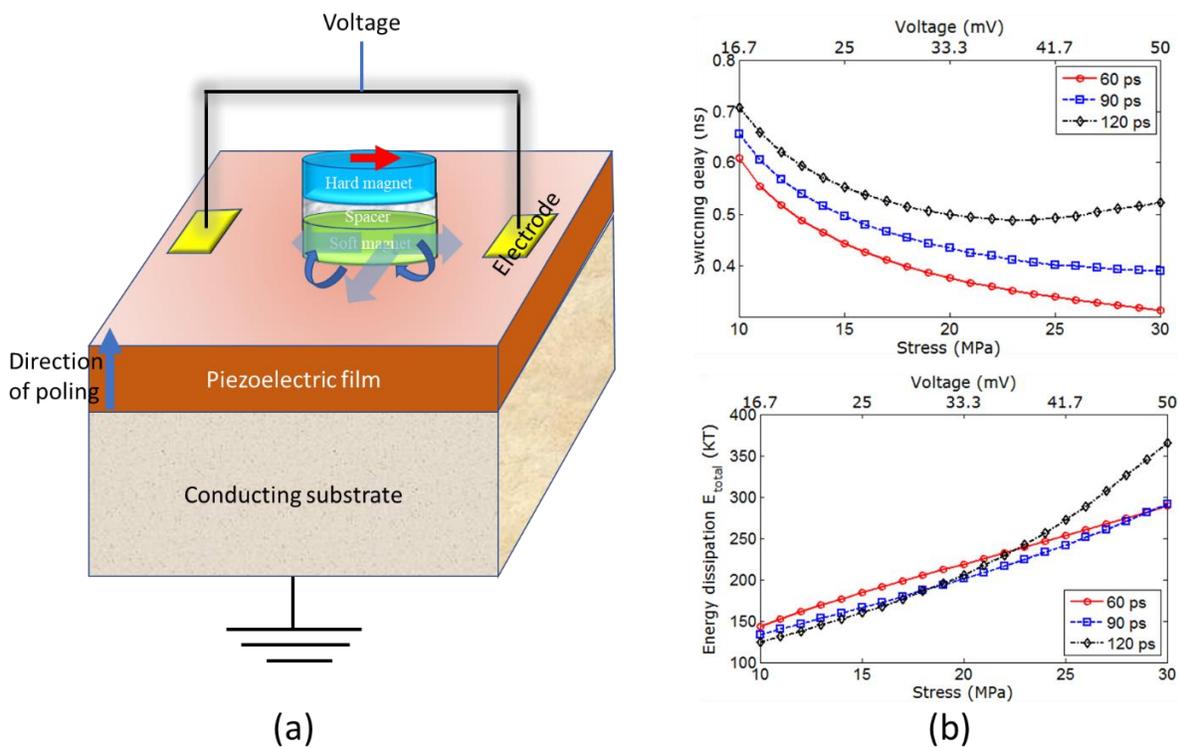

**Figure 14.** (a) Straintronic switching: flipping the magnetization of a magnetostrictive elliptical soft layer of a magneto-tunneling junction, elastically coupled to an underlying piezoelectric film, with a precisely timed strain pulse generated with the applied voltage. (b) Calculated switching delay and energy dissipation as a function of the stress (and the corresponding gate voltage needed to generate the required stress) at room temperature for three different stress ramp times (60, 90, 120 ps). The piezoelectric film thickness is 40 nm and the soft layer (made of Terfenol-D) thickness is 10 nm. Sub-figure (b) is reproduced from [166] with permission of the American Institute of Physics.

If, *as soon as* the 90° rotation is completed and the magnetization aligns along the minor axis, the strain is withdrawn, then the magnetization will continue to rotate further under an inertial torque and complete 180° rotation or full reversal [167]. This is shown in figure 14 (a). Therefore, in order to write a desired bit, we will first read the already stored bit. If it is the same as the desired bit, we will do nothing. Otherwise, we will flip the magnetization as just described and this will write the desired bit. A memory cell based on this kind of writing scheme is called "toggle memory" since all we can do is toggle the magnetization. As a result, it is always necessary to read the previously stored bit first and then take (or not take) the action to toggle. There is a non-toggle version of straintronic memory as well [168], but that needs the use of an in-plane magnetic field. The idea there is to apply the magnetic field (of the right strength) along the minor axis of the elliptical soft layer which will bring the two stable orientations out of the major axis and make them point in two directions that are mutually perpendicular as shown in figure 15. In that case, if the magnetostriction of the soft layer is positive, then applying tensile stress along one of the two directions will orient the magnetization along that direction, while compressive stress will orient the magnetization along the other direction. The opposite will be true if the magnetostriction is negative. Thus, we can write the desired bit (i.e., orient the magnetization along either direction) by simply choosing the sign of the stress applied along one of the two directions, without having to read the stored bit first.

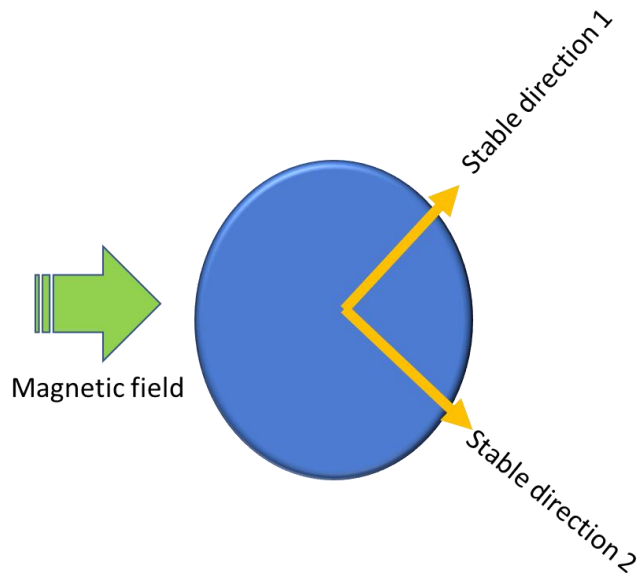

**Figure 15.** The basic principle of "non-toggle" straintronic memory. The in-plane magnetic field brings the two stable magnetization directions out of the major axis. They lie in the plane of the nanomagnet and subtend 90° angle with each other. If the magnetostriction is positive, then compressive stress along direction 1 will align the magnetization along direction 2 and tensile stress will align it along direction 1. The opposite will be true if the magnetostriction is negative.

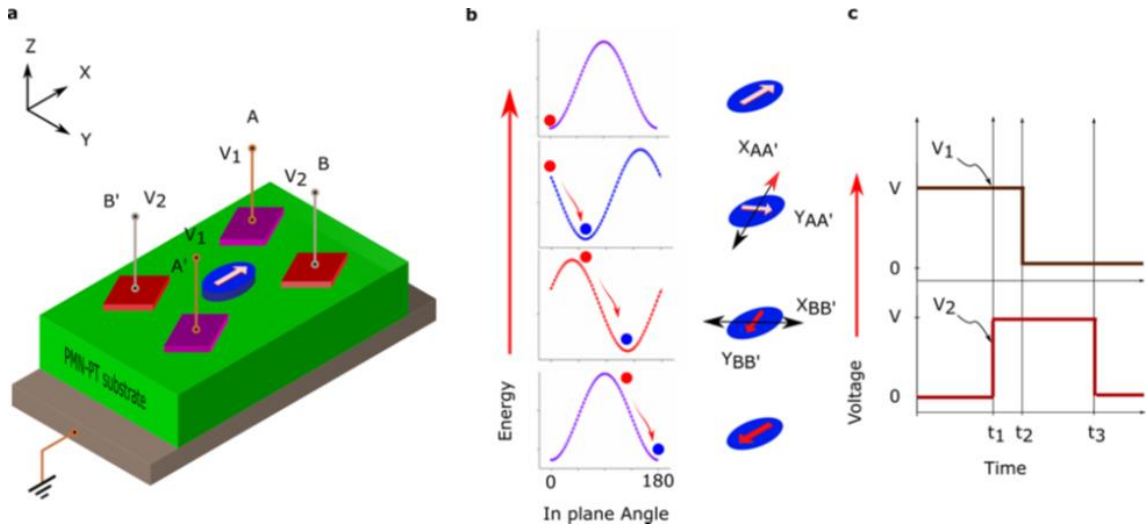

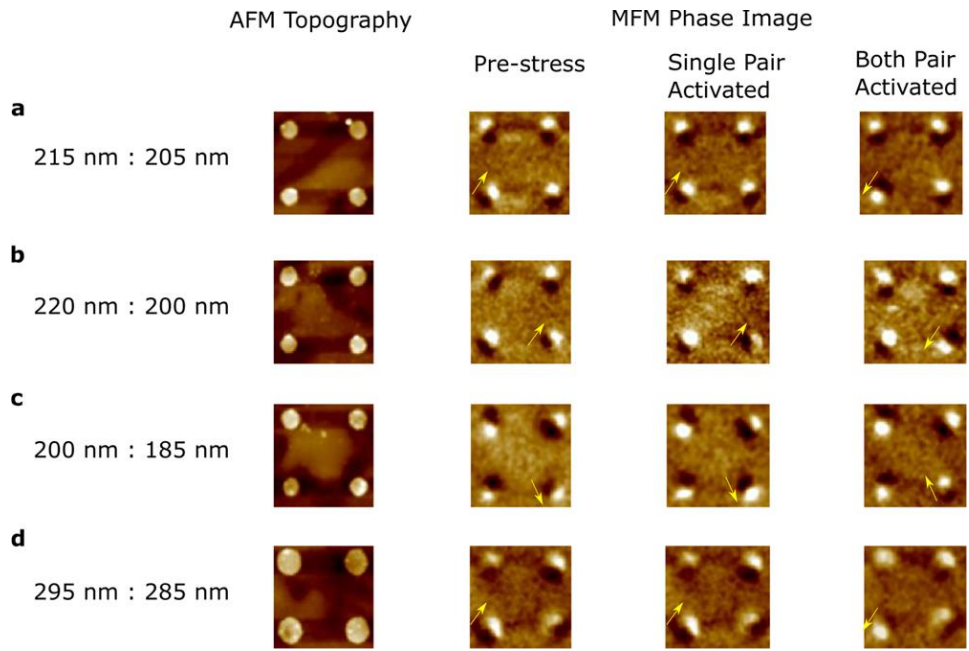

**Figure 16.** (Top panel) a. Arrangement for applying uniaxial stresses to the elliptical nanomagnet in two different directions (neither collinear with a principal axis) by sequentially activating the electrode pairs AA′ and BB′. b. Potential energy profile of the nanomagnet as a function of the angle subtended by the magnetization with the major axis (direction of the arrow shown) - with no electrode activated, only AA′ activated, BB′ activated and AA′ de-activated, and all electrodes de-activated. c. Timing sequence of the voltage $V_1$ applied to electrode pair AA′ and $V_2$ applied to electrode pair BB′. (Bottom panel). Atomic and magnetic force micrographs of four different sets of nanomagnet assemblies (with different major and minor axes dimensions) subjected to this stress sequence. The magnetic force micrographs are shown at three different stages of electrode activation. One out of four nanomagnets flip completely (180° rotation) upon completion of the stress cycle. Reproduced from [169].

One serious problem with the first approach (whereby stress is removed as soon as 90° rotation has been completed) is that timing the stress pulse accurately is a challenge. Because of thermal noise, there is a spread in the time it takes for the magnetization to complete the 90° rotation, which makes it difficult to time the stress/strain pulse precisely. A better strategy is to apply two uniaxial stresses in two different directions (neither of which is along the easy or hard axis) *sequentially* and that can complete the 180° rotation in two steps [170]. Two antipodal gate pads shown in the top panel of figure 16 are first activated by applying a voltage to them. This would generate biaxial stress, but we can approximate the effect by assuming that uniaxial stress is generated along the line joining the two activated gate pads. Such a stress will rotate the magnetization away from the major axis (easy axis) of the soft layer and roughly stabilize it in a direction perpendicular to the line joining the activated pair if the product of the stress and magnetostriction has a negative sign. The other pair is then activated (followed by deactivation of the first pair) and this rotates the magnetization further, bringing the total rotation to an angle θ, where $90^0 < \theta < 180^0$. Finally, when the second gate pair is deactivated, the magnetization relaxes to the *nearest* stable state along the major axis, which is opposite to the initial direction. That completes 180° rotation, or complete magnetization reversal. This strategy would require four gate pads as shown in the top panel of figure 6, but it eliminates the requirement of having to time the stress pulse precisely, which is a very difficult proposition when thermal noise is present. This latter strategy has been demonstrated experimentally [169] (see figure 16).

Several groups have experimentally studied the control of magnetization in magnetostrictive films using voltage generated strain in a piezoelectric film [171], demonstrating reversible control of nanomagnetic domains [172], repeatable reversal of perpendicular magnetization in the absence of a magnetic field in regions of a Ni film [173], and strain assisted reversal of perpendicular magnetization in Co/Ni multilayers [174]. Others have demonstrated the use of strain control of magnetization orientation in LSMO films [175, 176], iron films [177], TbCo$_2$/FeCo multilayers [178] and strain control of magnetic properties of FeGa/NiFe multilayer films [179] and FeGa films [180].

Strain has been shown to reorient magnetization in magnetostrictive Ni rings [181, 182] and Ni squares of 2 μm side [183] and the soft layer of MTJs of lateral dimensions 20 μm × 40 μm [184]. The magneto-electric effect has also been used to read the magnetization orientation in a composite multiferroic heterostructure [N×(TbCo$_2$/FeCo)]/[Pb(Mg$_{1/3}$Nb$_{2/3}$)O$_3$]$_{1-x}$[PbTiO$_3$]$_x$ [185].

There are many reports of demonstrated control of magnetization in nanomagnets deposited on piezoelectric substrates. For example, an electric field induced stress mediated reversible control of magnetization orientation in nanomagnets of nominal lateral dimensions 380 nm × 150 nm deposited on a 1.28 µm thick PZT film was demonstrated with the application of 1.5 V to the PZT film [185]. Furthermore, building on individual control of magnetoelectric heterostructures with localized strain to reorient the magnetization in a Ni ring of 1000 nm outer diameter, 700nm inner diameter, and 15 nm thickness, deterministic multistep reorientation of magnetization in a 400 nm wide Ni dot of 15 nm thickness has been reported [186].

Uniform magnetization rotation through 90° has also been demonstrated through imaging with x-ray photoemission electron microscopy (X-PEEM) and x-ray magnetic circular dichroism (XMCD) in elliptical nanomagnets of nominal lateral dimensions ~100 nm ×150 nm [187]. The field of magneto-elastic switching (or "straintronics") is now a mature field where the physics is well understood and engineering applications appear to be extremely promising.

There are also reports of switching the resistance states of MTJs with electrically generated mechanical strain [188, 189], making this methodology mainstream. In [189], the experimentally measured high/low resistance ratio, sometimes referred to as "tunneling magnetoresistance ratio" (TMR) exceeded 2:1 at room temperature, which is very respectable.

### 8.7.1 Dynamic straintronics

In the earlier section, we discussed switching the magnetization of a magnetostrictive nanomagnet's magnetization with static strain and how energy-efficient it is. In this section, we will discuss the effect of time-varying strain produced by a (surface) acoustic wave. Time-varying strain results in an equally (if not more) energy efficient modality of switching the magnetization of nanomagnets. Surprisingly, it appears to be more reliable than switching with static strain, although the reason behind the increased error resilience is not well understood.

### 8.7.2 SAW enhanced spin transfer torque (STT) switching of magneto-tunneling junctions for energy-efficient memory

We start with a discussion of where SAW-induced switching can find an application. One salient drawback of STT switching mechanism for MTJs is that it is not particularly energy-efficient since a relatively large current density is required to accomplish the switching. Considerable amount of research has been carried out in an effort to reduce the current density, and current densities as low as 2.1 MA cm$^{-2}$ in an MTJ with a resistance-area product of 16 Ω µm$^2$ have been reported [190]. In an MTJ whose cross-sectional area is 1 µm$^2$, the power dissipated to switch would be ~ 7 mW, which is extremely high. This is why small cross-sectional areas are needed for STT switching. Attempting to reduce the switching current further by thinning the magnetic layers or the spacer layer results in dramatic reduction of the high- to low-resistance ratio, or TMR, since it is governed by spin-dependent tunneling between the magnetic layers. Typically, the energy dissipated to switch with STT is several fJ. There have been some recent efforts to reduce the energy dissipation by using spacer layers that have smaller bandgap, such as ScN, which would offer a lower tunneling resistance and hence a lower resistance-area product, but this may be counter-productive since the lower barrier to tunneling may increase the thermionic emission over the barrier. Since thermionic emission is not spin-dependent unlike tunneling, the overall effect will be to reduce the TMR even further.

As discussed earlier, there have been proposals to replace STT switching with SOT switching which involves passing a spin current instead of a charge current through the MTJ in order to switch the magnetization of the soft layer and thus switch the MTJ resistance. The spin current is generated by passing a charge current through a heavy metal layer (e. g. Pt, β-Ta, etc.), or a topological insulator, placed underneath the soft layer, which converts the charge current to a spin current by virtue of the GSHE in the heavy metal layer [136-138] or the spin-momentum locking effect in a topological insulator [191]. The "spin Hall angle" is typically less than unity in the case of the heavy metal layer. However, because the charge current is passed through a *metal* layer which has a much smaller resistance than an MTJ, the power and energy dissipations are reduced because the current path is no longer though the MTJ. The reduction in the energy dissipation depends on the thickness of the metal layer and an analysis can be found in ref. [192] which shows how the energy dissipation depends on different geometric features. The topological insulator, on the other hand, may not have a low resistance unlike the heavy metal, but it may produce an effective spin Hall angle that exceeds unity. The salient drawback of these approaches is that they will all result in a *three-terminal* MTJ, which is unattractive for memory applications since it will have a much larger footprint than a two-terminal MTJ used in conventional STT-random-access-memory (STT-RAM). On the flip side, the advantage is the reduced energy dissipation and the physical separation of the read and write paths, which avoids "read-disturb" (corrupting the stored bit during the reading operation) and reduces damage to the MTJ since the charge current path is not through the MTJ but though a different layer. This improves the memory's endurance.

We proposed a different approach where the memory remains two-terminal. Our approach is a *bimodal* switching mechanism in the sense that two different switching mechanisms are pressed into service at the same time to reduce the energy dissipation [193]. In our hybrid approach, we use a magnetostrictive soft layer placed atop a piezoelectric substrate. A surface acoustic wave (SAW) is launched in the substrate and flows underneath the soft layer, straining it periodically. At the same time, we synchronously pass a charge current through the MTJ during the appropriate cycle of the SAW to generate STT and drive the magnetization of the soft layer to the desired orientation. The SAW rotates the magnetization by 90° during the cycle when the product of the magnetostriction and strain is negative. If during that cycle, a charge current is introduced to produce STT, then a complete 180° rotation can be achieved with *reduced charge current* since the SAW lends a helping hand to the STT. In fact, SAW does the "heavy lifting" and since it is much more energy efficient than STT, the overall energy dissipation is reduced, perhaps by an order of magnitude [82]. Two conditions however must be fulfilled for reliability: (1) the probability of switching the magnetization of a magnet to the desired orientation (writing of bits) must be ~100% at room temperature when the STT charge current is injected, and (2) the probability of unintentionally switching the magnet due to the SAW alone is ~0% at room temperature when no STT current is injected. This will ensure that bits are written reliably in the target memory cells and data already stored in other cells are not corrupted. We showed in ref. [193] that both conditions can be fulfilled with proper design. The structure for this bimodal switching is shown in figure 17. A very small amount of energy is required to generate the global SAW that acts on all MTJs on the wafer, and when that energy is amortized over all the MTJs, the energy cost per MTJ is miniscule. We found that this approach can reduce the energy dissipation by approximately an order of magnitude. Further reduction may be possible with design optimization.

Periodic switching of magnetization between the hard and easy axis of 40 μm × 10 μm × 10 nm Co bars sputtered on $LiNbO_3$ has been shown [194]. Other authors have studied acoustically induced switching in thin films [195] including focusing surface acoustic waves (SAWs) to switch a specific spot in an iron-gallium film [196]. Several proposals suggest a complete 180° rotation with an appropriately timed acoustic pulse [197]. Stroboscopic x-ray techniques have been used to study strain waves and magnetization at the nanoscale [198].

Excitation of SW modes in GaMnAs layers by a picosecond strain pulse [199] as well as magnetization dynamics in GaMnAs [200] and GaMn(As,P) [201] have been demonstrated. In in-plane magnetized systems, SAWs have been utilized to drive ferromagnetic resonance in thin Ni films [202, 203]. Resonant effects have also been studied by spatial mapping of focused SAWs [204]. There are theoretical studies of the possibility of complete magnetization reversal in a nanomagnet subjected to acoustic wave pulses [197]. Interestingly, for high frequency excitation of extremely small nanomagnets, the Einstein De Haas effect seems to dominate as has been proposed [205] and experimentally demonstrated [206, 207].

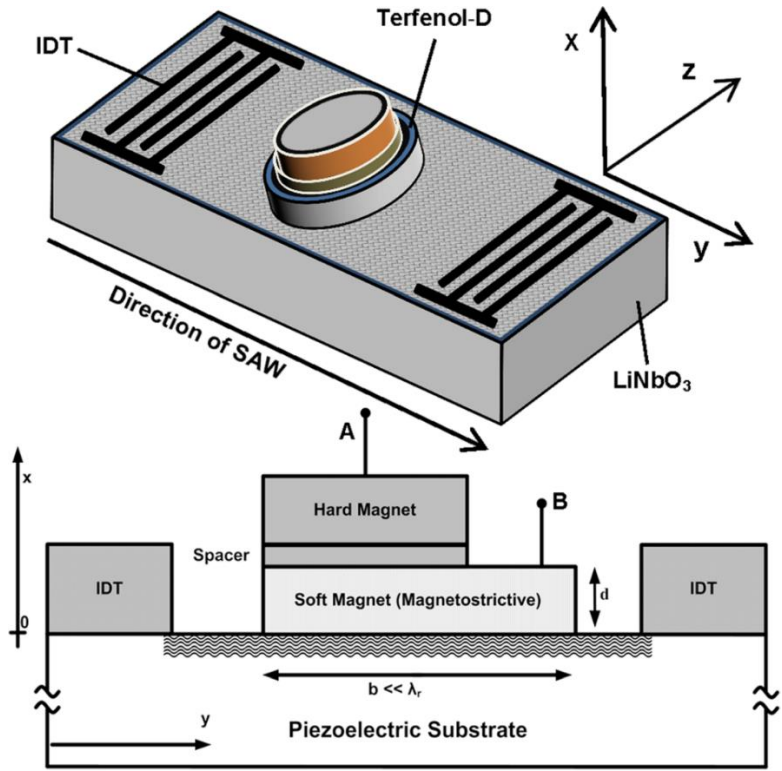

**Figure 17.** (Top) Schematic illustration of the hybrid system with interdigitated transducers (IDTs) to launch the SAW and an MTJ, serving as a bit storage unit, placed between IDTs on a piezoelectric substrate. The soft layer of the MTJ is in contact with the substrate and is periodically strained by the SAW. The resistance between the terminals A and B is used to read the bit stored (we assume that both magnets are metallic). For writing, a small spin polarized current is passed between the same two terminals during the appropriate cycle of the SAW, when the magnetization rotates out of the easy axis. In this configuration, the reading and writing currents do not pass through the highly resistive piezoelectric, so the dissipation during the read/write operation is kept small. Bits are addressed for read/write using the traditional crossbar architecture. Reproduced from [193] with permission of the American Institute of Physics.

### 8.7.3 Clocking a straintronic inverter (Boolean NOT gate) with a surface acoustic wave (SAW)

The simplest Boolean logic gate for Boolean computing is the inverter or NOT gate. It is a single input-single output gate in which the output bit is always the logic complement of the input bit. Consider the system shown in figure 18(a) consisting of two closely spaced elliptical magnetostrictive nanomagnets, one of which is more elliptical than the other. Each nanomagnet's magnetization can point along two opposite directions along its major axis (easy axis), and because of dipole interaction between the two, the magnetizations will be mutually antiparallel when the system is in the lowest energy (ground) state.

Imagine that the bit value is encoded in the magnetization orientation; for example, magnetization pointing up could be bit 1 and pointing down could be bit 0. Say, the left nanomagnet hosts the input bit and the right one the output bit. In this case, as long as we can drive the system to the ground state, the output will be the logic complement of the input, thereby implementing the NOT gate.

Next, suppose that we change the input bit from 0 to 1 with some external agent (e.g. STT or SOT), which will flip the input bit up, temporarily placing the duo in the "parallel" configuration shown in the top row of figure 18(b). We might naively expect that the right nanomagnet will subsequently flip spontaneously, because of dipole interaction, to realize the NOT operation, but this does not happen, since the right nanomagnet will have to overcome its own internal shape anisotropy energy barrier (due to its elliptical shape) in order to flip. Dipole interaction is not strong enough to beat the shape anisotropy energy barrier and hence the system will be stuck in the metastable state shown in the top row (i. e. the "parallel" state), and never reach the ground state which corresponds to the inverter configuration.

To trigger, the NOT operation, we apply uniaxial stress of the correct sign along the major axis of both nanomagnets (the stress is global and hence affects both nanomagnets). Because the left nanomagnet is extremely elliptical, it has a very high shape anisotropy energy barrier and consequently stress cannot budge its magnetization. However, the right nanomagnet is less elliptical and hence its magnetization is rotated $90^0$ by the stress, as shown in the second row of figure 18(b). Finally, when we remove the stress, the magnetization of the right nanomagnet must settle along one of the two stable orientations along the major axis, but because of the dipole interaction, it will prefer to settle in a direction antiparallel to magnetization of the left nanomagnet, thereby implementing the NOT operation, as shown in the bottom row of figure 18(b). The stress therefore acts as a "clock" to trigger the NOT gate operation.

In figure 18(c), we show the scanning electron micrograph of two such nanomagnets. The major axes of both nanomagnets are 200 nm. The minor axis of the left one is 80 nm and that of the right one is 130 nm. All pairs are forcibly magnetized in the same direction by a strong external magnetic field as shown in the magnetic force micrograph (MFM) in the pre-stress condition. This corresponds to the top row of figure 18(b). We then apply uniaxial stress along the nanomagnets' major axes and relax. The stress triggers the NOT action in one out of nine pairs as shown (the pair is identified with a yellow arrow).

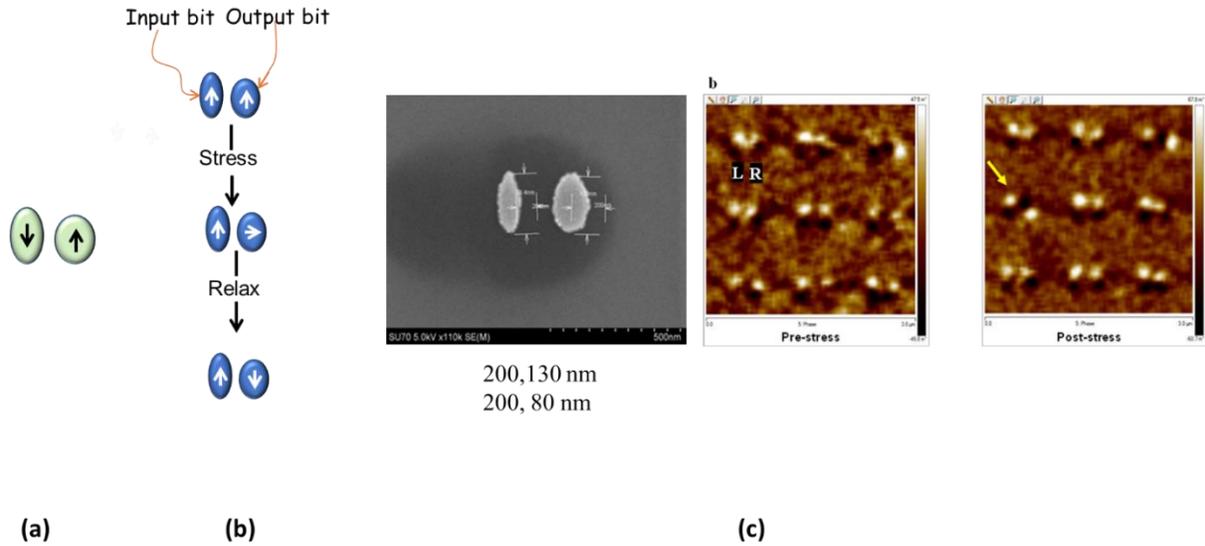

**Figure 18.** (a) Two dipole-coupled elliptical nanomagnets implement a NOT gate. (b) If the input is flipped, the output does not lip in response because dipole interaction cannot overcome the internal shape anisotropy energy barrier within the output nanomagnet. If stress is applied, the input nanomagnet is unaffected since it is too elliptical and has a very high shape anisotropy energy barrier, but the output nanomagnet's magnetization rotates by $90^0$. Upon removing stress, the output nanomagnet's magnetization must assume one of the stable orientations along the major axis, but now, because of dipole interaction, it will prefer to assume the orientation antiparallel to the magnetization of the input nanomagnet. This realizes the NOT operation. (c) Scanning electron micrograph of a dipole-coupled pair and magnetic force micrographs of pre-stressed pairs post-stressed pairs which have been initially magnetized with a global magnetic field to orient their magnetizations in the direction of the field. The nanomagnets are made of Co and are delineated on a poled PMN-PT substrate where the poling direction coincides with the major axes of the nanomagnets. Stress is generated by applying a high dc voltage between two ends of the substrate such that an electric field is produced along the major axis of the nanomagnets. Since the electric field is along the direction of poling, tensile stress is generated along the major axis of the Co nanomagnets (Co has negative magnetostriction) and that should rotate the magnetization of the less elliptical partners by $90^0$ away from their major axes. Figure 18(c) is reproduced from ref. [208].

The obvious question is why is the statistics so poor that only 1 out of 9 pairs responds? There are many possible reasons for this: Co is only weakly magnetostrictive, there are pinning sites within the nanomagnets due to defects which prevent rotation, etc. However, when the experiment was repeated with time-varying stress generated by a SAW, the statistics improved. While static stress switches 1 out of 9, time-varying stress switched 4 out of 4 pairs as shown in figure 19. It appears that repeated cycles of stressing coaxes the nanomagnets to respond better, but this remains to be investigated further before this can be confirmed.

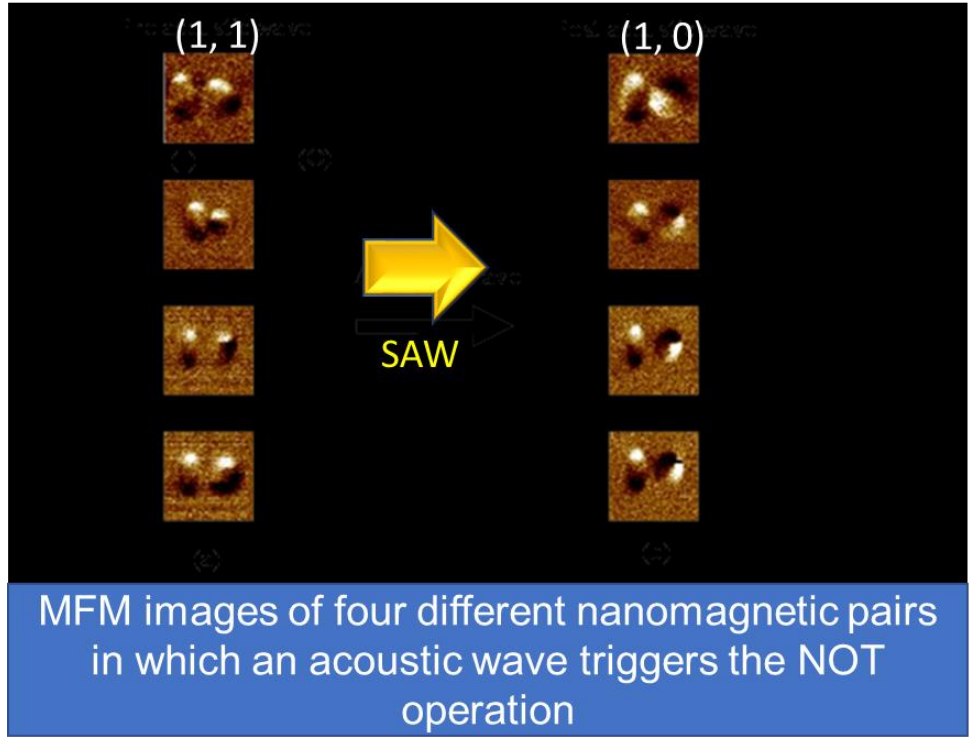

**Figure 19.** (a) Magnetic force micrographs of four different Co nanomagnet pairs delineated on a $LiNbO_3$ substrate. All pairs are initially magnetized in the same direction with a global magnetic field and then subjected to a surface acoustic wave (SAW) launched with interdigitated transducers. The SAW triggers the NOT action, making the magnetizations of the left (input) and right (output) nanomagnets mutually antiparallel. Adapted from [209] with permission of the American Institute of Physics.

### 8.7.4 Hybrid magneto-dynamical modes excited in a single magnetostrictive nanomagnet excited by high-frequency surface acoustic waves

In the previous subsection, we examined the interaction of nanomagnets with low frequency SAWs via magneto-elastic coupling. The existence of this coupling begs the question if high frequency precession can be induced by high frequency (microwave frequency) SAW. SAW of such high frequency would be difficult to generate with traditional interdigitated transducers (IDT) since the small spacing between the IDT fingers (one-quarter of the SAW wavelength) would be a challenge to maintain over many fingers. However, it has been known for some time that ultrashort laser pulses can also induce (polychromatic) SAW waves with microwave frequency components. This technique can be brought to bear on exciting microwave frequency SAW waves in a piezoelectric that is elastically coupled to an array of nanomagnets and then study the precessional dynamics induced in them by the SAW.

In order to study this phenomenon, we investigated the ultrafast magneto-dynamics associated with straintronic switching in a single quasi-elliptical magnetostrictive Co nanomagnet deposited on a piezoelectric Pb(Mg$_{1/3}$Nb$_{2/3}$)O$_3$-PbTiO$_3$ (PMN-PT) substrate using time-resolved magneto-optical Kerr effect (TR-MOKE) measurements [210]. The pulsed laser pump beam in the TR-MOKE plays a dual role: it causes precession of the nanomagnet's magnetization about an applied bias magnetic field and it also generates SAWs in the piezoelectric substrate that produce periodic strains in the magnetostrictive nanomagnet and modulate the precessional dynamics. This modulation gives rise to intriguing *hybrid magneto-dynamical modes* in the nanomagnet, with rich SW texture. The characteristic frequencies of these modes are 5-15 GHz, indicating that strain can affect magnetization in a magnetostrictive nanomagnet in time scales much smaller than 1 ns (~100 ps). This can enable ~10 GHz-range magneto-elastic nano-oscillators that are actuated by strain instead of a spin-polarized current, as well as ultrafast magneto-electric generation of SWs for magnonic logic circuits, holograms, etc.

In our TR-MOKE setup, a femtosecond laser pump beam excited the magnetization of the Co nanomagnet to precess about an applied bias magnetic field (see figure 20(a)). At the same time, the alternating electric field in that same beam also generates periodic (compressive and tensile) strain in the PMN-PT substrate from $d_{33}$ and/or $d_{31}$ coupling. This happens because the laser electric field periodically reconfigures the charge distribution on the surface of the PMN-PT substrate and that, in turn, modulates the electric field within the substrate via the Poisson equation. The frequency with which the electric field oscillates is the frequency with which the surface charges can renormalize (this will be much lower than the frequency of the laser light). Since PMN-PT is piezoelectric and has been poled, the periodically modulated electric field within the substrate will produce a periodic strain due to $d_{33}$ and $d_{31}$ coupling. The strain will alternate between tensile and compressive. Additional periodic strain is generated in the substrate (underneath the nanomagnet) from the differential thermal expansions of the nanomagnet and the substrate due to periodic heating by the pulsed pump beam [211-213]. This thermally generated strain is however always tensile in the substrate in our experiment (its magnitude varies periodically, but the sign does not change) because the thermal coefficient of expansion of Co ($13 \times 10^{-6}$/K) is greater than that of PMN-PT ($9.5 \times 10^{-6}$/K). Note that the former mechanism requires a piezoelectric substrate while the latter does not. The periodic strain sets up a *high frequency* surface acoustic wave (SAW) in the substrate.

Figure 20(b) shows the reflectivity signal as a function of time when the laser spot in the TR-MOKE set-up is focused on the piezoelectric substrate, away from the nanomagnet. The oscillations seen in the reflectivity signal are signatures of the high frequency SAW that is produced in the substrate by the laser pulses. The fast Fourier transforms (FFT) of these oscillations show that there are multiple frequency components of the SAW – at 2, 8 and 15 GHz. These SAW waves generate high-frequency time varying stresses in the magnetostrictive nanomagnet which causes their magnetization vectors to precess around a bias magnetic field applied to the nanomagnet. In addition to these precessions, there is another precessional mode due to Kittel precession [214]. The two modes "mix" to produce hybrid magneto-dynamical modes within a nanomagnet, with very rich SW texture.

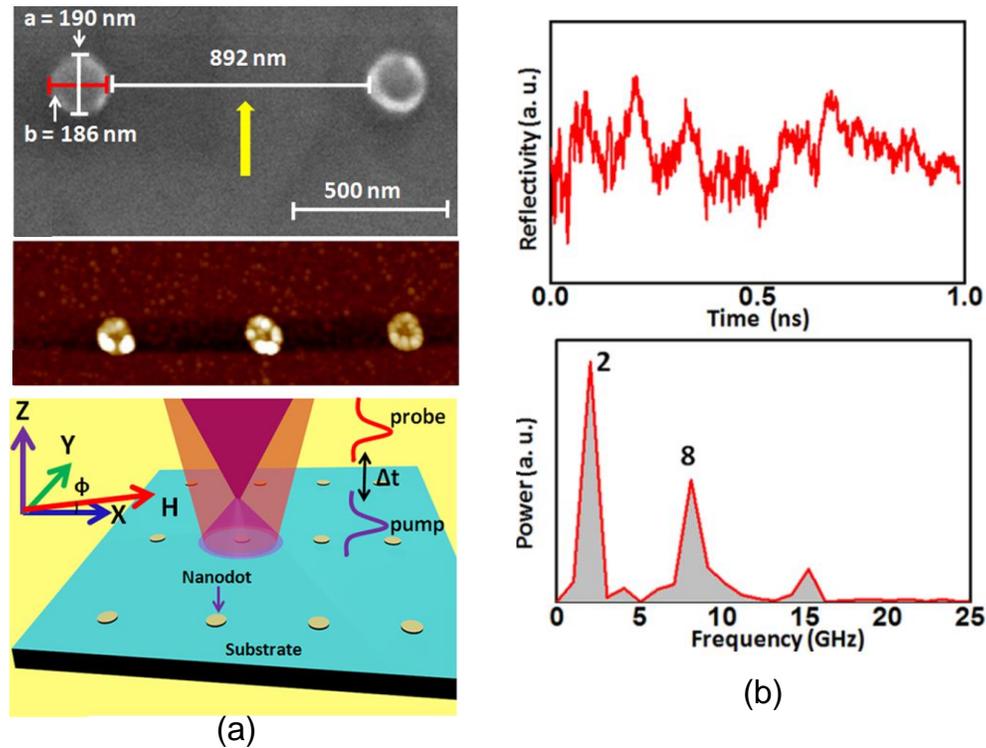

**Figure 20.** (a) TR-MOKE set up showing the scanning electron and atomic force micrographs of the nanomagnets fabricated on PMN-PT substrate in the top panel. (b) The time-dependence of the reflected signal when the laser spot is focused on the PMN-PT piezoelectric substrate and the Fourier transform of the signal. These oscillations are due to the high frequency surface acoustic waves of frequency components at 2, 8 and 15 GHz that are generated in the substrate by the 100 fs laser pulse. Reproduced from [210].

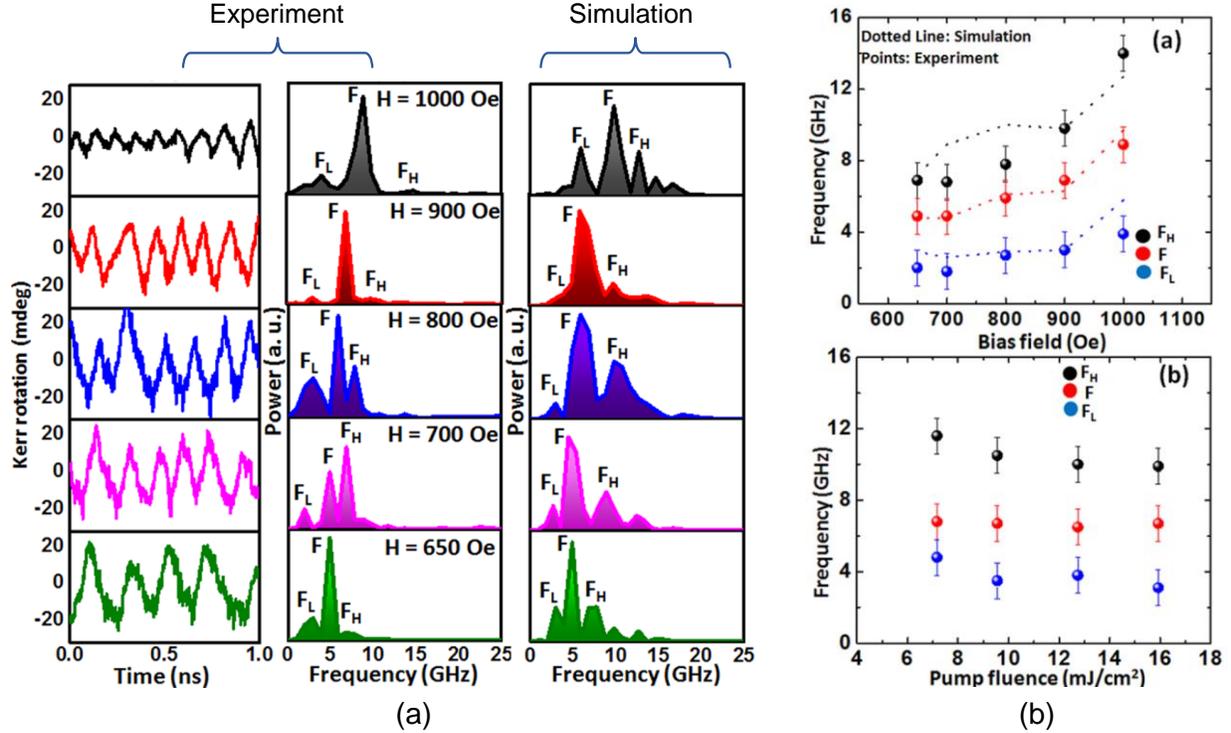

**Figure 21.** (a) Measured Kerr oscillations from a single magnetostrictive nanomagnet excited by the high frequency SAW and their Fourier transform. These are the hybrid magneto-dynamical modes at different bias magnetic fields of 650, 700, 800, 900 and 1000 Oe. There are three primary frequency components at low, intermediate and high-frequencies. Also shown are the results of theoretical simulations. (b) Frequencies of the hybrid magneto-dynamical modes as a function of bias magnetic field and pump fluence. In the bias field dependence plot, we show both the experimental data points with error bars and the simulated results (broken lines). Reproduced from [210].

Figure 21(a) shows the Kerr oscillations (oscillations in the polarization of the light reflected from a single nanomagnet) and their FFTs at bias magnetic fields of 650, 700, 800, 900 and 1000 Oe. We typically see a low frequency peak, an intermediate frequency peak and a high frequency peak in the FFTs. We have also simulated the Kerr oscillations using the micromagnetic simulator MuMax3 [215], where the effect of the laser is modeled with an out-of-plane toggle magnetic field and the SAW is modeled with a time-varying stress anisotropy energy. The far right panel in figure 21(a) shows the results of the simulations. In figure 21(b), we show the experimentally measured frequencies of the hybrid magneto-dynamical modes (the low-, intermediate- and high-frequency modes) as a function of the bias magnetic field and compare them with the results of theoretical simulations. We also show the experimentally measured frequencies of these modes as a function of the laser pump fluence. Note that the frequencies are all in the several GHz range, indicating that magnetization precession induced by high frequency SAW can be very fast and occur in even sub-100-ps time scale.

In figure 22, we show the calculated power and phase profiles of the confined SW modes associated with the hybrid magneto-dynamical modes at different bias magnetic fields. The top row shows the power and phase profiles of the pure Kittel modes in the absence of any SAW, whereas the rest of the rows show the profiles in the presence of the SAW. Note that the SAW changes the profiles dramatically from the pure Kittel modes. Whereas the pure Kittel modes concentrate power at the center (center mode) or edges (edge mode), the hybrid dynamical modes have much more complex profiles showing quantized behavior. These SW modes are all high frequency modes in the several GHz range, showing that SAW can generate high frequency complex SW modes in a single magnetostrictive nanomagnet.

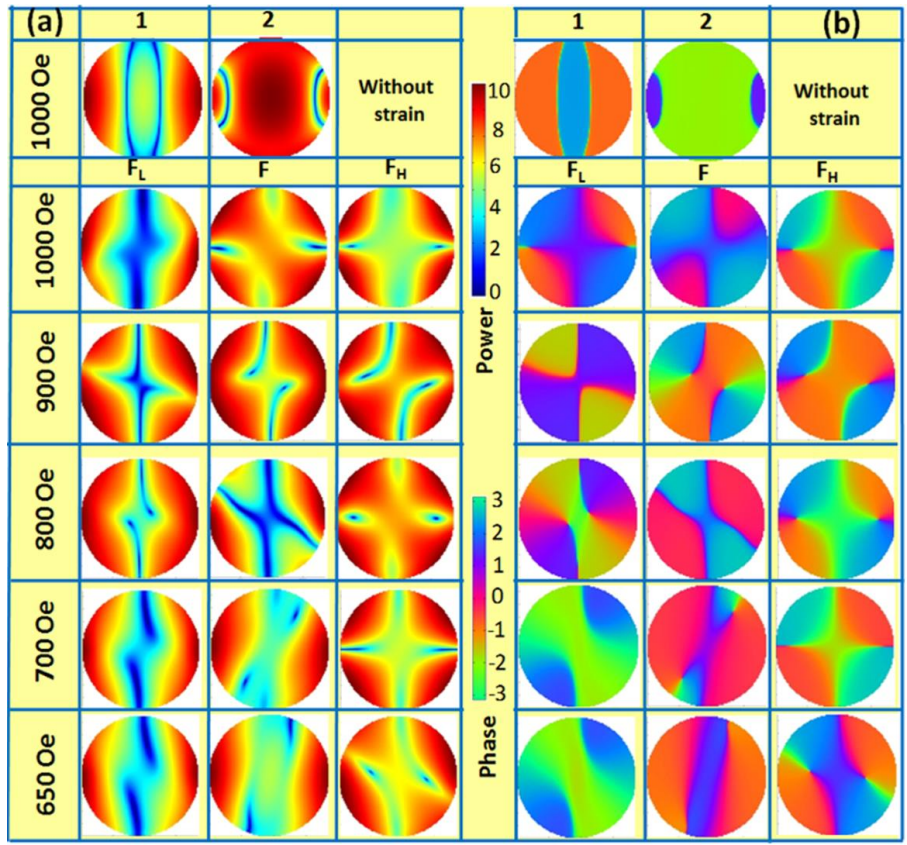

**Figure 22.** (a) Power and (b) phase profiles of the confined SW modes at different bias magnetic fields. The top row shows the profiles for the pure Kittel mode that occurs in the absence of any SAW, while the rest of the rows show the profiles in the presence of the SAW. These are the power and phase profiles of the SWs associated with hybrid magneto-dynamical modes. Reproduced from [210].

## 8.8 Precessional magnetization switching

The magnetization vector in a ferromagnet tends to switch much slower than the Néel vector in an antiferromagnet where the strong exchange coupling between sublattices facilitates ultrafast switching. This shortcoming has stymied the application of ferromagnets in situations where fast switching is desired. Magnetization reversal at picosecond and femtosecond timescale is required for fast access in future ferromagnetic memory and storage devices. Precessional switching is considered as one of the promising mechanisms for this purpose. The reversal time is one-half of the precessional period, which is much shorter than the time it takes to reverse magnetization via domain wall motion (of the order of ns) if the real or effective magnetic field that the magnetization precesses around is sufficiently strong. Precessional motion is at the heart of VCMA based switching of magnetic memory cells (e. g. MTJs) as shown in figure 12.

To investigate precessional switching dynamics, ultrashort pulse generation is a prerequisite, which is a nontrivial challenge. Investigation of picosecond magnetization dynamics started with the availability of femtosecond laser sources, which were used to generate ultrashort excitation. One important issue in precessional switching is known as *ringing*, i.e., continued precession even after reversal, which must be suppressed immediately after half a period of precession has been completed [216, 217]. We discuss some remarkable developments in this area below.

Choi *et al.* demonstrated fast magnetization reversal through domain wall rotation [89, 90] in a 10 μm × 2 μm Py strip placed on a transmission line using stroboscopic scanning Kerr microscopy. By manipulating the bias magnetic field and pulsed magnetic field, fast reversal was achieved. Application of a transverse magnetic field leads to faster reversal (1.2 ns) due to domain wall motion, while the absence of it leads to slower reversal (5 ns) by domain wall nucleation. Schumacher *et al.* demonstrated precessional switching with sub-ns magnetic field pulse applied along the hard axis of a nanomagnet by compensating the easy axis offset field [218]. A 360 ps duration field pulse with an amplitude of 170 Oe led to switching. They also showed precessional reversal using a current pulse of 120 ps duration and $5 \times 10^{11}$ A/m$^2$ current densities in a spin valve stack [219]. About 2% change in the GMR indicated full reversal for the applied current pulse. Precessional switching was confirmed by the observed periodic variation of switching and non-switching when the current pulse duration ($T_{pulse}$) was varied over a wide range. Switching was found to occur when $T_{pulse}= (n+1/2)T_{precession}$, where $T_{precession}$ is the precession period and *n* is an integer. Back *et al.* showed precessional reversal of magnetization in Co/Pt thin film with a 2-ps field pulse having an amplitude of 184 kA/m when applied transverse to the magnetization [220, 221].

A flurry of reports appeared in the literature dealing with magnetization reversal in different systems. Lederman *et al.* reported spontaneous thermal switching in single domain $Fe_2O_3$ particles [222]. Ju *et al.* reported magnetization reversal using sub-picosecond laser excitation resulting in large modulation of the unidirectional exchange bias field across NiFe/NiO bilayers [223]. Pulsed electron beams from the Stanford linear accelerator were used by Siegmann *et al.* to generate ultrashort magnetic field pulses of 6 ps duration for magnetization reversal of CoPt film [224]. Doyle *et al.* [225] and Hiebert *et al.* [226] generated sub-nanosecond magnetic field pulse using a microstrip line. Shaping of the magnetic field pulse was required for suppression of precession (ringing) after the reversal. Gerrits *et al.* addressed the issue of ringing after precessional switching by engineering the field pulse profile [227]. They triggered two GaAs photoconductive switches independently with femtosecond laser excitation, while the desired field pulse profile was achieved by superimposing these two pulses. One pump pulse was used to initiate the precession whereas the other pump pulse acted as a stop pulse. The time delay between them was precisely varied by using a delay line. Reversal time of 200 ps (switching rate of 5 GHz) was achieved by coherent suppression of the ringing by setting the delay as half the period of precession (figure 23).

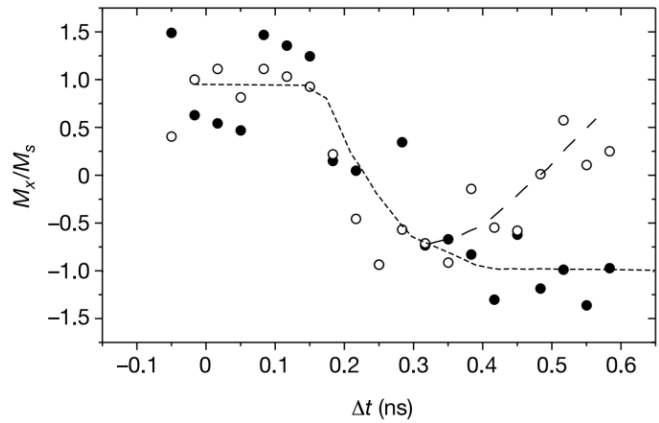

**Figure 23.** Switching by large-field excitation and suppression of ringing. Without a stop pulse, the system switches back to its initial state (open circles). After sending the stop pulse, the suppression of the ringing of the magnetization can clearly be observed (solid circles). The lines are guides to the eye. The low signal-to-noise ratio in the $M_x$ component results from the very weak longitudinal MSHG signal with an incoming polarization parallel and second harmonic polarization perpendicular to the plane of incidence. Reprinted from [227] with the permission of Springer Nature.

Hiebert *et al.* [228] used spatial, temporal and vector resolved Kerr microscopy to demonstrate large angle precessional magnetization dynamics in a stadium shaped $Ni_{80}Fe_{20}$ (permalloy, Py hereafter) element. A static magnetic field of 100 Oe was applied along the +*x*-direction (short axis) and a magnetic field pulse of -160 Oe (along -*x*-direction) was produced with a transmission line. The shape of the sample generated an extra magnetic field along its long axis (*y*-direction). When the field pulse varied from -60 Oe to +100 Oe, the magnetization experienced a torque along the *y*-direction and started a nearly uniform reversal process. The torque drove the magnetization out of the *x-y* plane, which, in turn, generated a demagnetizing field. The magnetization showed a damped resonant oscillation around the +*x* axis with decreasing demagnetizing field. High frequency SW generation was attributed to the reduction of the magnetization vector length. Precessional switching in a spin valve element, pinned along its easy axis, was reported by Kaka *et al.* [229] using a magnetoresistive technique. A bias field was applied along the easy axis and a pulsed field of 216 Oe peak amplitude and 230-325 ps duration was applied along the hard axis of the sample. The duration and the amplitude of the pulse field was found to be crucial for reversal via large angle precession, and the probability of switching decreased with increasing pulse duration.

In 2004, Kent *et al.* theoretically proposed the notion of spin-current pulse driven rapid magnetization reversal of a nanomagnet. The switching speed is determined by the precession frequency of the thin film element. Micromagnetic simulations showed that this switching occurs above a threshold pulse current, which can be faster than 50 ps [94]. Lee *et al.* demonstrated reliable spin-torque-driven ballistic precessional switching in a spin-valve device having both in-plane and out-of-plane spin polarizers using 50 ps current impulses. Varying the threshold currents as a function of switching direction and current polarity enabled the final orientation of the magnetic free layer to be steered by the sign of the pulse. This can eliminate the need for read-before-write toggle operation [230]. In Pt/Co/AlO$_x$ nanodots of 90 nm width, STT were used for deterministic magnetization reversal in timescales ranging between 180 ps to ms in a switching geometry based on in-plane current injection. The results established two distinct regimes: a short-time intrinsic regime, where the critical current for STT, $I_c$, scales linearly with the inverse of the pulse duration, and a long-time thermally assisted regime, where $I_c$ varies weakly [231] with pulse duration. Baumgartner reported direct observation of SOT-driven magnetization dynamics in Pt/Co (1 nm)/AlOx (2 nm) dots (having 500 nm diameter) during a current pulse injection, using time-resolved x-ray imaging with 25 nm spatial and 100 ps temporal resolution. The switching was achieved with a sub-nanosecond current pulse through the fast nucleation of an inverted domain at the edge of the dot and propagation of a tilted domain wall across the dot. The combined action of the damping-like and field-like SOTs and the Dzyaloshinskii–Moriya interaction broke the magnetic symmetry and reproducible switching events were observed over $10^{12}$ reversal cycles [232].

A heat assisted route involving nonlinear magnetization precession, triggered by a transient thermal modification of the growth-induced crystalline anisotropy in the presence of a fixed perpendicular magnetic field, was reported in a dielectric bismuth-substituted yttrium iron garnet. During the switching process, the damping became anomalously large in contrast to the usual small damping [233].

In a more recent development, Jhuria *et al.* demonstrated generation of electrical pulse as short as 6 ps using photoconductive switches which was injected into a Ta(5 nm)/Pt(4 nm)/Co(1 nm)/Cu(1 nm)/Ta(4 nm)/Pt(1 nm) heterostructure [234]. A complete reversal of Co magnetic moment by SOT was achieved by manipulating the polarity of the injected current pulses with respect to an in-plane magnetic field.

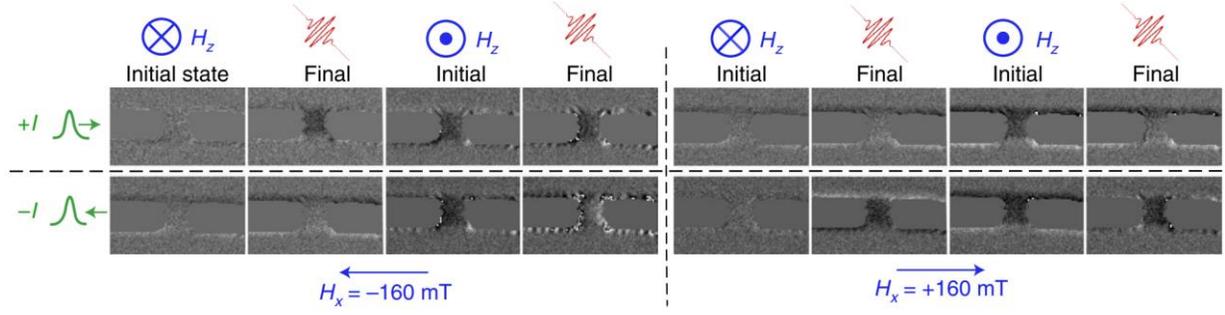

**Figure 24.** MOKE micrographs of single 6-ps electrical pulses switching the magnetization via SOT. The four quadrants show two before-pulse and two after-pulse images under different in-plane field and current directions. The inversion of the final state with current or in-plane field is a clear signature of SOT switching. Bias voltages used for switching were slightly above the critical threshold ($\Delta V \approx 40$ V). Light and dark grey indicate magnetization down and up, respectively. Reproduced from [234] with permission of Springer Nature.

Figure 24 shows the polar MOKE micrographs of initial and final configurations after injection of a single 6-ps electrical pulse [234]. By investigating the effects of various combinations of current ($I$) and in-plane magnetic field ($H_x$) directions, as well by repeating the experiment for a number of times, the final state was found to be independent of the initial magnetic state. Additionally, the switching probability was better than 91% with a 95% confidence interval. By further probing the generated magnetization torques with sub-picosecond resolution, the reversal process was found to be energy-efficient.

## 8.9 All-optical switching

The thermodynamic potential of an isotropic, non-absorbing and magnetically ordered medium with static magnetization $M(0)$ and magneto-optic susceptibility $\alpha_{ijk}$ in presence of a monochromatic light field $E(\omega)$ has the form

$$F = \alpha_{ijk} E_i(\omega) E_j(\omega)^* M_k(0) \quad (4)$$

The electric field of light at frequency $\omega$ acts on the magnetization as an effective magnetic field directed along the wavevector of the light **k** given by:

$$H_k = -\frac{\partial F}{\partial M_k} = -\alpha_{ijk} E_i(\omega) E_j(\omega)^* \quad (5)$$

In an isotropic media, $\alpha_{ijk}$ is a fully antisymmetric tensor with a single independent element $\alpha$ and equation (5) may be rewritten as:

$$\mathbf{H} = \alpha [\mathbf{E}(\omega) \times \mathbf{E}(\omega)^*]. \quad (6)$$

Thus, right- and left-circularly polarized light should act as magnetic fields of opposite signs and there can be an inverse Faraday effect caused by the susceptibility $\alpha$.

Complete non-thermal or opto-magnetic excitation and control of magnetization dynamics was first reported by Kimel *et al*. in DyFeO$_3$, in the hundreds of gigahertz frequency range [235]. In DyFeO$_3$ the antiferromagnetically coupled Fe spins are slightly canted due to the Dzyaloshinskii-Moriya interaction, leading to a small spontaneous magnetization of 8Gauss, while having a giant Faraday rotation of 3000 ° $cm^{-1}$. These properties enabled the detection of optically induced magnetization by measuring the direct magneto-optical Faraday effect. The transient magnetic signal is followed by magnetization oscillation with about 200 GHz frequency, while its sign depends on the helicity of the pump pulse indicating a direct coupling between photons and spins in $DyFeO_3$. A linear dependence of the photo-induced spin oscillation on the pump intensity indicates the photo excitation of magnons to be a two-photon process as predicted by Eq. #6. This portends a coherent control of spin precession by using two pump-pulses separated by a controllable optical delay. There, a pump-pulses of helicity $\sigma^+$ is used to trigger the antiferromagnetic spin precession in $DyFeO_3$ followed by another pump pulse of same helicity but time delayed by an odd multiple of half precession period to shift the magnetization farther away from the effective magnetic field causing an amplification of precession amplitude. On the contrary, if the second pump-pulse is time delayed by an integer number of full precession period the precession stops.

All-optical helicity dependent switching (AO-HDS) was investigated in GdFeCo thin films [236]. The laser pulses were incident normal to the sample surface, to induce an effective optically generated magnetic field directed along the magnetization, similar to a conventional recording scheme. The laser beam was swept at high speed across the sample, to land each pulse at a different spot. The switching of domains with opposite magnetization by pulses of opposite helicity unambiguously demonstrated all-optical magnetization reversal by single ultrashort circularly polarized laser pulses without the aid of an external magnetic field. It was envisaged that high-density recording may also be achieved by employing near-field antenna structures similar to those being developed for heat assisted magnetic recording. These experiments, combined with emerging development of ultrafast lasers and nanotechnology, may lead to the realization of a new generation of magnetic recording devices.

The development of AO-HDS was initially limited to ferrimagnetic systems, e.g., rare earth – transition metal (RE-TM) alloys and synthetic ferrimagnets, where two different sub-lattices are antiferromagnetically exchange coupled. The mechanism of AO-HDS was thought to be due to the existence of an effective field created by the circularly polarized light via the inverse Faraday effect or by the angular momentum transfer from the photons to the magnetic system. Claims of formation of a transient ferromagnetic state due to different demagnetization times for RE and TM sublattices as the reason behind AO-HDS were also made, where the helicity of light plays a secondary role [237]. Laser-induced super-diffusive spin currents flowing in heterogeneous systems may potentially contribute to the AO-HDS process [238-242]. More recently, Mangin *et al.* showed AO-HDS in a much broader variety of materials [243], such as rare earth-free Co-Ir-based synthetic ferrimagnetic heterostructures and Co/Tb multilayers.

Subsequently, Lambert *et al.* demonstrated optical control of ferromagnetic thin films and multilayers [244], including [Co(0.4 nm)/Pt(0.7)]$_N$, [Co($t_{Co}$)/Pt($t_{Pt}$)]$_N$, [Co($t_{Co}$)/Pd($t_{Pt}$)]$_N$, [Co$_x$Ni$_{1-x}$(0.6 nm)/Pt(0.7 nm)]$_N$, and [Co/Ni]$_N$ multilayer structures, where several material parameters ($t_{Co}$, $t_{Pt}$, N, and Ni concentration) were varied that change magnetic properties such as magnetization, Curie temperature $T_C$, anisotropy, and exchange interaction. The AO-HDS was further studied in high-anisotropy FePt-based HAMR media, which are FePtAgC and FePtC granular films forming high-anisotropy FePt grains separated by C grain boundaries. The average grain size of ~9.7 and ~7.7 nm for the FePtAgC and FePtC correspond to the room-temperature magnetic anisotropy and coercive fields of 7 T and 3.5 T, respectively. A helicity dependent net magnetization was achieved for the circularly polarized light, whereas no change was observed with linear polarization, demonstrating clear control of magnetization by the polarization of the light. The observation of AO-HDS switching on single Co films and Co/Pt, Co/Ni multilayers suggested that super-diffusive currents coupling different magnetic regions in a heterogeneous sample for AO-HDS is unlikely. However, heating near the Curie point is important for the AO-HDS in ferromagnetic materials, because the threshold intensities for both AO-HDS and thermal demagnetization (TD) generally follow the expected trends for $T_C$ and do not scale with parameters such as interlayer exchange or anisotropy. Near $T_C$, the inverse Faraday effect resulting in transfer of angular momentum from the light to the magnetic system is expected to be most effective. Moreover, the experimental geometry makes the opto-magnetic field parallel to the magnetization, ruling out the possibility of precessional switching well below $T_C$. Furthermore, if demagnetization and thermal energies are too large, then the sample will demagnetize during cooling.

However, achieving AO-HDS in a single nanomagnet has been a daunting challange. El-Ghazaly *et al.* revealed intriguing physics of AO-HDS in a single nanomagnet which can lead to direct and fast data writing. To this end, smaller nanoelements settling to their final magnetization states faster (~2 ps) after switching than larger elements was an exciting observation [245]. The faster switching speed was attributed to the electron-lattice and spin-lattice interactions with higher spin temperatures for smaller nanoelements [144]. Lalieu *et al.* experimentally demonstrated combined effects of both thermal single-pulse AOS and SHE induced domain wall motion in a Pt/Co/Gd racetrack with PMA. This exploits the chiral DW Neél structure associated with the coherent and efficient motion of the optically written domains. They claimed Pt/Co/Gd racetrack to be an ideal candidate to facilitate the integration of AOS with spintronics. The thermal nature of the AOS suggests that the final downsizing of the AOS towards the nanometre scale can be done using plasmonic antenna to heat the recording material very locally by reducing the laser-pulse size to sub-50 nm. This might pave the way towards integrated photonic memory devices [246].

# 9. EXPERIMENTAL METHODS FOR EXCITING AND DETECTING MAGNETIZATION DYNAMICS IN NANOMAGNETS

## 9.1 Experimental techniques for exciting magnetization dynamics

There are several ways to excite precessional magnetization dynamics in nanomagnets and their arrays. Microwave induced Oersted field is one of the most popular ones. For this purpose, coplanar waveguides are fabricated on top of the nanostructures and the RF current induced Oersted field is employed to excite ferromagnetic resonance (FMR) (figure 25(a)) [48]. In some cases, the nanostructures are fabricated between the ground and the signal lines of the coplanar waveguide. Conventional microstrip antennae are also used for exciting FMR and SWs (figure 25(b)) [247]. When a microwave current is passed through the microstrip antennae, the current induced Oersted field couples with the SW modes and excites propagating SWs. In some cases, arrays of nanomagnets and thin films are directly deposited on a wide metal strip line for exciting magnetization dynamics (figure 25(c)) [248]. The FMR technique is a robust method for the characterization of the dynamic properties of nanostructured magnetic materials [249]. In this case, a magnetic material is placed inside a microwave cavity having a single resonant frequency with high Q-factor. When the applied bias magnetic field satisfies the resonant condition, the FMR of the magnetic materials under study is excited, and the absorption of microwave power in the cavity reaches a peak. The absorption is measured as a function of the bias magnetic and typically exhibits a Lorentzian line shape, from whose linewidth, one can obtain information about the damping or loss in the material [250].

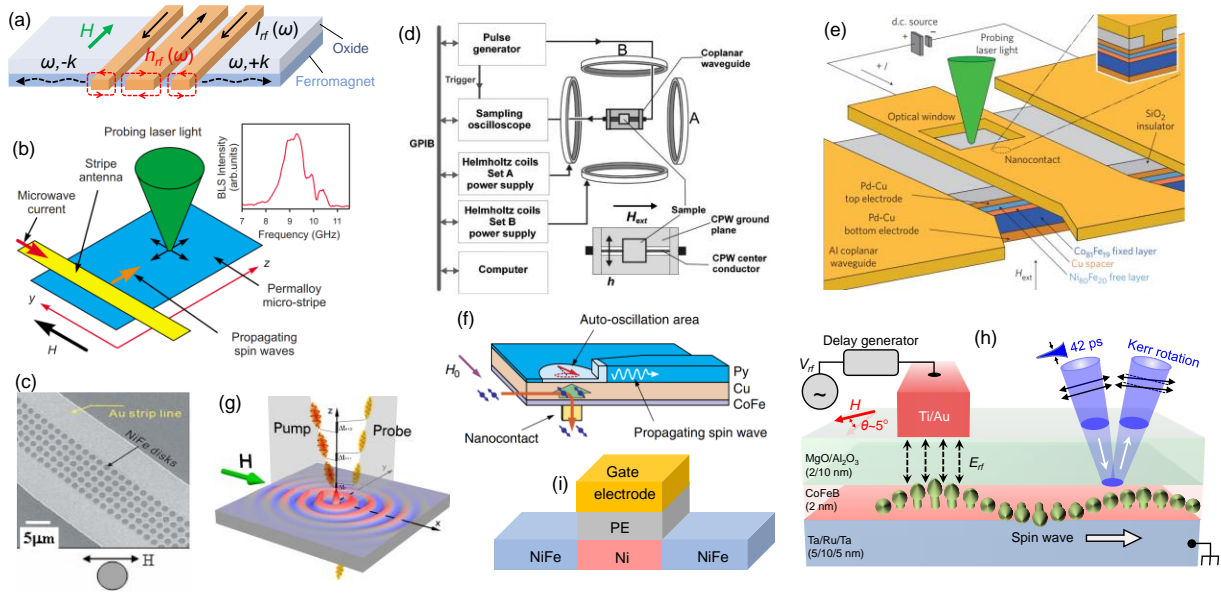

**Figure 25.** (a) Schematic diagram shows SW excitation by radio frequency (rf) current induced Oersted field. Current is passed through a microwave antenna to generate spatially nonuniform Oersted field, which inductively couples to the SWs having same frequency (ω) as the rf current. (b) Schematic diagram of a device for the excitation of SWs by microstrip antennae. The excited SWs can be probed by optical means, e.g., Brillouin light spectroscopy (BLS). Reproduced from [247] with permission of the American Physical Society. (c) SEM micrographs of the arrays of Py microdots deposited on a Au microstrip line. Reproduced from [248] with permission of the American Institute of Physics. (d) Schematic illustration of a pulsed inductive microwave magnetometer (PIMM). The sample is placed on the coplanar waveguide (CPW) structure. Reproduced from [249] with permission of the American Institute of Physics. (e) Schematic diagram shows the cross-section of a sample used for spin-transfer torque induced SW excitation. Reproduced from [251] with permission of the Springer Nature. (f) Schematic diagram of a device used for the excitation of SWs by pure spin current. Reproduced from [252] with permission of the Springer Nature. (g) Schematic diagram shows the excitation of SWs by femtosecond laser pulses. Reproduced from [253] with permission of the Springer Nature. (h) Schematic diagram of a device and experimental set up for the excitation of SWs by VCMA and detection by optical means [254]. (i) The schematic diagram shows the excitation of SWs by magneto-electric cell (ME cell).

Dynamical properties of magnetic materials are often measured with a pulsed inductive microwave magnetometer (PIMM). It uses a coplanar waveguide as a source of fast pulsed magnetic fields provided by a pulse generator, and also an inductive flux sensor which can measure the fast pulse data using a digital sampling oscilloscope (figure 25(d)) [249, 255]. Oersted fields are not suitable for nanostructures owing to their spatially non-localized nature. However, some techniques enable spatially uniform microwave Oersted fields to excite SWs with shorter wavelength without localizing it. For instance, SWs with shorter wavelength can be excited by creating magnetic nonuniformities in a sample either naturally or artificially. Furthermore, a localized dynamic dipolar field created by the FMR of a rod-like magnetic nanostructure placed on top of a SW waveguide can excite shorter wavelength SWs [256]. Even the smooth interface between two magnetic films can excite short wavelength coherent SWs very efficiently [257]. Spin current induced torques (STT/SOT) is another potential method for exciting magnetization dynamics in nanostructured magnetic elements [251]. For exciting FMR using STT, one patterns the magnetic material into sub-micrometre size elements, or injects the spin current from a sub-micrometre sized nanocontact. By changing the sample geometry, it is also possible to create pure spin current and excite FMR in magnetic thin films by spin-orbit torque (SOT) (figure 25(f)) [252]. The key advantages of exciting FMR with STT and SOT are that local excitations are possible and the excited FMR can also be amplified [258]. However, all the above-mentioned current induced methods suffer from significant power dissipation owing to inherent Joule heating. Irradiation with a femtosecond pulsed laser beam is yet another efficient way for exciting FMR and coherent SWs in nanomagnets (figure 25(g)) [253, 259].

Femtosecond laser pulses can partially or fully demagnetize an ordered magnetic material within sub-picosecond time scales and efficiently excite coherent propagating SWs in the process [253]. The key advantages of this method are the local nature of the excitation technique and the ability to choose the propagation direction of excited SWs by shaping the laser pulse. Owing to the thermal agitation the spins in magnetic materials always oscillate in an incoherent manner, which excites very low intensity SWs even in absence of other external perturbations [260]. The periodic heating and cooling by a laser can thermally excite SWs which are studied for characterizing various dynamics parameters [261]. This method obviates the need for complex device fabrication.

VCMA is an emerging technique for exciting picosecond magnetization dynamics of magnetic thin films and nanostructures. Since VCMA at FM/oxide interfaces relies upon the modification of electronic occupation states in 3$d$ orbitals of FM, it is suitable for microwave applications. Nozaki *et al.* [262] and Zhu *et al.* [263] demonstrated that FMR can be excited by VCMA in magnetic nanostructures with sub-micrometer dimensions accompanied by ultralow power consumption which is at least two orders of magnitude lower than that in current induced STT excitation. VCMA is also suitable for excitation of coherent propagating SWs in ultrathin FM films (figure 25(h)) [254]. In this case, a metal gate electrode is placed on top of an oxide layer deposited on top of the FM layer, i.e., a waveguide. An alternating voltage of microwave frequency is applied at the gate electrode and that launches a microwave electric field into the oxide layer. The microwave periodically modulates the PMA of the FM layer underneath the gate electrode via the VCMA effect. At resonance frequency, coherent SWs are excited and propagate along the waveguide.

Use of a magneto-electric cell (ME cell) [264] is another approach for exciting magnetization dynamics with an electric field. As shown in figure 25(i), a FM spin wave bus (NiFe) has a magnetostrictive layer made of embedded (Ni), which, in turn, has a piezoelectric layer (PE) on top. When a voltage is applied across top metal gate electrode, it causes deformation in the piezoelectric layer, which is then transferred to the Ni layer. As a result, the magnetization of the Ni layer is perturbed and SWs are created at the resonance condition [265].

It is also possible to excite magnetization dynamics in nanomagnets indirectly. When they are placed sufficiently close to each other, they are coupled via magnetostatic interactions which can transfer magnetization dynamics induced in one nanomagnet to the next, and so on. When a spin-torque nano-oscillator (STNO) is delineated over a small portion of an extended thin film, the oscillatory motion of the nano-oscillators can emit SW in the FM film [266]. Likewise, oscillatory motion of magnetic vortices can also emit SWs into an adjacent magnetic channel [267].

**9.2 Experimental techniques for detecting magnetization dynamics**

There are numerous methods for detecting magnetization dynamics occurring in nanomagnets. Whenever the dynamics is excited with microwave excitation involving coplanar waveguide and micro strip lines, the resonance signals are typically measured by inductive coupling using coplanar waveguides in the manner of antennas [268]. In magnetic heterostructures, the magnetization dynamics of magnetic layers are detected via spin pumping and the inverse spin Hall effect (ISHE) or the inverse Edelstein effect (IEE) in the form of a dc voltage. Homodyne detection is another sensitive technique for detecting resonance signals from sub-hundred nanometer magnetic elements with sub-nanometer thickness. In this case, the probed magnetic layer is fashioned into the free layer of a magnetic tunnel junction (MTJ). The time varying MTJ resistance due to the oscillation of the soft layer's magnetization is mixed with the time varying current through the junction, resulting in a rectified dc voltage. In STT or SOT induced FMR, the resonance signals are also detected via a dc voltage. In this case anisotropic magnetoresistance due to the oscillation of magnetization is mixed with the RF current passing through the magnetic layer giving rise to a finite dc voltage. Recently, inverse VCMA has been reported as another interesting technique for detecting resonance signals [269]. The anomalous Hall effect (AHE) is another technique to detect FMR signals.

Optical techniques are best known for the detection of FMR signals locally in a non-invasive way. Time-resolved magneto-optical Kerr effect (TR-MOKE) is one of the most sensitive and widely used optical technique for locally detecting dynamic magnetic signals from various types of magnetic nanostructures [270]. In Brillouin light spectroscopy (BLS), inelastic scattering of light from the magnons (quanta of SWs) is utilized for detecting SW signals [271]. Recently, the local, quantitative, and phase-sensitive detection of SWs has also been demonstrated by single nitrogen-vacancy (NV) centers in diamond [272]. However, this method can increase the complexity of device fabrication since FM films need to be deposited directly on a diamond film containing NV centers. In ME cells, the inverse of the above discussed process is used for detecting SW signals. Some of the non-conventional methods for detecting FMR signals are magnetic resonance force microscopy (MRFM) [273], magnon-induced heat [274], nuclear resonant scattering [275], and time resolved x-ray magnetic circular dichroism [276].

## 10. SPIN WAVES IN CONFINED MAGNETIC MEDIA

Investigation of magnetization dynamics in patterned microstructures providing spatial confinement to the SWs started in the late 1990s. In a seminal work, Heibert *et al.* reported the measurement and imaging of magnetization precession and relaxation dynamics in an 8 μm wide Py disk [226]. The Py disk was patterned at the centre of a lithographically patterned gold coil which was connected with a biased photoconductive switch made of GaAs. The photoconductive switch was triggered by using a pulsed laser of 2 ps pulse-width producing a pulsed magnetic field perpendicular to the plane of the Py disk, which was biased by an in-plane magnetic field. The time-resolved precessional oscillation was measured with a time-resolved scanning Kerr effect microscope (TRSKEM) having 0.7 μm spatial resolution. The bias-field dependent precession frequency was fitted with the Kittel formula to extract various magnetic parameters, including a damping constant of 0.008 for Py. The space-time evolution of magnetization was imaged by fixing the time delay and then scanning the disk under a focused probe laser spot. This yields a non-uniform spatial distribution, indicating the presence of modal frequencies in this confined structure. In another interesting work Jorzick *et al.* used BLS spectroscopy to measure the SW spectra in rectangular elements having lateral dimensions of $1 \times (1 - 2)$ μm$^2$ and stripes with 1 μm × 90 μm area [277]. Thermal magnons were measured in the backscattered geometry. The spectra revealed a number of eigenmodes in the system consisting of Damon-Eshbach (DE), backward volume (BV) and perpendicular standing SW (PSSW) geometry. In particular, they observed the existence of a new, spatially localized SW mode of exchange nature in these elements with the localization caused by the inhomogeneity of the internal field. In 2003, Barman *et al.* used TRSKEM to study the precessional magnetization dynamics of a 10 μm square $Ni_{81}Fe_{19}$ element [278]. Time-resolved Kerr rotation at the center of the element revealed the presence of a fourfold anisotropy in the precession frequency due to the internal field generated by the nonuniform static magnetization. The damping was much larger when the static magnetic field was applied parallel to the diagonals of the square. The dynamic images revealed this to be associated with spatial nonuniformity at the centre of the element. The dynamic magnetization was initially nonuniform at the edges of the element, with the nonuniformity then extending towards the centre [279]. These authors thoroughly investigated the dephasing of SW modes by extensive imaging of the dynamic magnetization combined with micromagnetic simulations. The measured transient Kerr rotation at different positions along an axis parallel to the bias field revealed a number of different modes. The number of antinodes in modal oscillation from dynamics images was found to be larger along the diagonal than along the edge of the square explaining the observed increase in damping due to dephasing of modes (figure 26) [280]. The authors later systematically studied the anisotropy in the apparent damping in samples with varying shape [281] and aspect (width/thickness) ratio and found it to become stronger with decreasing aspect ratio [282]. Extensive micromagnetic simulation study revealed the spatial profile of each resonant mode, which revealed a uniform centre mode, localized edge modes, and other localized modes with nodal planes perpendicular to the bias magnetic field [283]. Belov *et al.* studied modal oscillations in a 4 μm $Ni_{80}Fe_{20}$ element with a central pinhole to show that the spatial pattern of the magnetization oscillation response depends sensitively on weaker variations of the static magnetization. They also observed a spatially nonuniform damping in this sample as a result of conversion of energy into shorter wavelength modes in the vicinity of the domain boundaries [284].

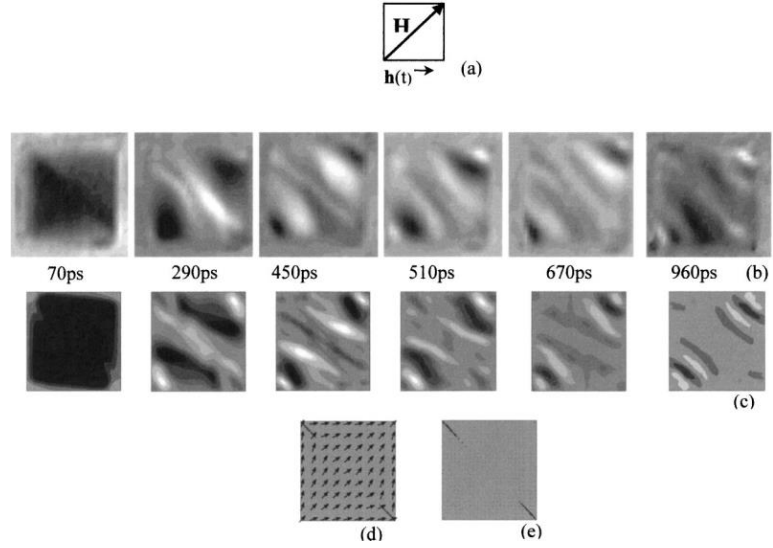

**Figure 26.** The experimental dynamic images (b) and the simulated dynamic images (c) obtained in the experimental geometry (a) are shown. The static magnetization (d) and the total internal field (e) in the static configuration are shown. The greyscale represents the out-of-plane components of the magnetization and internal field. Reproduced from [280] with the permission of American Physical Society.

Barman *et al.* also studied coherent suppression of precession dynamics in circular microdots of $Ni_{81}Fe_{19}$ with varying aspect ratio with a shaped magnetic field pulse. Coherent suppression was achieved by adjusting the static field so that there was no net torque acting upon the magnetization after one cycle of precession. The coherent suppression became more spatially nonuniform in samples with smaller aspect ratio. Time-resolved Kerr images and simulations of the demagnetizing field within the samples suggested that increased nonuniformity resulted from confinement and dephasing of multiple SW modes affecting the coherent suppression [285]. With continuing advancement of nanofabrication capabilities, the focus in this field gradually shifted to studying the dynamics of single and arrays of nanomagnets as described in the following sections.

**10.1 Spin waves in single nanomagnets**

Investigation of the intrinsic dynamics of single nanomagnet in the deep nanoscale regime and at extremely fast timescale is of pure fundamental interest. The spatial confinement should suppress the domain formation in the nanoscale according to Brown's fundamental theorem. An individual nanomagnet is expected to behave like a macrospin comprising of large number of isotropic spins aligned together. However, nanomagnets with non-ellipsoidal shapes exhibit shape anisotropy arising due to geometric configuration besides magnetocrystalline and strain anisotropies. Furthermore, the magnetic microstate may vary with relative orientation between the applied magnetic field and the anisotropy field. Size reduction from micrometer to nanometer scale results in transitions between various magnetic microstates, namely multidomain states, magnetic vortex, quasi-single domain states and single domain. Consequently, the magnetization dynamics shows a variety of behaviour, such as domain wall oscillation, vortex core gyration, radial and azimuthal modes of SWs, confined, localized and standing SW modes etc.

Despite the experimental studies of SW spectra arrays of submicron magnetic structures in early 2000 using FMR [286] and BLS [287] techniques, the study of intrinsic dynamics of isolated nanomagnets was missing. In 2003, Kiselev *et al.* addressed an outstanding question of the type of magnetic motions that can be generated by STT. By using electrical measurement on a magnetic-multilayer nanopillar structure of 130 nm × 70 nm lateral dimensions (80 nm Cu/40 nm Co/10 nm Cu/3 nm Co/2 nm Cu/30 nm Pt fabricated on an oxidized silicon wafer), they showed that the nanopillar acts like a nanoscale motor, i.e. it converts energy from a dc electrical current into high-frequency magnetic rotations that might serve as nanoscale microwave sources or oscillators, tunable by current and magnetic field over a wide frequency range [5]. In 2004, Demidov *et al.* showed that a sub-micrometer sized individual Py magnet fabricated on top of a CoFe microstructure separated by copper spacer with memory-element like geometry can act as a genuine source of microwave radiation [288]. Using micro-focused BLS measurement they showed radiation of SWs into the surrounding magnetic film, which occurred at two discrete frequencies (around 10 to 12 GHz) corresponding to the frequencies of the quantized SW modes of the element. Koch *et al.* investigated time-resolved spin-transfer-induced (STI) magnetization reversal in nanosecond timescale in a current-perpendicular spin-valve nanomagnetic junctions subjected to a pulsed current bias [289]. In another important work Krivotov *et al.*, in 2005, investigated STT driven time-resolved magnetization dynamics of a Cu/IrMn/Py/Cu/Py/Cu nanopillar system of 130 nm x 60 nm lateral dimensions [290]. Their experimental results were in excellent agreement with the predictions of the Slonczewski spin-torque model [291], i.e. STT reduces the effective magnetic damping for applied charge currents less than a critical current. At higher currents, STT was able to drive phase-coherent precessional dynamics. STT could drive precessional switching with switching times less than 1 ns having narrow statistical distributions, which is a very exciting result for magnetic-memory applications. Later, in 2006, x-ray microscopy technique was used to image the magnetization switching process in a spin transfer structure [292]. Switching was found to occur by lateral motion of a magnetic vortex across a nanoscale element instead of coherent magnetic reversal.

Optical measurement of time-resolved magnetization in single nanomagnets well below the diffraction limit prompted the development of cavity-enhanced MOKE (CEMOKE) technique. In 2006, Barman *et al.* reported the ps dynamics of individual Ni nanomagnet with different sizes by exploiting time-resolved CEMOKE technique [270]. A drastic and non-monotonic variation of the resonance frequency from sub-GHz to GHz range was observed as the cylindrical magnet underwent transition in microstates from multidomain- to single-domain state. Additionally, with decreasing diameter, the damping of the coherent precessional mode of these cylindrical nanomagnets decreased rapidly to finally settle down to its intrinsic value in the single domain regime. The bias field dependence of the precessional frequency confirmed an extrinsic contribution to damping in micrometer sized magnets [293]. Around the same time, Laraoui *et al*. reported ultrafast demagnetization, precession [294] and relaxation [295] of $CoPt_3$ single sub-µm dots using an all-optical TR-MOKE system having femtosecond temporal resolution and high spatial resolution (~300 nm) achieved by the use of a reflective confocal Kerr microscope. A fast relaxation time of few ps, where electrons and spins exchange energy with the lattice, and a slow relaxation time of hundreds of ps, when electrons and the lattice exchange energy with the environment, were found in these nanomagnets. In 2008, Liu *et al*. measured the time-resolved magnetization dynamics of single 160 nm wide Py disks using time-resolved Kerr microscopy. While sweeping the bias magnetic field, a distinct hysteresis of fundamental mode frequency as a function of the in-plane bias field was observed owing to the variation of the internal spin configuration from vortex to quasi-single domain states. The critical fields to trigger the vortex annihilation and nucleation processes were determined in this study [296]. In 2011, remarkable advances in studying single nanomagnet dynamics were reported, when Rana *et al*. measured the time-resolved precessional dynamics of an isolated 50-nm-wide Py dot showing a dominant edge mode (EM) without the appearance of a center mode (CM) [297]. Liu *et al*. showed the ability to distinguish the higher-frequency dynamics of a single 150 nm wide nanomagnet from the lower frequency background of 500 nm wide nanomagnets by placing them in the same array. The optical diffraction limit could be avoided as the characteristic frequency of the 150 nm dot was sufficiently different from those of its neighbors [298]. Naletov *et al*. used MRFM to study the radial and azimuthal eigenmodes in a Py/Cu/Py spin-valve-like nanopillar by applying spatially uniform rf field or rf current flowing through the nanopillar. They established a selection rule for exciting different resonant modes by adjusting the excitation geometry [299]. Keatley *et al*. demonstrated controlled suppression of EM in an individual nanomagnet by excitation of larger amplitude coherent precession of CM. This is necessary for nanoscale STT oscillators and bi-stable switching devices where more uniform spin dynamics is desirable [300].

In further study, the antidamping torque from a pure spin current was found to set the magnetization precession to auto-oscillation by suppressing its intrinsic damping [266]. Investigation showed that SOT, generated from an adjacent heavy metal layer in a ferromagnet can switch the nanomagnet in a deterministic way, which can be used to construct a field-free clocking and nanomagnetic logic analogous to current CMOS technology [301]. Hybrid magnonics arising from strong coupling of magnon with photon, phonon or other magnon has emerged as interesting topic for quantum transduction. To this end direct observation of strongly coupled magnon–phonon dynamics by varying the direction of the applied magnetic field was important development [302]. Strain-induced switching also established a successful Boolean operation recently [208]. More recently, hybrid magneto-dynamical modes was observed in single magnetostrictive

Co quasi-elliptical dots deposited on a piezoelectric substrate using TR-MOKE microscope as described in detail in section 8.7.4 [210].

**10.2 Spin waves in 1D arrays of nanomagnets**

The study of picosecond magnetization dynamics in periodic arrays of nanomagnets is motivated by many potential applications. Such arrays may be thought of as artificial "crystals", known as *magnonic crystals* (MCs), in analogy with a material crystal that provides a periodic potential landscape for an electron wave or a photonic crystal that provides a periodic refractive index landscape for an electromagnetic wave. Just as there are electronic band structures in a material crystal and photonic band structures in a photonic crystal, there are magnonic band structures in an MC, which can be tailored by choosing the pitch (or periodicity) of the array, as well as the shapes and materials of the nanomagnets. The study and control of magnonic band configurations in nanomagnet arrays is very important for developing various types of magnonic devices such as microwave filters, attenuators, sensors, transistors, logic gates and so on. The dynamic behaviour of a single nanomagnet depends upon the ground state of magnetization, which, in turn, is determined by the shape, size, material parameters, strength and orientation of any external magnetic field, strain, etc. Interestingly, the dynamic behaviour of nanomagnet arrays is discernibly different from that of the individual nanomagnets making up the array owing to the magnetostatic interaction among the nanomagnets. The latter again depends upon the shape, size, material parameters, interelement separation, as well as the strength and orientation of any external magnetic field. Magnonic bands can be engineered by manipulating these parameters. In this topical review, we will restrict our discussion only to arrays of magnetic stripes and dots. Discussion of other shapes can be found in reference [46].

Gubbiotti *et al.* experimentally investigated collective magnetization dynamics in 1D periodic arrays of Py nanowires of width 175 nm and pitch 35 and 175 nm. For closely spaced nanowires, the strong dipolar coupling among the magnetic stripes led to the formation of collective modes in which the lowest frequency modes were found to be dispersive (i.e. the SW frequency depended on the wavevector). This indicated that the SWs had a non-zero group velocity because the dipolar coupling allowed the SWs to travel through the array. For nanowires spaced farther apart, the SW modes were found to be dispersionless owing to absence of dipolar coupling among the stripes [303]. In this case, the group velocity was zero indicating that the weak dipolar coupling does not allow a SW to travel though the array (from one nanomagnet to the next). In ensuing studies, they investigated the collective magnetization switching behaviour and collective SW modes in 1D arrays of nanowires of alternating width [304, 305]. The switching mechanism was observed to be very sensitive to the thickness to width ratio of the nanowires. A number of collective and dispersive SW modes were observed whose frequency periodically oscillated with respect to the wavevector encompassing several Brillouin zones, induced by the artificial periodicity of the array. The amplitude of oscillation was larger for the low-frequency mode and smaller for the high frequency modes. Interestingly, each mode existed in a range of frequency separated from the neighboring mode by a prohibited frequency zone, reminiscent of "bands" and "bandgaps" [305]. Notably, neighboring nanomagnets could be coerced into either parallel or anti-parallel magnetizations by varying a bias magnetic field. The magnonic bands for "parallel" ground state were markedly different from those of the antiparallel state [306].

Topp *et al.* utilized microwave assisted switching and minor-loop measurement to prepare various ground state of magnetization such as ferromagnetic state, multidomain state and antiferromagnetic states in periodically arranged magnetic nanowires [307]. They investigated collective SW modes in various ground states of magnetization and found that the SW dispersion character can be significantly tuned by this method of arranging nanomagnets into 1D arrays. Kostylev *et al.* showed that in these types of 1D arrays, the frequency gaps or band gaps are partial, i.e. the stop bands for SW propagation along the major axis of nanowires overlap with the passbands for SW propagation perpendicular to it [308]. However, this feature can be suppressed by using 2D arrays of nanomagnets. Wang *et al.* investigated magnonic band structures in 1D magnonic crystals formed by magnetic nanostripes made of alternating materials: namely cobalt and Py. A number of dispersive SW modes separated by magnetic-field-tunable band gaps was observed [309]. Interestingly, it was observed that the bandgap increases with the increase of cobalt stripe width, whereas the bandgap decreases with the increase of Py stripe width [310]. The center frequency bandgap is more sensitive to the stripe width of Py than that of cobalt. The reason behind this is the variation of effective saturation magnetization and exchange length which increases with the increase of cobalt width but decreases with the increase of Py width. Gubbiotti *et al.* investigated collective SWs in another type of 1D magnonic crystals where each stripe is made of two magnetic layers either with same (rectangular cross section) or with different width (L-shaped cross section) [311]. For the rectangular cross section, the lowest frequency fundamental mode originates from the in-phase precession of the magnetization in the two layers, whereas the higher frequency mode is spawned by the out-of-phase precession of the magnetization in the two layers. In the case of L-shaped cross section, two dispersive modes with a sizable magnonic band width were observed. One of the modes was the fundamental mode in the thick segment and another was the fundamental mode in the thin segment of each nanowire. Moreover, by carefully tuning the overlayer thickness, the mode frequencies and their spatial profiles could be significantly tuned [312].

In a very recent study, it has been demonstrated that if the coupling between the magnetic layers in a multilayered ferromagnetic structure can be changed from exchange to dipolar by inserting a thin metallic layer between consecutive layers and systematically varying its thickness, then the collective magnetization dynamics can be significantly modified [313]. In fact, it is possible to stabilize the magnetizations of two neighbouring layers either in the parallel or in the antiparallel configuration. Several modes can be excited depending upon the static magnetization configuration as well as the relative phase (in-phase or out-of-phase) of dynamic magnetizations between the two layers within the same nanowire. Saha *et al.* investigated collective magnetization dynamics of 1D arrays of magnetic nanostripes with varying width down to 50 nm [314]. For the narrowest stripe with 50 nm width, the inter-stripe magnetostatic interaction became negligible, which enabled the detection of magnetization dynamics of a single stripe. When a magnetic field was applied along the long axes of the nanostripes, the 50 nm wide stripes showed a uniform mode and a pure BV-like standing SW mode, whereas the wider stripes showed a uniform mode, a combination of BV- and DE-like standing SW mode, and a pure BV-like standing SW mode. However, the collective mode frequencies and spatial profiles were significantly changed when the bias magnetic field was rotated along the width of the nanostripes.

Another interesting type of 1D magnonic crystal is periodically arranged corrugated magnetic stripes [315]. The advantage of these types of magnonic crystal is that the SW frequencies and corresponding band gaps can be significantly tuned just by rotating in-plane bias magnetic field which essentially modifies the internal magnetization configuration. Especially, the edge modes are more sensitive to the degree of corrugation.

**10.3 Spin waves in 2D arrays of nanomagnets**

The study of magnetization dynamics in two dimensional (2D) periodic arrays of nanomagnets begun with theoretical investigations of the effect of inter-dot dipolar coupling on the dynamical behaviour in different array geometries and under different magnetic field orientations [316, 317]. Soon thereafter, it was experimentally demonstrated that a single SW mode, observed in an unpatterned magnetic thin film, is split into a number of modes when the film is patterned into a 2D array of square magnetic dots, because of the presence of a demagnetization area inside each dot [318]. Notably, it was observed that the mode frequencies strongly depend on the interdot separation and orientation of the external magnetic field [286].

A flurry of experimental investigations followed the early work, which demonstrated the presence of multiple SW modes in magnetic nanodots with uniform and nonuniform ground states of magnetization [287, 319, 320]. It was observed that the magnetization dynamics became more nonuniform with reduction in the dot diameter (in the sub-micrometer size scale) owing to nonuniform ground state of magnetization [321, 322]. Interestingly, low frequency SW modes were found to originate from the demagnetized edges of the magnetic dots and became more prominent with reduction of dot size. The resonance linewidth associated with the edge mode is generally higher than the linewidth of centre mode since the edge mode is highly sensitive to imperfections, and variations in the size and shape of the dots [323]. The dynamic modes in arrays of magnetic dots also show anisotropic behavior in their characteristic frequency as one varies the direction of applied magnetic field. This behaviour originates from the combined effect of static and dynamic effective magnetic fields inside the magnetic dots [324-327]. The number of symmetries in the anisotropy is determined by the shape of individual dots and the symmetry of their lattice arrangement [326-329]. The collective magnetization dynamics were extensively studied for various shapes of the magnetic dots [330-332] such as square [297, 325, 333], circular [327, 328, 334, 335], triangular [336], rectangular, elliptical, diamond, stadium [337], cross [326, 338-340] and were found to be significantly affected by the shape of the individual dots. In 2010 Kryuglyak *et al.* first imaged collective modes of magnetization dynamics in 2D arrays of magnetic dots with time-resolved scanning Kerr microscopy [337]. It was observed that the SWs were confined within the whole array as if the latter behaved like a single element made of a continuous material. Micromagnetic simulation has been widely used to analyze the nature of the collective magnetization dynamics of magnetic dot arrays. These simulations revealed that various types of modes appear because of the collective oscillations of different groups of dots in an array with well-defined phase relationships which are determined by the structure and the geometry of the array and its constituent nanoelements [259, 325-328, 330-332, 335, 339, 341]. Rana *et al.* studied magnetization dynamics of 2D arrays of magnetic dots by systematically varying the interdot separation (figure 27(a)). A smooth transition from strongly collective magnetization dynamics to non-collective dynamics was observed with the increase in the interdot separation (figures 27 (b-d)) [297, 333]. Surprisingly, strongly collective to non-collective transition was also observed just by changing the orientation of the in-plane bias magnetic field from $0°$ to $45°$ with respect to the symmetry axis of the dot array [325]. This is because the static and dynamic magnetostatic stray fields, which control the collective precession dynamics, are affected by the orientation of bias magnetic field as well. Arrays of magnetic dots, made of the same element but with different sizes, also show very interesting collective dynamics. A significant variation in the number of dynamics modes, their profiles and slopes of the magnonic bands are observed when the in-plane orientation of the bias magnetic field is varied [342-344]. In nanometer scale magnetic dots, defects in the form of edge roughness, local thickness variation, etc. can generate new resonant modes and quench the existing modes in isolated as well as arrays of magnetic dots [345].

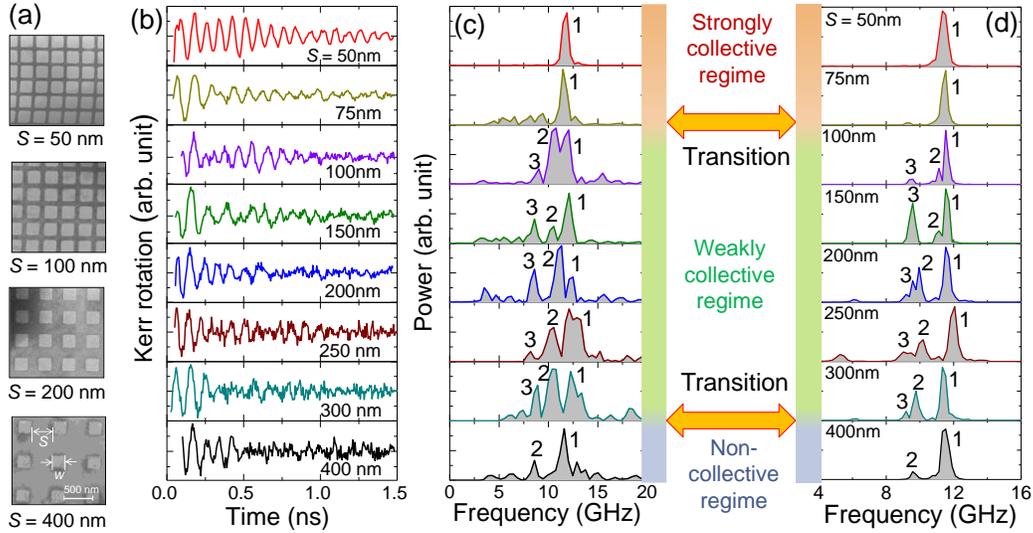

**Figure 27.** (a) Scanning electron micrographs of arrays of permalloy dots with width (W) = 200 nm, thickness = 20 nm, and varying interdot separation (S). (b) Experimental time-resolved Kerr rotations and (c) corresponding fast Fourier transform (FFT) spectra are shown for arrays of permalloy square dots. (c) FFT spectra of simulated time-domain magnetization for 7 × 7 array of dots with same specifications as in the experiment. Adapted from [333] with permission of the Institute of Physics.

### 10.4 Spin waves in connected nanomagnets

Connected nanomagnets have attracted interest over the last two decades because of their inherently large shape anisotropy aiding efficient transfer of energy in space [341, 346, 347]. A magnonic crystal is said to be a *connected dot* (CD) if all the magnetic nanodots are physically connected. A CD system can display properties characteristic of dots and/or antidots (AD). Antidot arrays typically exhibit higher SW propagation velocity and hence longer propagation wavelength at a given frequency. However, the primary difference between the CD system and an AD array is that the latter generally does not have regular shaped connector channels. The CDs have rarely been studied in the context of magnonics. The 1D bicomponent magnonic CD system with uniform width has been investigated experimentally as well as theoretically [309, 348]. SW dynamics in 1D magnonic quasicrystals arranged in a Fibonacci sequence with Py NWs of two different widths have been used to precisely tune SW spectra. This system shows a two-step hysteresis loop and its field dispersion shows two discontinuities at the transition points between ferromagnetic order and antiferromagnetic order. The observed properties of the Fibonacci structures are promising for applications in magnonic devices for data processing and wave-based computing [349].

More recently, an analytical model of a graded-index (GRID) 1D slab, with a finite spatial gradient of saturation magnetization or anisotropy, predicted phase coherent refraction of SW at desirable angles. The fundamental SW mode can be guided coherently in the GRID slab along a bend in a magnonic waveguide [350]. Subsequently, an alternative way was discovered where the 1D stripe or nanowire was replaced by a connected pseudo-1D magnonic crystal with tilted rectangular and/or rounded rectangular shape [341, 346]. This system is capable of magnonic band structure reconfiguration via subtle variation of the bias magnetic field angle. It has the potential for implementing logic gate functionality by virtue of the anisotropy in SW dispersion and bandgap. At the first Brillouin zone boundary, the magnonic bandgap can show an asymmetric dependence on the azimuthal angle of the bias magnetic field [351]. In another connected tilted rectangular nanodots, the magnonic band structure and bandgap were found to be strongly dependent on the magnetic history driven spin texture [341]. Depending upon the magnetic history, leaf-, S-, and shifted-core vortex states could be stabilized, which presented drastically different potentials to the propagating SWs passing through the nanochannels connecting the dots. More importantly, at the same bias field value, two different spin textures were obtained as a consequence of magnetic field-induced bifurcation giving rise to a reconfigurable magnonic band structure. In square nanodots connected by rectangular nanochannels, the SW modes are affected by magnetic coupling as well as by selective channeling of SW modes [352]. The selective channeling of SW modes through nanochannels may be further exploited to construct an electronic demultiplexer [353]. Artificial spin-ice structures in the form of connected nanobars or nanoelements arranged in a square or honeycomb (Kagome) lattice are also interesting due to their metastable ground-state spin configuration. As SWs are very sensitive to internal fields, SW dynamics can also be affected by the creation of monopole-anti-monopole pairs and Dirac strings in these system [347, 354].

**10.5 Dispersion Relations of Spin-waves in Nanomagnet Arrays**

During the last two decades, reconfigurable magnonics has turned attention to 1D and 2D magnonic crystals (MCs), including pseudo-1D MC [309, 346, 348, 355]. The SW dynamics in these systems can be efficiently reconfigured by tuning active parameters such as magnetic field strength and orientation, magnetic history, electric field, strain, spin texture, etc., as well as passive parameters such as size, shape, lattice constant, lattice symmetry, and constituent materials. SW dispersion is usually measured with BLS [356] and analytically investigated with the help of plane wave method (PWM) [357], dynamical matrix method [358], micromagnetic simulation software such as OOMMF [359] and Mumax3 [215]. The magnetostatic SW dispersion relation is primarily governed by the long wavelength magneto-dipole interaction and the short wavelength exchange interaction. In the following subsection, we illustrate the PWM for calculation of SW dispersion in some model nanostructures.

**10.5.1 Plane wave method (PWM):**

In this method, the Landau-Lifshitz (LL) equation is solved for a magnetic structure (MC in this case) using Bloch's theorem [357, 360]. The temporal evolution of the magnetization vector ($M(r,t)$) in the structure is described by:

$$\frac{\partial M(r,t)}{\partial t} = -\gamma\mu_0 M(r,t) \times H_{eff}(r,t). \tag{7}$$

Here, $H_{eff}(r,t)$ is the effective magnetic field given by:

$$H_{eff}(r,t) = H + H_d + H_{ex}, \tag{8}$$

where $H_d$ is the total demagnetizing field, which can written as:

$$H_d = H_d(r) + h_d(r)e^{i2\pi\nu t} \tag{9}$$

The quantities $H_d(r)$ and $h_d$ are the static and the dynamic components of the dipolar field, which satisfy the magnetostatic Maxwell's equations. The quantities $H$ and $H_{ex}$ are the applied magnetic field and the exchange field, respectively, where $H_{ex} = (\nabla . l_{ex}^2(r)\nabla)m(r,t)$; $l_{ex}(r) = \sqrt{\frac{2A(r)}{\mu_0 M_s^2(r)}}$ is the exchange length, and $A(r)$ is the exchange stiffness constant. In the linear approximation, the component of the magnetization vector, $M_s(r)$ parallel to the static magnetic field (*i.e.* z-axis in our case) is constant in time, and its magnitude is much greater than that of the perpendicular component $m(r, t)$, so that:

$$M(r, t) = M_s(r)\hat{z} + m(r, t). \tag{10}$$

To obtain a solution of the LL equation, we can use Bloch's theorem with lattice constant $a$, so that: $m(r) = \sum_G m_k(G)e^{i(k+G).r}$, where $k = (k_x, k_z)$ is the wavevector in the first Brillouin zone (BZ), and $G = (G_x, G_z) = \frac{2\pi}{a}(n_x, n_z)$ denotes the reciprocal lattice vector of the periodic structure. Next, we perform Fourier transformation of $M_s(r)$ and $l_{ex}^2(r)$ to find the relevant quantities in reciprocal space using:

$$M_s(r) = \sum_G M_s(G) e^{iG.r} \tag{11}$$

$$l_{ex}^2(r) = \sum_G l_{ex}^2(G) e^{iG.r} \tag{12}$$

$$\text{Here} \quad M_s(G) = \begin{cases} M_{S,A}t + M_{S,B}(1-t) & G = 0 \\ (M_{S,A} - M_{S,B})I(G) & G \neq 0 \end{cases} \tag{13}$$

where, $t$ is the filling fraction of magnetic material in the lattice and $M_{S,A}$ and $M_{S,B}$ are saturation magnetization of two magnetic materials $A$ and $B$, respectively. Finally, $I(G)$ is a function that is specific to the nanodot structure used in this calculation.

### 10.5.2 Calculation of spin-wave dispersion in two-demensional magnonic crystals:

In the case of a 1D MC, there is only one symmetry axis with the corresponding lattice symmetry. However, in the cases of a 2D MC and a pseudo-one-dimensional MC, the lattice symmetry can be varied by arranging nanomagnets in different periodic patterns. The 2D MC category can be divided into three sub-categories: a) physically isolated ferromagnetic dot lattice, b) antidot lattice and c) bi-component MC.

There are numerous reports of the study of 2D MC with varying magnetic properties and geometric architectures. Recently, ferromagnetic nano-cross structures have attracted interest owing to some remarkable properties, including magnonic mode softening, mode splitting and strong magnon-magnon coupling induced avoided crossing of SW branches in $Ni_{80}Fe_{20}$ nano-cross arrays [340]. Hence, we focus on arrays of nano-cross structures as a case study for calculation of SW dispersion.

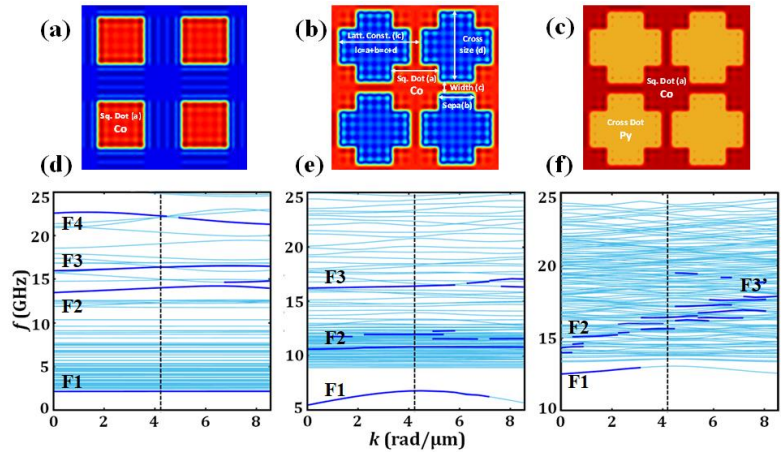

**Figure 28.** (a)-(c) represent the structure of three model 2D MCs, namely, square dot array, connected dot (CD) array and bi-component magnonic crystal (BMC). (d)-(f) Calculated magnonic band structures for the above three different MCs, i.e. square dot array, CD array and BMC, respectively.

The parameters used in the above calculation are: $M_{s,Co} = 1.15 \times 10^6$ A/m, $A_{Co} = 28.8$ pJ/m, $M_{s,Py} = 0.78 \times 10^6$ A/m and $A_{Py} = 13.0$ pJ/m in the ferromagnetic regions, whereas very small $M_s$ and $A$ values were assumed for the nonmagnetic regions to avoid nonphysical SW frequencies. We have calculated SW dispersion in DE geometry for a) square dot array, b) CD structure, and c) BMC structure for a fixed lattice constant of 700 nm and square dot width of 400 nm as shown in figures 28(a-c). The calculated dispersion of the Co dot array shows four SW modes, of which three modes F2, F3 and F4 are weakly dispersive in nature. To get deeper insight into the nature of these modes, we have calculated the SW mode profiles at $k = 1.6$ rad/µm as shown in figure 29. The nondispersive mode F1 has its power concentrated at the centre of the square dot. The modes F2 and F3, which show weakly positive dispersion, exhibit different spatial characters. The power of mode F2 is primarily concentrated at two horizontal edges of the dots (edge mode), while the F3 shows standing wave pattern in the DE geometry with quantization number $m = 5$ as shown in figure 29. The F4 mode is a negatively dispersive mode with standing wave character in the BV geometry with quantization number $n = 3$. In the CD array, we have considered Co square dots of size = 400 nm, lattice constant = 700 nm and connector dimensions (length × width) of $300 \times 100$ nm². In this case, three modes are observed, out of which the mode F1 is weakly dispersive in nature and one band gap is opened at the 1st Brillouin zone (BZ), as shown in figure 28(e). This mode shows DE character with $m = 4$. The modes F2 and F3 are nondispersive in nature and the power of F2 is concentrated in the horizontal connector channel with $m = 2$. The power in mode F3 is mainly concentrated at the four corners of the square dot with quantization number $(n, m)$ of (2, 2). The BMC structure is identical to the CD array except for the fact that the blank regions are now filled with Py. In this case, two dispersive modes are observed and an indirect type band gap is present between F1 and F2 modes, as shown in figure 28(f). Figure 29 shows that the two lower frequency dispersive modes F1 and F2 are located inside the Py regions with mode quantization numbers $(n, m)$ being (3, 1) and (7, 3), respectively. A higher frequency dispersive mode F3′ also appears with prominent intensity in the 2$^{nd}$ BZ, which again shows power in the Py region with mode quantization number (2, 7).

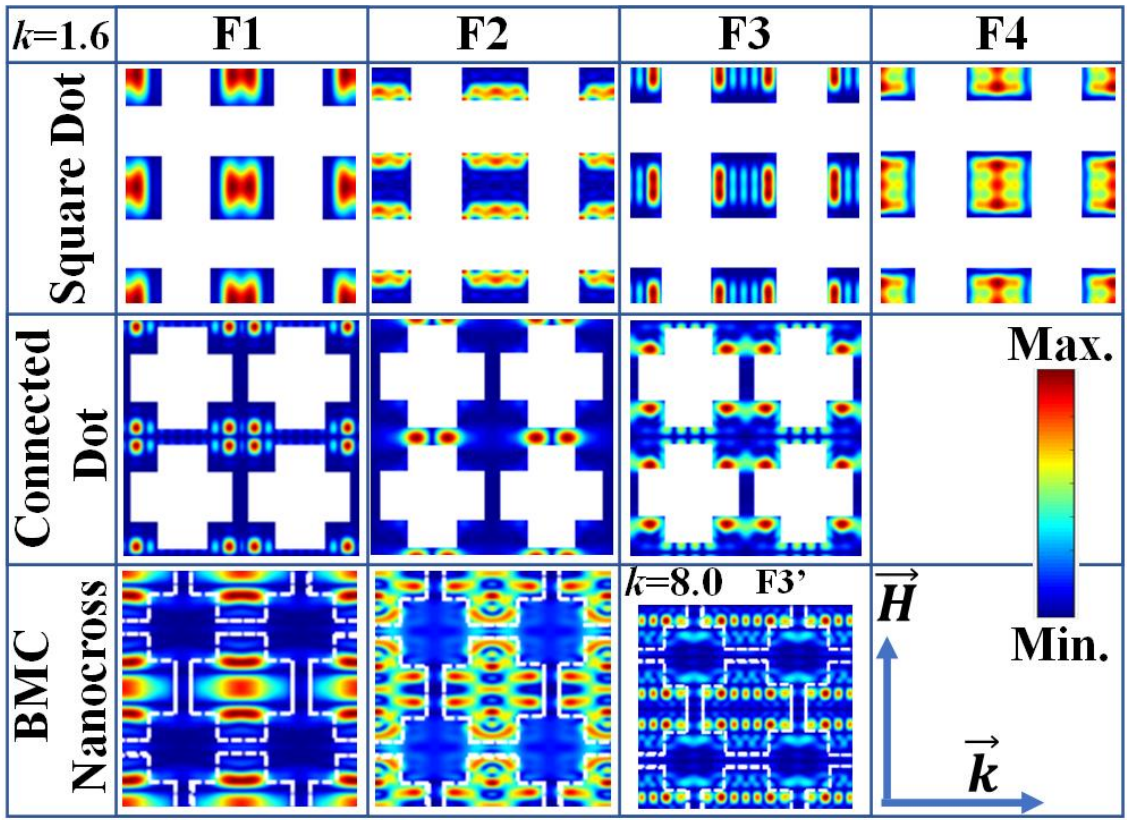

**Figure 29.** Spatial profiles of selected SW modes at $k = 1.6$ rad/μm for three different magnonic crystals, namely, square dot array, connected dot array and bi-component magnonic crystal.

**10.6 Effects of geometric parameters on spin waves in nanomagnet arrays**

Geometric parameters of magnetic dots such as shape, size, aspect ratio (thickness/diameter) and the arrangement of the dots in lattices, have significant impact on the isolated and collective magnetization dynamics of magnetic dots. These geometric parameters basically control the magnetostatic energy inside individual dots and the stray field distribution within the array. Therefore, controlling isolated and collective magnetization dynamics by varying the geometrical parameters have been thoroughly investigated over the last two decades. Barman *et al.* investigated the magnetization dynamics of isolated magnetic dots with sizes varying from 5 μm down to 125 nm as described in section 10.1 [270]. A significant dependence of the effective damping with the dot size was also observed [293]. Gubbiotti *et al.* have shown that in 1D periodic arrays of magnetic stripes, the periodic variation of stripe width can be a very useful tool for producing rich magnonic spectra [305]. Later, Tacchi *et al.* also reported that the magnetization ground state in a similar type of stripe array can be either tuned to the ferromagnetic or to the anti-ferromagnetic type of arrangement by judiciously choosing the applied magnetic field. The magnonic spectra can be drastically different in these two cases of magnetic ground states [306]. The precessional magnetization dynamics in 2D arrays of magnetic dots made of two different dots sizes have also been studied [343, 344]. Effort has been made to prepare 1D nanomagnet array by step modulated thickness either made of a single material [361] or two different materials [312, 362]. In the latter case, the height of the top layer strongly affects the magnonic band structures. Saha *et al.* have shown that the precessional magnetization dynamics in 1D arrays of stripes strongly depends on the width of the stripes as a result of the variation of interdot magnetostatic interactions. In the case 2D arrays of magnetic dots, the magnetization dynamics become more nonuniform with reduction of dot diameter [321, 322] and low frequency SW modes originating from the demagnetized edges of the magnetic dots become more prominent with the reduction of dot size. Zivieri *et al.* [363] and Kostylev *et al.* [364] investigated the magnetization dynamics in 1D arrays of rectangular magnonic dots arranged in various ways: such as along long axes, along short axes, in a non-collinear fashion, etc.

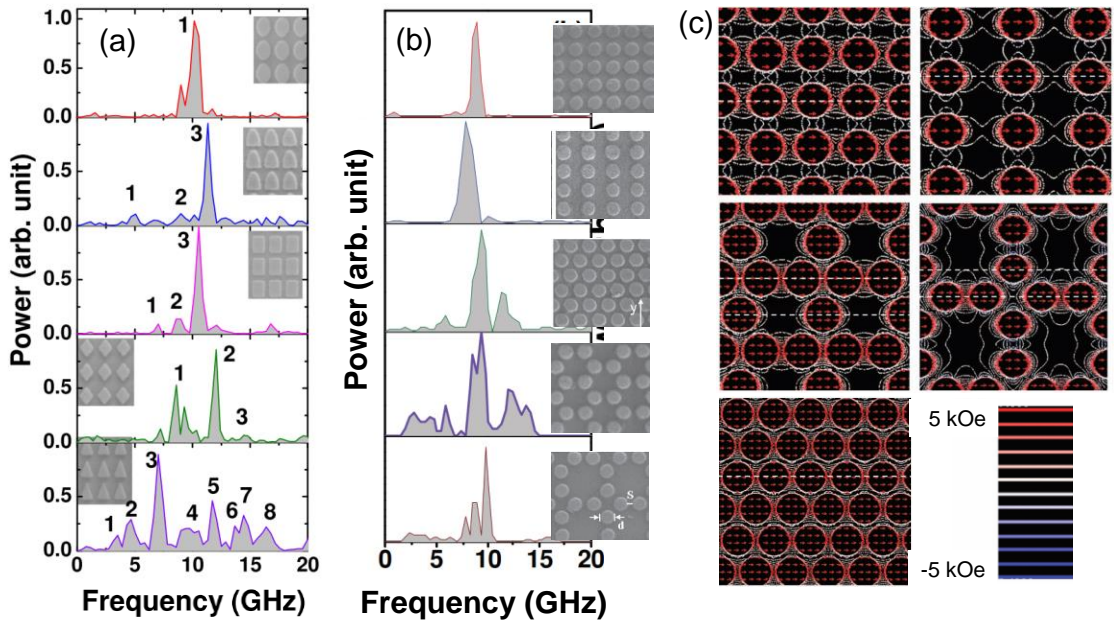

**Figure 30.** (a) Experimental SW spectra for arrays of Ni$_{80}$Fe$_{20}$ (Py) nanodots with varying shapes as shown in the inset. Reproduced from [330] with permission of the American Institute of Physics. (b) Experimental SW spectra for circular shaped Py nanodots arranged in varying lattices [327]. (c) Simulated magnetostatic field distributions are shown for arrays of Py nanodots arranged in variable lattices at $H$ = 1.3 kOe. The small arrows inside the dots denote the magnetization distribution within the dots, while the strengths of the stray magnetic fields are represented by the colour bar given at the lower right corner of the figure. Adapted from [327] with permission of the Wiley-VCH.

Mahato *et al.* investigated collective magnetization dynamics of square arrays of magnetic dots made of different shapes [330, 331]. The dot shapes not only affect the dynamics of individual dots, but also significantly modify the collective dynamics of the array (figure 30(a)). The number of resonant modes and the configurational anisotropy strongly depend upon the shape of the dots [330, 331]. Saha *et al.* have demonstrated that when circular magnetic dots are arranged in various lattice geometry the collective SW spectra is significantly modified with the variation of lattice arrangement (figure 30(b)) [327, 328]. This is solely due to the distribution of the magnetostatic stray field among the dots, which significantly vary with the lattice arrangement of the dots [327, 328] and also interdot separation [334, 365] (figure 30(c)).

## 11. EXTERNAL CONTROL OF NANOMAGNET DYNAMICS

The magnetization dynamics in nanomagnets is mainly decided by the internal magnetization configuration, which can be potentially controlled by the strength and orientation of an external magnetic field. As a result, a bias magnetic field can significantly control the precessional magnetization dynamics of isolated nanomagnets. When the nanomagnets are arranged in a periodic manner, they are coupled via magnetostatic stray field if the interelement distance becomes comparable to the nanomagnet dimension. The magnetostatic coupling can suppress the dynamics of individual nanomagnets leading to the collective dynamics of the array. Therefore, the bias magnetic field strength and orientation can also affect the collective dynamics of the array. A significant number of reports are found in the literature about bias-field-dependent magnetization dynamics in single magnetic dot, as well as 1D arrays and 2D arrays of magnetic dots. Barman *et al.* demonstrated that the precessional frequency of circular shaped single magnetic dots monotonically increases with increasing bias magnetic field, and the nature of increment of precessional frequency is different for dots of different size owing to different ground states of magnetization [270]. The effective damping constant, on the other hand, either decreases or remains unchanged depending upon the size of the dots [293]. Notably, in the case of circular dots, the precessional frequencies may also show an abrupt change due to the change in the magnetization state [296]. In isolated magnetic dots, two types of resonant modes are generally observed – one of which originates from the central part of the dot and other from the oscillation of magnetization taking place at the demagnetized edges perpendicular to the bias magnetic field direction. However, the edge mode can be suppressed by exciting large amplitude magnetization dynamics [300]. Kostylev *et al.* have shown a systematic variation of the frequencies of the precessional modes in 1D arrays of magnetic stripes with the magnetic field orientation [308]. Similar types of studies were carried out for 1D magnonic crystals made of periodic arrays of nanostripes with alternating widths [306], and nanostripes whose dimensions vary from 200 nm to 50 nm [314]. For 2D arrays of magnetic dots, an anisotropic variation in precessional frequencies was observed where the degree of anisotropy was determined by the lattice symmetry in the magnetic dot array as well as the symmetry in the shape of the individual dots (figure 31(a-b)) [46, 324-327, 331, 336, 344, 366]. As a general rule, the in-plane magnetic field dependent anisotropy in the resonance frequency is dominated by the dot shape when the dots in the array are well separated, and by the lattice symmetry of the array when the dots are placed sufficiently close to each other to have strong magnetostatic coupling. The magnetic field can not only systematically tune the collective mode frequencies [319, 321, 327, 331, 333, 334, 336-340, 344, 361, 362, 365, 366], but also the magnonic band structures, as well as the band gap widths in 1D [309, 311] and 2D arrays [287, 319] of magnetic dots. Moreover, frequencies of isolated and collective dynamics can change dramatically with the bias magnetic field because of the change in ground state of magnetization, which can show a transformation from nonuniform state to uniform state with increasing bias magnetic field [296, 319, 338, 341]. Rana *et al.* observed that the uniform collective magnetization dynamics in closely shaped square magnetic dot array is gradually transformed to the non-collective dynamics when the in-plane bias magnetic field is rotated from 0° to 45° with respect to the symmetry direction of the square lattice (figures 31 (c-d)) [259, 325].

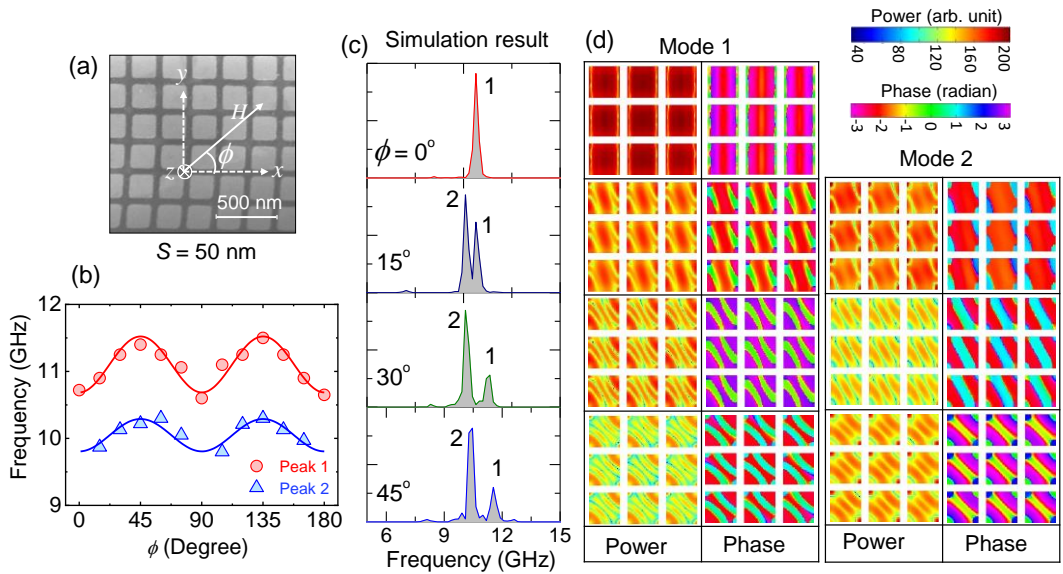

**Figure 31.** (a) Scanning electron micrograph (SEM) of square permalloy dots with 200 nm width, 20 nm thickness and 50 nm interdot separation. (b) Evolution of the collective mode frequencies with the in-plane orientation of the bias magnetic field. (c) Simulation results show the evolution of collective modes with gradually varying azimuthal

angle of the bias magnetic field. (d) Simulated spatial profiles of the collective resonant modes. Adapted from [325] with permission of the American Institute of Physics.

## 11.1 Electric field based control

In this section we will mainly confine our discussion to the control of nanomagnet dynamics by voltage (i.e., electric field) controlled magnetic anisotropy (VCMA) followed by a short description of other electric field-based methods. The mechanisms of VCMA and other electric field-based methods have been discussed earlier in section 8.6. The SW frequency, damping, and band structures in ultrathin ferromagnetic (FM) films strongly depend on the interfacial PMA. Therefore, an electric field applied at FM/oxide interfaces can significantly tune the FMR and SW properties of the FM via VCMA. While studying electric field induced modification of PMA, Nozaki *et al.* [367], Skowroński *et al.* [368] and Kanai *et al.* [369] have experimentally demonstrated a significant modulation of FMR frequency in ultrathin ferromagnetic films upon the application of an electric field (figure 32(a)). The VCMA induced modulation of SWs was first reported by Nagaoka *et al.* [370]. It was shown that the resonance fields of propagating magnetostatic surface SWs in a 5 nm thick Fe film possessing PMA can be controlled using VCMA. When a thicker FM film was used, the change in the resonance field was smaller, probably because the film exhibited weaker PMA. Later on, Rana *et al.* experimentally demonstrated the manipulation of FMR and dipole-exchange SWs frequency by VCMA in ultrathin CoFeB films with thicknesses down to 1.6 nm [371]. A significant modulation in the SW frequency was observed, particularly in a 1.6-nm-thick CoFeB film, where the interfacial PMA is almost compensated by the demagnetizing field (figure 32(b)). Okada *et al.* demonstrated that an electric field applied at a FM/oxide interface not only modulate the PMAs, but also modulates the damping constant of FMR and the latter modulation is linearly proportional to the applied electric field [372]. Interestingly, a negligible modulation of the damping constant by electric field was observed for CoFeB films thicker than 1.5 nm. However, the reason behind this remains unknown. Later, Rana *et al.* reported nonlinear variation of damping constant with electric field, especially in ultrathin CoFeB films (figure 32(c)) [373]. It was theoretically explained that the presence of the Rashba SOC at a CoFeB/MgO interface and the electric field dependence of the Rashba coefficient is responsible for the observed nonlinear behavior (figure 32(d)), which was confirmed by studying other reference samples. Recently, Fulara *et al.* have demonstrated the modulation of spin Hall nano-oscillator effective damping by VCMA. Basically, VCMA tunes the auto-oscillation frequency of the nano-constriction in the oscillator device such that its mode volume and coupling to propagating SWs are drastically modified, resulting in a giant variation (42%) of the effective damping over a moderate range of applied gate voltage [374].

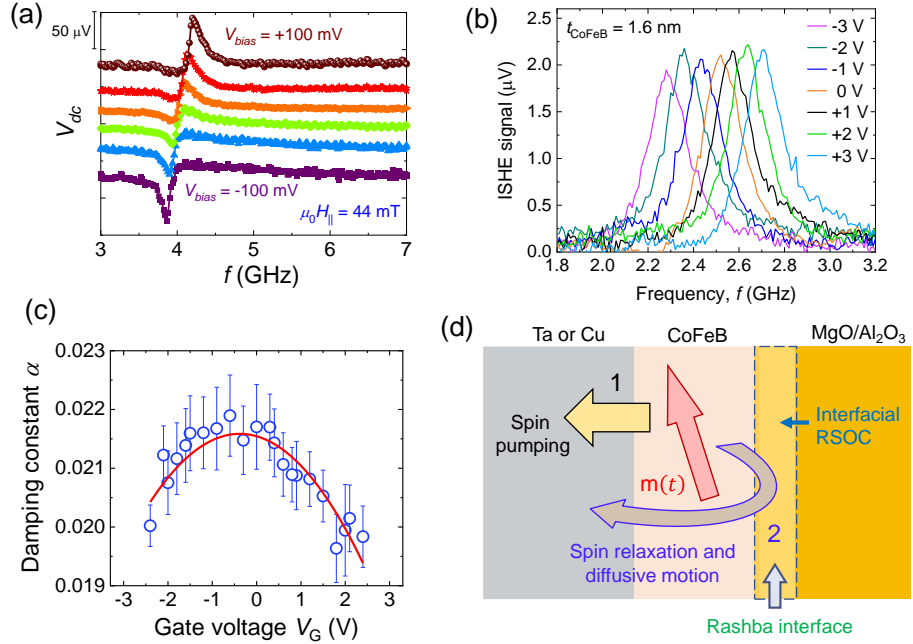

**Figure 32.** (a) Homodyne-detected FMR spectra measured from an ultrathin CoFeB film under the application of an in-plane magnetic field of 44 mT under bias voltage steps of 40 mV. Adapted from [369] with permission of the American Institute of Physics. (b) SW signals measured from a 1.6 nm thick CoFeB film with inverse spin Hall effect and spin pumping technique at various values of gate voltage. Adapted from [371] with permission of the American Physical Society. (c) Dependence of damping constant on gate voltage in a Ta/CoFeB/MgO heterostructure with 1.6 nm CoFeB film thickness. The solid curve represents fitting with a quadratic function. (d) Schematic diagram showing the relaxation mechanism of spin angular momentum in Ta(or, Cu)/CoFeB/MgO/Al$_2$O$_3$ heterostructure through two major processes. Adapted from [373] with the permission of the American Physical Society.

VCMA can also be used to form magnonic crystals (MCs) by periodically modulating the PMA of an ultrathin FM film. Wang *et al.* proposed that reconfigurable MCs can be created with VCMA by placing periodic arrays of stripe-like metal gate electrodes on top of an ultrathin FM film [38]. When a dc voltage is applied across gate electrodes and the FM waveguide, the PMA of the FM is modulated only underneath the gate electrodes, which then results in an artificial MC with a lattice constant equal to that of the gate electrodes. These VCMA induced MCs significantly modify frequency versus wavevector band structures of SWs, i.e., magnons, opening a magnonic band gap at the BZ boundary, where the propagation of SWs is prohibited. The key advantage of VCMA controlled MCs is that the width of the band gap and transmission of SWs at BZ boundary can be tuned by varying the gate voltage magnitude. In a pioneering work, Chaudhury *et al.* experimentally demonstrated the formation of periodic arrays of parallel nanochannels (NCs) by VCMA [39]. These NCs were created by applying dc voltages across the periodically arranged stripe-like gate electrodes delineated on an ultrathin CoFeB film (figure 33(a)). When no gate voltage is applied, a single SW mode with monotonic variation of frequency with the wavevector is observed (figure 33(b)). This SW dispersion looks similar to the SW dispersion of a stand-alone CoFeB film without any gate electrodes. When a moderate gate voltage (-4 V) is applied, i.e., when a periodic array of NCs is formed, two SW modes appear with a bandgap (figure 33(c)). Interestingly, the bandgap can be tuned just by changing the gate voltage. After performing a PWM calculation and analysis, it was found that the higher frequency mode (i.e., mode 2) is confined within channel 2 underneath the gate electrodes, while the lower frequency mode (i.e., mode 1) is confined within channel 1 interposed between the gate electrodes (figures 33(d-e)).

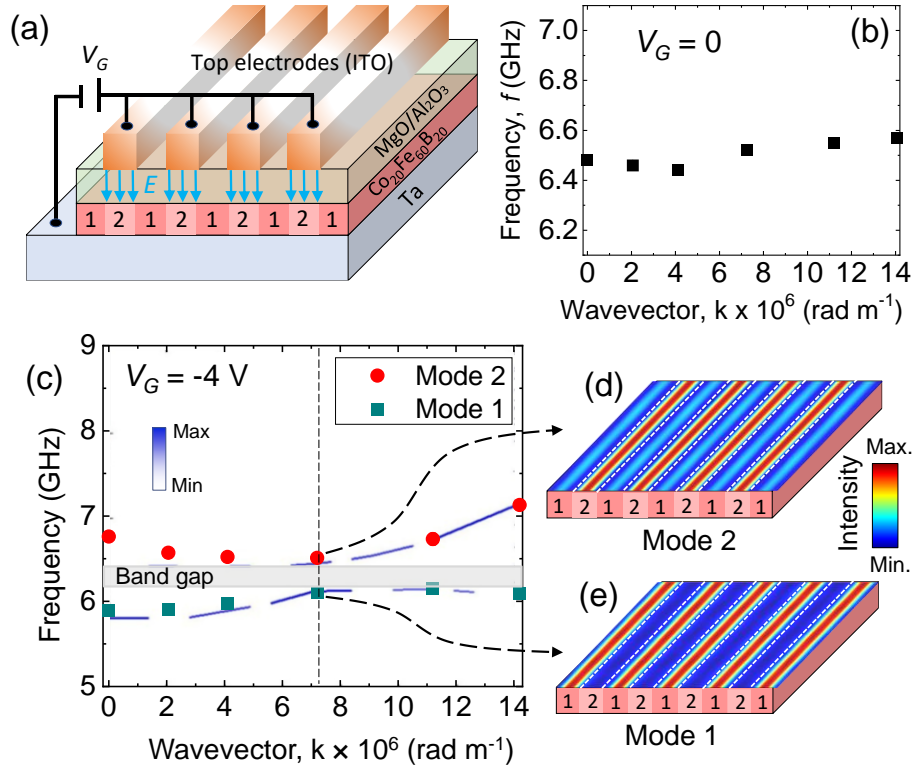

**Figure 33.** (a) Periodic nature of electric field applied at CoFeB/MgO interface through periodically arranged metal gate electrodes, giving rise to two periodic regions: region 1 (outside of top electrodes) and region 2 (underneath the top electrodes). Experimentally measured frequency versus wavevector dispersion at $V_G = 0$ V (b), and - 4 V (c) under the application of 200 mT bias magnetic field [39]. (c) Symbols represent measured SW frequencies, while blue lines denote SW intensities as calculated by PWM. The corresponding colour map is given inside. The dashed vertical line indicates the position of anticrossing, and the corresponding magnonic band gap is shown by the shaded region. (d-e) Calculated spatial profiles of the SW modes for wavevector $k = 7.1 \times 10^6$ rad m$^{-1}$ under the application of $V_G = - 4$ V at $\mu_0 H = 200$ mT. Low frequency mode, i.e., mode 1 is confined in region 1 and high frequency mode, i.e., mode 2 is confined in region 2. Adapted from [39] with permission of the American Association for the Advancement of Science.

## 12. SPINTRONIC EFFECTS IN NANOMAGNETS

Magnetism arises from the quantum mechanical spin degree of freedom of charge carriers and nuclei in a solid. In fact, a *single domain* nanomagnet, consisting of ~$10^4$ electron spins, behaves like one single giant *classical* spin in which all the spins together rotate coherently in unison when the magnetization rotates [85]. It therefore behooves us to discuss some spintronic effects that are germane to many observed effects in nanomagnets.

Spintronics, or spin electronics, is a multidisciplinary research field that deals with the study of active control and manipulation of spin degrees of freedom in solid-state systems to perform myriad tasks such as computing, information storage, communication, sensing, etc. The gamut of spintronics spans such disparate topics as nitrogen-vacancy centres in diamonds to act as spin qubits [375], topological spin textures [376], spin based classical computing [377], SW-based information processing [378], spin transistors [379, 380], spin batteries [381-383], spintronic photodetectors [384], spin current generation [385], spin transport [142] and spin dynamics in macroscale systems [386]. Generation of spin polarized current, spin relaxation and spin detection are three major tasks in modern spintronics, and they often involve nanomagnets.

Generation of spin polarized electrons requires creation of a nonequilibrium spin population. This can be achieved by many ways. Circularly polarized photons have been used to transfer their angular momenta to electrons and produce a spin-polarized population of electrons in semiconductors [387], but for device applications such as spin transistors, electrical spin injection is preferred. The latter is usually achieved by connecting a ferromagnetic contact to a sample and then injecting electrons from the contact into the sample by connecting a voltage source across the interface between the contact and the sample. A ferromagnet has a spin-split conduction band where the densities of states of electrons at the Fermi level are different for spins parallel and antiparallel to the direction of magnetization. Since the electrons mostly come from around the Fermi level in the contact, the injected current naturally has majority spins and minority spins. The spin injection ratio is defined as $\eta = \frac{n_{maj}-n_{min}}{n_{maj}+n_{min}}$, where $n_{maj}$ is the population of majority spins and $n_{min}$ is the population of minority spins in the injected current. Other methods, which do not require magnets but require such constructs as quantum point contacts, have also been demonstrated for spin injection [388]. In the ferromagnetic contact method, the efficiency of spin injection is increased by interposing a tunnel barrier between the ferromagnet and the non-magnetic sample to overcome the so-called resistivity mismatch problem if the resistivities of the contact and the sample are significantly different [389]. A spin current can also be generated by microwave excitation and by harnessing various spin-orbit effects such as spin pumping, spin Hall effect, spin caloric effects (e. g. spin Seebeck and Nernst effect), etc.

Historically, generation of spin polarized current was first demonstrated by Julliéré in 1975 in an experiment involving tunneling of electrons between two ferromagnetic films separated by a spacer layer interposed between them [390]. This tri-layer construct is called a "spin valve" and it led to the concept of tunneling magnetoresistance TMR = ($R_{AP}$-$R_P$)/$R_P$ = $2P_1P_2/(1 - P_1P_2)$; $P_1$ and $P_2$ being the spin polarizations of the two ferromagnetic layers, while $R_{AP}$ is the resistance of the spin valve when the polarizations of the two layers are mutually antiparallel and $R_P$ is the resistance when they are parallel. The notion of TMR is a key concept in modern spintronics and is a metric used to gauge the degree of spin selective injection from one layer, and spin-polarized extraction by the other layer, in a spin valve structure. In other words, TMR is a measure of the quality of the "spin polarizer" layer and the "spin analyzer" layer. The MTJ is also a spin valve and there, TMR is again a measure of the quality of spin selectivity in the tunneling between the hard and the soft layers.

With the observation of significant room temperature TMR in spin valve structures in 1995 [391, 392], this field has grown substantially leading to the development of magnetoresistive random access memory (MRAM) based on MTJ [393]. Introduction of perfectly ordered (100) oriented MgO tunnel barrier as the spacer layer interposed between CoFeB electrodes as the ferromagnetic layers showed giant TMR value of 220% at room temperature (RT) [394] and even higher value of 604% in Co0.2Fe0.6B0.2/MgO/Co0.2Fe0.6B0.2 at RT demonstrated by Ikeda *et al.* [2].

Afterwards, the giant magnetoresistance (GMR) effect was discovered in 1988 by Fert [395] and Grunberg [396] in metallic Fe/Cr multilayers. The effect is manifested as a giant change in the electrical resistance of a device in a magnetic field. The GMR typically occurs due to spin dependent scattering of electrons at the interface, although it can also occur because of other reasons [394]. In analogy with TMR, one defines GMR = ($R_{AP}$-$R_P$)/$R_P$, where $R_{AP}$ and $R_P$ correspond to the resistances of the multilayer structure in antiparallel and parallel magnetization configuration of the two ferromagnetic layers, respectively. GMR is observed both in current-in-plane (CIP) and current- perpendicular-to-plane (CPP) geometries, i. e. when current flows parallel and perpendicular to the heterointerfaces, respectively. Its magnitude depends both on the ferromagnet FM (e. g. Fe) and normal metal NM (e. g. Cr) layer thicknesses as given by the Valet-Fert model [397]. In 1998, IBM introduced a low field GMR sensor which is a key component for data reading in magnetic hard disk drives today.

In early 1993, a pioneering work on lateral spin valve was reported [398], which spurred intense interest in early 2000 when Jedema *et al.* demonstrated diffusive spin-injection in a non-local geometry from a Py nanowire electrode followed by spin accumulation in a Cu nanowire and subsequent spin detection by another Py nanoelectrode [399]. An estimated spin diffusion length in Cu of 350 nm at room temperature was obtained. Later they studied controlled spin precession of electrically injected and detected electrons in a diffusive metallic conductor, using Al-O tunnel barriers in combination with metallic ferromagnetic electrodes as spin injector and detector [400]. An increasing spin signal due to ballistic spin injection was reported. Efforts on increasing spin injection efficiency continued and an efficiency up to 27% was reported using half-metallic $Co_2FeSi$ FM electrode [401], later exceeded by a demonstration of 70% spin injection efficiency at room temperature in a CoFe/MgO/GaAs device [402].

The STT effect, mentioned earlier in this review, requires spin polarized injection of current into or extraction of current from the soft layer of a MTJ to reverse its magnetization. Here, the MTJ's hard layer acts as the spin polarizer. STT was demonstrated first in a point contact [403] and subsequently in CPP-GMR [404] and TMR nanopillars. There is a critical current density in STT below which magnetic reversal does not take place in the soft layer. Jiang *et al.* reduced the critical current density for STT by an order of magnitude in a CPP pseudo-spin-valve nanopillar by inserting a spin-scattering Ru ultrathin layer and an additional FM layer [405, 406]. More recently, Vautrin *et al.* studied the thickness and angular dependence of the magnetocurrent of hot electrons in a magnetic tunnel transistor (MTT) with crossed magnetic anisotropies. They showed an increase in magnetocurrent with ferromagnetic layer thickness which attained a value up to 85%, close to the theoretical maximum value of 100% for MTTs with crossed magnetic configurations [407]. Spin-polarized current driven magnetic switching in nanoscale MTJ with Al-O barrier showed TMR up to 30% at current densities less than $10^7$ A/cm$^2$ [408]. Introduction of a MgO barrier and post-processing annealing in low-resistance $Co_{40}Fe_{40}B_{20}$/MgO/$Co_{40}Fe_{40}B_{20}$ MTJs showed reduction of critical current density $J_C$ to as low as $7.8\times10^5$ A/cm$^2$ (TMR = 49%) and increase of TMR ratio to as high as 160% ($J_C = 2.5\times10^6$ A/cm$^2$) [409]. The improved scalability of ultrasmall out-of-plane magnetized MTJs for STT-MRAM cells has been made possible by using magnetic shape anisotropy in combination with interface anisotropy [410]. Recently, insertion of a double tunneling barrier in a double MTJ has led to the formation of a resonant tunnelling junction, which showed TMR oscillation for the FM layer thickness up to 12 nm [411]. More recently, quantum resonant tunneling has been observed in a double MTJ incorporating a double quantum well [412]. In recent years spintronics has branched off into new areas by merging with various ancillary fields, namely magnon spintronics [413], spin-orbitronics [414], antiferromagnetic spintronics [415], topological spintronics [416], flexible spintronics [417], organic spintronics [418], spin-based quantum [419, 420], neuromorphic [421] and probabilistic computing [422], and so on. The overriding goals are: further miniaturization, faster operation, greater functionality, energy efficiency, integration on flexible substrate etc.

## 12.1 Spin orbitronics

In STT, the exchange interaction between injected electron spins and resident spins in magnetic materials mediates transfer of spin angular momenta from the injected spins to the resident spins. Spin-orbitronics is a new avenue of spintronics, which exploits the spin-orbit coupling (SOC) in nonmagnetic materials instead of the exchange interaction in magnetic materials to generate, detect or use pure spin currents. This opens the pathway to building spin-based devices made of only nonmagnetic materials which can be operated without magnetic fields. A number of effects originating from SOC are important in the fields of nanomagnetism and spintronics. These range from the fundamental magnetic anisotropy energy, namely magnetocrystalline anisotropy to more advanced topics such as anisotropic magnetoresistance, spin Hall effect, Rashba interface, Edelstein effect, SOT, spin pumping effect, Dzyaloshinskii-Moriya interaction, spin caloric effects etc. Two of the most important aspects arising from the above effects are the generation of pure spin current, i.e. the so-called spin-charge conversion and appearance of chiral spin textures.

The SOC is a fundamental interaction occurring between an electron's spin and its orbital motion around the nucleus. An electron moving around a nucleus experiences an electric field due to the positively charged nucleus and this electric field Lorentz transforms into a magnetic field in the rest frame of the electron. The interaction of this magnetic field with the spin angular momentum of the electron gives rise to spin-orbit interaction or coupling. The SOC increases as $Z^4$ ($Z$ is the atomic number) in hydrogen-like atom [142]. The general derivation of SOC from the Dirac equation for an electron of mass $m$ and charge $-e$ in an external electrical field $E(r) = -\nabla\phi(r)$ yields:

$$H_{SO} = \frac{e\hbar}{4m^2c^2}\hat{\sigma} \cdot \left[\vec{E}(r) \times \vec{P}\right] \tag{14}$$

where $\vec{P}$ is the momentum operator and $\hat{\sigma}$ is the Pauli spin matrix. Two major contributions to SOC in solids are: a) the Dresselhaus contribution occurring in crystals with bulk inversion asymmetry, leading to a net electric field for certain crystal directions [423] and b) the Rashba SOC occurring in systems with net electric field due to structural inversion asymmetry (SIA) [424]. The Rashba SOC can produce momentum-dependent splitting of spin bands in bulk material and low-dimensional condensed phase systems such as surface and interface states. The splitting occurs due to the SOC and asymmetry of the crystal potential, perpendicular to the two-dimensional plane. It can also give rise to the spin-momentum locking. The spin–momentum locking in 2D geometries can influence the interplay between the charge and spin transport. Rashba-Edelstein effect is an intrinsic charge to spin conversion mechanism. When a charge current flows through the spin-polarized surface states, it causes a spin accumulation [425]. In the case of a 2D Rashba gas, where this band splitting occurs [426], this effect is called Rashba–Edelstein effect.

Topological insulators (TI) exhibit a spin-split linear dispersion relation on their surfaces, while having a band gap in the bulk. In a TI, spin-split surface states occur due to the surface topology, irrespective of the Rashba effect. Besides, they also exhibit spin-momentum locking [427]. When a charge current flows in the spin-polarized surface states of a TI, a spin accumulation is generated and the effect is called the Edelstein effect [425].

The spin Hall effect (SHE) acts as the leading mechanism for the conversion of charge to transverse spin current inside a nonmagnetic metallic system with large spin-charge conversion ratio (spin Hall angle or SHA). Dyakonov and Perel first proposed this phenomenon in 1971 [428] which was later expounded by Hirsch [429]. Subsequently, optical and electrical measurement techniques were used to study SHE and inverse SHE in different non-magnetic systems [430-432]. Its origin lies in the intrinsic effect, namely Berry curvature effect associated with the spin dependent band structure of the crystalline material and extrinsic mechanisms associated with the asymmetric scattering of conduction electrons from impurity potential, namely skew scattering and side-jump mechanism. There is also an intrinsic universal spin Hall effect that is due to spin-orbit interaction in a 2D system which yields a universal spin Hall coefficient of $q/8\pi$ ($q$ = electron charge) [433]. Among nonmagnetic metals, heavy metals like Pt, Ta, Ir, Hf, W show large SHA with W and its oxides exhibiting the highest value of ~0.5 for the SHA [434]. On the other hand, TIs show giant SHA with spin Hall angles well above unity associated with the topological surface state and unique valley degree of freedom [435].

Spin pumping is another efficient method for generating pure spin current. It is a non-local effect, which refers to the flow of pure spin current generated by the precession of magnetization, from a ferromagnet (FM) into a normal metal (NM). The magnetization precession results in spin accumulation at the NM/FM interface. The accumulated spins carry angular momentum to the adjacent NM layer. The NM layer acts as a spin sink by absorbing the spin current through spin-flip scattering. This leads to an enhancement of the Gilbert damping parameter of the FM layer. Tserkovnyak, Brataas and Bauer theoretically demonstrated the spin pumping induced enhancement via damping in NM/FM heterostructures in 2002, using time-dependent adiabatic scattering theory [436]. There, the magnetization dynamics in the presence of spin pumping can be expressed by a modified Landau-Lifshitz-Gilbert (LLG) equation as [437, 438]:

$$\frac{d\boldsymbol{m}}{dt} = -\gamma(\boldsymbol{m} \times \boldsymbol{H}_{eff}) + \alpha_0(\boldsymbol{m} \times \frac{d\boldsymbol{m}}{dt}) + \frac{\gamma}{VM_s}\boldsymbol{I}_s, \qquad (15)$$

where $\gamma$ is the gyromagnetic ratio, $\boldsymbol{I}_s$ is the total spin current, $\boldsymbol{H}_{eff}$ is the effective magnetic field, $\alpha_0$ is the intrinsic Gilbert damping constant, $V$ is the volume of the ferromagnet and $M_s$ is the saturation magnetization of the ferromagnet. The spin current $\boldsymbol{I}_s$ generally consists of a direct current contribution $\boldsymbol{I}_s^0$, spin current $\boldsymbol{I}_s^{pump}$ due to pumped spins from the FM to NM, and spin current backflow $\boldsymbol{I}_s^{back}$ to the FM due to reflection from the NM/substrate interface which is assumed to be a perfect reflector. This can be writted as:

$$I_s = I_s^0 + I_s^{pump} + I_s^{back} \qquad (16)$$

The contribution $I_s^{back}$ is determined by the spin diffusion length of the NM layer. The spin transport through NM/FM interface depends upon the spin-mixing conductance, which can be classified into two types: (a) intrinsic spin-mixing conductance, $G_{\uparrow\downarrow}$, which ignores the contribution of backflow of spin angular momentum, and (b) effective spin-mixing conductance $G_{eff}$, which includes the backflow contribution [439]. Spin-mixing conductance describes the conductance property of spin channels at the interface between the NM and the FM. The spin transparency, $T$, of an NM/FM interface takes into account various effects that lead to electrons being reflected from the interface instead of being transmitted during transport [440]. Further, $T$ depends on both intrinsic and extrinsic interfacial factors, namely band-structure mismatch, Fermi velocity, interface imperfections, etc [441].

Another important effect arising from SOC is the Dzyaloshinskii-Moriya interaction (DMI). This has its root in the seminal work in 1958, when the phenomenological theory of antisymmetric exchange coupling between spins to explain the phenomenon of weak ferromagnetism in antiferromagnetic compounds [442] was developed. In 1960 Moriya derived this interaction as a spin–orbit coupling between electrons within the framework of super-exchange theory [443, 444]. This anisotropic exchange interaction (DMI) arises in materials that lack inversion symmetry and where strong SOC effects are present. This spans non-centrosymmetric bulk ferromagnets, multiferroics, perovskites and cuprates. Interfacial DMI is observed in ferromagnetic thin film heterostructures. This helps to stabilize chiral spin textures such as magnetic helices, skyrmions, skyrmion lattices, and chiral domain walls in ferromagnetic materials. Early theoretical effort considered the RKKY interaction, i.e. spin–orbit scattering of the conduction electron gas from the heavy metal impurities to explain DMI [445]. Later, proximity induced magnetic moment in heavy metal was claimed to be responsible for DMI [446], which was immediately contradicted and was found to have no role to play in Co/Pt [447]. The sign and magnitude of DMI are related to the degree of 3d–5d orbital hybridization around the Fermi level [448]. It depends on filling of 5d orbitals (electronegativity) in half-metals (HM) [449]. DMI is driven by spin-flip transitions between 3d states (in the FM) involving intermediate states (from the adjacent layer) with strong SOC strength [236] and is correlated with spin-mixing conductance.

The SOC related effects are generally studied in HM systems but other candidates like TI and 2D materials are gradually making inroads. For example, the SOC magnitude for the *sp*2 bonded structure of pristine graphene is small (about 10 μeV) [450]. However, SOC can be greatly enhanced by proximity and hybridization with adjacent materials. In-plane and out-of-plane deformations that mix the *sp*2 and *sp*3 orbitals in strained or buckled graphene, as well as interactions with adatoms and electric fields, can also lead to enhancements in SOC. SOC-induced splitting of graphene bands leads further to spin–momentum locking and chiral spin orientations; for example, the large SOC splitting (about 100 meV) for graphene on Au is caused by the strong hybridization between the Dirac-cone states and *d* states of Au [451].

**12.2 Spin mechatronics**

Spin mechatronics utilizes mechanical motion of nanomagnets for generation and control of spin currents. Fundamentally, spin current can be generated by using angular momentum conversion among magnetization, photons, orbital motion of electrons, and spin angular momentum. Likewise, angular momentum conservation between mechanical motion and spins triggers the spin current generation by mechanical motion [452]. To understand this phenomenon, a spinning top driven by external torque can be considered. When mechanical rotation is applied to this system, the rotation axis of a spinning top aligns with the axis of external torque due to the Coriolis force given by: $H_{Cor} = -\mathbf{L} \cdot \mathbf{\Omega}$, where $\mathbf{L}$ is the mechanical angular momentum of the spinning top and $\mathbf{\Omega}$ is the angular velocity of the applied rotation field. Analogously, classical Coriolis force can be replaced by quantum mechanical spin-rotation coupling (SRC), i.e., coupling between spin angular momentum $\mathbf{S}$ and mechanical rotation given by: $H_{SRC} = -\mathbf{S} \cdot \mathbf{\Omega}$. In 1915, Barnett discovered a gyromagnetic effect known as Barnett effect, which stipulates that when a magnetic element is rotated, the spins inside the element are aligned along the rotation axis, i.e. the element is magnetized (figure 34(a)) [453]. This is caused by the Zeeman coupling of $\mathbf{S}$ with the emergent magnetic field $B_\Omega = \Omega/\gamma$ created by the rotation of the element, where $\gamma$ is the gyromagnetic ratio.

The converse phenomenon of the Barnett effect is known as Einstein–de Haas effect. Einstein and de Haas experimentally demonstrated that a freely suspended ferromagnet starts to rotate when an external magnetic field is applied (figure 34(b)) [454]. When the magnetic field is applied, the magnetic moment in the ferromagnet is modulated. Because of the conservation of angular momentum, a mechanical angular momentum is induced to compensate for the modulation of the spin angular momentum of the ferromagnet. In the Barnett effect, the mechanical rotation not only couples to the electronic spin system, but also couples with the nuclear spin system. Chudo *et al.* measured the Barnett field in non-ferromagnetic system by observing the shift of the NMR frequency due to the SRC between nuclear spin and mechanical rotation [455]. By measuring the induced magnetic moment in a polycrystalline Gd (paramagnetic) sample, Ono *et al.* also estimated the Bernett field [456]. Interestingly, SRC not only modulates spin angular momentum, but also can generate spin current by two routes: first, mechanically induced SOC due to the rigid motion, and second, coupling between spin and vorticity, i.e., the local rotational motion of elastic or fluid materials. In the first case, SHEs can be induced by rigid [457] and vibrational [458] mechanical motion even in the absence of conventional SOC, which enables spin current generation. As an example, if a Pt film vibrates at 10 GHz frequency with an amplitude of 0.1 nm, then the amplitude of the generated ac spin current can be estimated as $10^5$ A m$^{-2}$ [458]. In the second case, i.e., in the presence of spin-vorticity coupling, the gradient of the vorticity gives rise to a spin-dependent force given by $F_S = \mathbf{S} \cdot \nabla\boldsymbol{\omega}$, where $\nabla\boldsymbol{\omega}$ is the gradient of vorticity. This shows that spin current can be generated along the vorticity gradient. When SAWs propagate through a nonmagnetic film, the atoms attached to the lattice points oscillate around their equilibrium position and a gradient of vorticity is established across the thickness of the film. This generates a spin current across the thickness [459-461]. Interestingly, this mechanism allows us to utilize materials with relatively small SOC (e.g., Cu, Al and even carbon nanotubes) to generate spin current with reasonable density. Takahashi *et al.* demonstrated that the spin-vorticity coupling caused by the gradient of vorticity field in a liquid metal flow, such as flow of Hg or GaInSn in a fine quartz pipe, can also generate spin current [462].

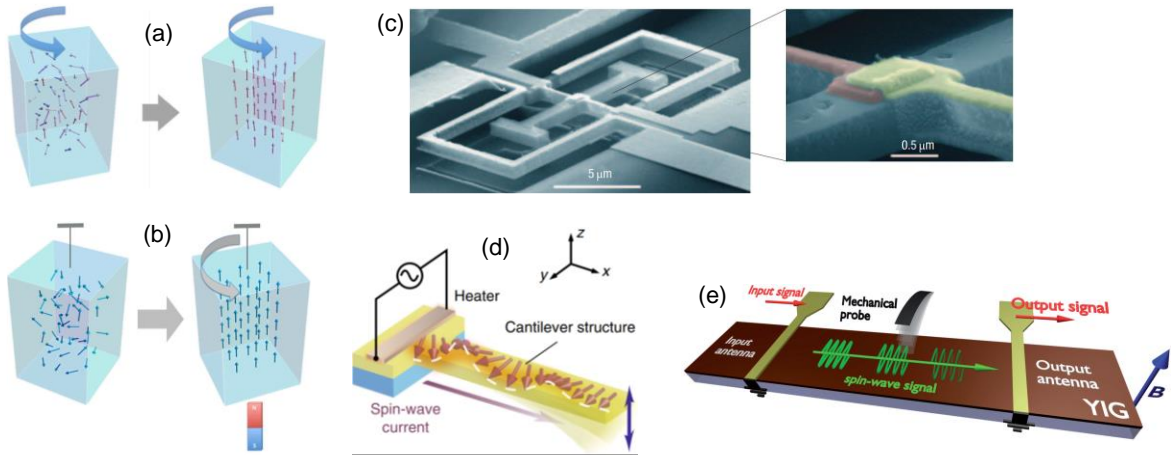

**Figure 34**. (a) Schematic illustration of the Barnett effect. The magnetic moments in a ferromagnetic material are aligned along the rotation axis of the external mechanical rotation due to spin-rotation coupling [452, 463]. (b) Einstein–de Haas effect, the converse of the Barnett effect, which causes mechanical rotation owing to magnetization, is illustrated. When a suspended magnetic material is magnetized, i.e., the spins are aligned along a particular direction by applying an external magnetic field, a mechanical angular momentum is induced to compensate for the modulation of the spin angular momentum. Reproduced from [463] with the permission of Frontiers Media. (c) Scanning electron microscopy image of a nanomechanical device with the enlarged image showing the FM/NM interface (cobalt, red; gold, yellow). The fully suspended structure is clamped rigidly at the large support pads. Reproduced from [464] with permission of Springer Nature. (d) The schematic illustration shows a cantilever made of YIG connected to an edge of a YIG film and a heater placed on the YIG film around the root of the cantilever. An electric current applied to the heater generates heat current, which creates SW (magnon) accumulation at the surface and the bottom of the YIG film. The accumulation of magnons injects spin current into the YIG cantilever. Reproduced from [465] with permission of Springer Nature. (e) The schematic diagram of the experimental setup that demonstrated the feasibility of utilizing mechanical coupling to propagate magnetostatic SWs through a YIG waveguide. The amplitude modulated SWs, excited by a microstrip inductive antenna, are used to drive out-of-plane oscillations of a magnetic micro-mechanical

probe (nickel, diameter 10 µm) suspended above the surface of YIG film. The displacements of the probe were measured using a laser interferometric vibrometer. Reproduced from [466] with permission of the authors.

In spintronics, the spin current, i.e., flow of spin angular momentum is utilized for transmission and processing information. The spin angular momentum can be carried either by translational motion of conduction electrons or by SWs. Since spin current carries angular momentum, a spin-flip or spin-transfer process always involves a change in angular momentum, which can induce a mechanical torque and hence mechanical oscillation to the nanoelement attached to the system. Based on this principle, Mohanty *et al.* proposed a spin-mechanical device to control and detect spin currents by mechanical torque [467], which was experimentally demonstrated later [464]. It was demonstrated that when a spin-polarized current passes through a metallic nanowire in which one half is ferromagnetic and the other half is nonmagnetic, the spins of the conduction electrons are 'flipped' at the interface between the two regions which then produces a nanomechanical torque (figure 34(c)). The spin torque was determined from a measurement of the magnetic field dependence of the magnetomotive voltage. This enables nanomechanical detection of spin flip phenomenon. This idea was later extended for nanomechanical detection of spin Hall coefficient. Recently, Harii *et al.* have experimentally demonstrated transmission of mechanical angular momentum and force generation due to SWs injected into a yttrium iron garnet (YIG) film by the spin-Seebeck effect [465]. The SW current, transmitted through a YIG micro cantilever, was observed to create a mechanical force on the cantilever as a non-local reaction to the spin-Seebeck effect (figure 34(d)). Karenowska *et al.* proposed a hybrid spin mechatronic structure in which information encoded in the amplitudes and phases of propagating SWs are written and read by resonant magno-mechanical elements [466]. The principle of operation is very similar to magnetic resonance force microscopy (MRFM), where short wavelength, highly spatially localized standing SWs are detected by a cantilever tipped with a ferromagnetic particle. They have proposed to use a YIG waveguide for SW transmission and a magnetic cantilever suspended above the film as a micromechanical probe (figure 34(e)). The probe is movable in the *x*-, *y*-, and *z*-directions. An optical vibrometer (resolution 0.1nm) is used to detect the displacements of the cantilever due to the stray field generated by propagating SWs. This magnetic cantilever movement can even perturb the SW amplitude and phase at sub-millimeter approach distances.

## 13. CONCLUSION

Research in nanomagnetism has seen an explosive growth over the last few decades. New insights were provided on how size reduction and spatial confinement affect spin configurations and how they, in turn, affect magnetization reversal and dynamics. A myriad of complex nanostructures comprising arrays of interacting nanomagnets of different shapes and sizes, multilayered nanomagnets, as well as patterned devices like MTJs, spin valves, spin torque nano-oscillators, etc. have revealed rich physics and continue to portend enticing applications. They will motivate ever more complex investigations. Patterning magnetic textures by external fields promises the development of on-demand and reconfigurable nanomagnetic devices with unprecedented flexibility in engineering their performance and adapting them to new and different system platforms. Progress in the development of elegant and sophisticated characterization techniques has afforded the opportunity to probe exotic magnetic textures as well as intriguing switching phenomena and dynamics with femtosecond temporal and sub-10 nm spatial resolution. X-ray and electron microscopy-based techniques combined with others offering high temporal resolution and polarization sensitivity have enabled exploration of the most fundamental interactions and dynamics in nanoscale spin structures. In terms of magnetization manipulation, magnetization switching, precession, damping, SW propagation and dispersion, enormous strides have been made. Exquisite control over the effects of thermal, microwave, and strain energies on magnetization switching, as well as electric and magnetic field-based tailoring of the associated dynamics, have yielded capabilities heretofore unimaginable. These capabilities have spawned ultrahigh performance magnetic storage, memory, logic, communication, and processing devices.

The field of magnonics offers the tantalizing possibility of replacing traditional charge-based logic, communication and information processing approaches with an alternate paradigm that may eclipse the traditional paradigm in many aspects. Hybridization of magnonics with spintronics, spin-orbitronics, phononics, straintronics, spin-mechatronics, etc. holds out the promise of embedding multiple functionalities in compact systems that are energy-efficient, fast and versatile. As nascent fields like strong coupling of magnons with other quasiparticles, antiferromagnetic nanostructures, nutation dynamics, topological effects, etc. begin to mature and burgeon, they may shed new light on fundamental physics and lead the development of advanced devices and system with significant technological and societal impact. Many challenges remain, but the excitement that permeates this field is likely to transcend the hurdles and take this field to new heights.

**Acknowledgement:** The work of SB in this field has been supported over the last decade by the US National Science Foundation under grants ECCS-1124714, CCF-1216614, CMMI-301013, ECCS-1609303, CCF-1815033, CCF-2001255 and CCF-2006843, the Semiconductor Research Corporation under task 2203.001, the State of Virginia CRCF Fund and the Virginia Commonwealth University Commercialization Fund. AB and SB acknowledge travel support from the Indo-US Science and Technology Fund Center Grant titled "Center for Nanomagnetics for Energy-Efficient Computing, Communications and Data Storage" (IUSSTF/JC-030/ 2018). AKM thanks the S. N. Bose National Centre for Basic Sciences for a senior research fellowship.

[18]  Abeed M A and Bandyopadhyay S 2020 Simulated annealing with surface acoustic wave in a dipole-coupled array of magnetostrictive nanomagnets for collective ground state computing *J. Phys. D: Appl. Phys.* **53** 445002

[19]  McCray M T, Abeed M A and Bandyopadhyay S 2020 Electrically programmable probabilistic bit anti-correlator on a nanomagnetic platform *Sci. Rep.* **10** 12361

[20]  Sutton B, Camsari K Y, Behin-Aein B and Datta S 2017 Intrinsic optimization using stochastic nanomagnets *Sci. Rep.* **7** 44370

[21]  Borders W A, Pervaiz A Z, Fukami S, Camsari K Y, Ohno H and Datta S 2019 Integer factorization using stochastic magnetic tunnel junctions *Nature* **573** 390-393

[22]  Debashis P, Faria R, Camsari K Y, Appenzeller J, Datta S and Chen Z 2016 Experimental demonstration of nanomagnet networks as hardware for Ising computing *IEEE International Elec. Dev. Meeting (IEDM)* 34.33.31-34.33.34

[23]  Bhanja S, Karunaratne D K, Panchumarthy R, Rajaram S and Sarkar S 2016 Non-Boolean computing with nanomagnets for computer vision applications *Nat. Nanotechnol.* **11** 177-183

[24]  Palecio J, Bhanja S and Sarkar S 2011 An experimental demonstration of the viability of energy minimizing computing using nanomagnets *11th IEEE International Conference on Nanotechnology, Portland, Oregon* 1038

[25]  Nasrin S, Drobitch J L, Bandyopadhyay S and Trivedi A R 2019 Mixed-mode magnetic tunnel junction based deep belief network *2019 IEEE 19th International Conference on Nanotechnology (IEEE-NANO)* **Macao** 443

[26]  Nasrin S, Drobitch J, Shukla P, Tulabandhula T, Bandyopadhyay S and Trivedi A R 2020 Bayesian reasoning machine on a magneto-tunneling junction network *Nanotechnology* **31** 484001

[27]  Khasanvis S, Li M, Rahman M, Biswas A K, Salehi-Fashami M, Atulasimha J, Bandyopadhyay S and Moritz C A 2015 Architecting for causal intelligence at nanoscale *Computer* **48** 54-64

[28]  Khasanvis S, Li M, Rahman M, Salehi-Fashami M, Biswas A K, Atulasimha J, Bandyopadhyay S and Moritz C A 2015 Self-similar magneto-electric nanocircuit technology for probabilistic inference engines *IEEE Trans. Nanotechnol.* **14** 980-991

[29]  Nasrin S, Drobitch J L, Bandyopadhyay S and Trivedi A R 2019 Low power restricted Boltzmann machine using mixed-mode magneto-tunneling junctions *IIEEE Elec. Dev. Lett.* **40** 345-348

[30]  Manasi S D, Al-Rashid M M, Atulasimha J, Bandyopadhyay S and Trivedi A R 2017 Skewed straintronic magnetotunneling-junction-based ternary content-addressable memory—part I and II *IEEE Trans. Elec. Dev.* **64** 2835 and 2842

[31]  Biswas A K, Atulasimha J and Bandyopadhyay S 2017 Energy-efficient hybrid spintronic–straintronic nonvolatile reconfigurable equality bit comparator *SPIN* **07** 1750004

[32]  Deac A M, Fukushima A, Kubota H, Maehara H, Suzuki Y, Yuasa S, Nagamine Y, Tsunekawa K, Djayaprawira D D and Watanabe N 2008 Bias-driven high-power microwave emission from MgO-based tunnel magnetoresistance devices *Nat. Phys.* **4** 803-809

[33]  Chen T, Dumas R K, Eklund A, Muduli P K, Houshang A, Awad A A, Dürrenfeld P, Malm B G, Rusu A and Åkerman J 2016 Spin-torque and spin-Hall nano-oscillators *Proc. IEEE* **104** 1919-1945
85

[70] Sreenivasan S V 2017 Nanoimprint lithography steppers for volume fabrication of leading-edge semiconductor integrated circuits *Microsyst. Nanoeng.* **3** 17075

[71] Karim W, Tschupp S A, Oezaslan M, Schmidt T J, Gobrecht J, van Bokhoven J A and Ekinci Y 2015 High-resolution and large-area nanoparticle arrays using EUV interference lithography *Nanoscale* **7** 7386-7393

[72] Ding J and Adeyeye A O 2013 Binary ferromagnetic nanostructures: fabrication, static and dynamic properties *Adv. Funct. Mater.* **23** 1684-1691

[73] Smith H and Schattenburg M 1992 Why bother with x-ray lithography? *Proc. SPIE 1671*

[74] Drobitch J L, De A, Dutta K, Pal P K, Adhikari A, Barman A and Bandyopadhyay S 2020 Extreme subwavelength magnetoelastic electromagnetic antenna implemented with multiferroic nanomagnets *Adv. Mater. Technol.* **5** 2000316

[75] Garcia R, Knoll A W and Riedo E 2014 Advanced scanning probe lithography *Nat. Nanotechnol.* **9** 577-587

[76] Chumak A V, Serga A A, Hillebrands B and Kostylev M P 2008 Scattering of backward spin waves in a one-dimensional magnonic crystal *Appl. Phys. Lett.* **93** 022508

[77] Zhou X, Hou Y and Lin J 2015 A review on the processing accuracy of two-photon polymerization *AIP Adv.* **5** 030701

[78] Takeshita H, Suzuki Y, Akinaga H, Mizutani W, Tanaka K, Katayama T and Itoh A 1996 Magneto-optical response of nanoscaled cobalt dots array *Appl. Phys. Lett.* **68** 3040

[79] Lee K L, Thomas R R, Viehbeck A and O'Sullivan E J M 1993 Selective electroless plating on electron beam seeded nanostructures *J. Vac. Sci. Technol. B* **11** 2204

[80] Kamata Y, Kikitsu A, Hieda H, Sakurai M and Naito K 2004 Ar ion milling process for fabricating CoCrPt patterned media using a self-assembled PS-PMMA diblock copolymer mask *J. Appl. Phys.* **95** 6705

[81] Masuda H, Yamada H, Satoh M, Asoh H, Nakao M and Tamamura T 1997 Highly ordered nanochannel-array architecture in anodic alumina *Appl. Phys. Lett.* **71** 2770

[82] Williams W D and Giordano N 1986 Experimental study of localization and electron-electron interaction effects in thin Au wires *Phys. Rev. B* **33** 8146-8154

[83] Zeng H, Zheng M, Skomski R, Sellmyer D J, Liu Y, Menon L and Bandyopadhyay S 2000 Magnetic properties of self-assembled Co nanowires of varying length and diameter *J. Appl. Phys.* **87** 4718-4720

[84] Menon L, Zheng M, Zeng H, Bandyopadhyay S and Sellmyer D J 2000 Size dependence of the magnetic properties of electrochemically self-assembled Fe quantum dots *J. Elec. Mater.* **29** 510-515

[85] Cowburn R P, Koltsov D K, Adeyeye A O, Welland M E and Tricker D M 1999 Single-domain circular nanomagnets *Phys. Rev. Lett.* **83** 1042-1045

[86] Kikuchi R 1956 On the minimum of magnetization reversal time *J. Appl. Phys.* **27** 1352-1357

[87] Ababei R-V, Ellis M O A, Evans R F L and Chantrell R W 2019 Anomalous damping dependence of the switching time in Fe/FePt bilayer recording media *Phys. Rev. B* **99** 024427

[88] Matsuzaki J, Tanaka T, Kurisu H and Yamamoto S 2009 Magnetization Switching Time for Hard/Soft Magnetic Composite Pillar *Trans. Mater. Res. Soc.* **34** 415-418

[89] Choi B C, Belov M, Hiebert W K, Ballentine G E and Freeman M R 2001 Ultrafast magnetization reversal dynamics investigated by time domain imaging *Phys. Rev. Lett.* **86** 728-731

[125] Ishida N, Soeno Y, Sekiguchi K and Nozaki Y 2013 Frequency dependence of critical switching asteroid of CoCrPt–SiO$_2$ granular film under 50-ns microwave impulse *J. Appl. Phys.* **114** 043915

[126] Okamoto S, Kikuchi N, Hotta A, Furuta M, Kitakami O and Shimatsu T 2013 Microwave assistance effect on magnetization switching in Co-Cr-Pt granular film *Appl. Phys. Lett.* **103** 202405

[127] Nistor C, Sun K, Wang Z, Wu M, Mathieu C and Hadley M 2009 Observation of microwave-assisted magnetization reversal in Fe$_{65}$Co$_{35}$ thin films through ferromagnetic resonance measurements *Appl. Phys. Lett.* **95** 012504

[128] Okamoto S, Kikuchi N, Kitakami O, Shimatsu T and Aoi H 2011 Microwave assisted magnetization switching in Co/Pt multilayer *J. Appl. Phys.* **109** 07B748

[129] Zhu J, Zhu X and Tang Y 2008 Microwave sssisted magnetic recording *IEEE Trans. Magn.* **44** 125-131

[130] Zhu J and Wang Y 2010 Microwave assisted magnetic recording utilizing perpendicular spin torque oscillator with switchable perpendicular electrodes *IEEE Trans. Magn.* **46** 751-757

[131] Suto H, Kanao T, Nagasawa T, Mizushima K and Sato R 2018 Magnetization switching of a Co/Pt multilayered perpendicular nanomagnet assisted by a microwave field with time-varying frequency *Phys. Rev. Applied* **9** 054011

[132] Okamoto S, Kikuchi N, Furuta M, Kitakami O and Shimatsu T 2015 Microwave assisted magnetic recording technologies and related physics *J. Phys. D Appl. Phys.* **48** 353001

[133] Slonczewski J C 1996 Current-driven excitation of magnetic multilayers *J. Magn. Magn. Mater.* **159** L1

[134] Berger L 1996 Emission of spin waves by a magnetic multilayer traversed by a current *Phys. Rev. B* **54** 9353

[135] Wang K L, Alzate J G and Khalili Amiri P 2013 Low-power non-volatile spintronic memory: STT-RAM and beyond *J. Phys. D: Appl. Phys.* **46** 074003

[136] Liu L, Pai C-F, Li Y, Tseng H W, Ralph D C and Buhrman R A 2012 Spin-torque switching with the giant spin Hall effect of Tantalum *Science* **336** 555-558

[137] Pai C-F, Liu L, Li Y, Tseng H W, Ralph D C and Buhrman R A 2012 Spin transfer torque devices utilizing the giant spin Hall effect of tungsten *Appl. Phys. Lett.* **101** 122404

[138] Niimi Y, Kawanishi Y, Wei D H, Deranlot C, Yang H X, Chshiev M, Valet T, Fert A and Otani Y 2012 Giant spin hall effect induced by skew scattering from bismuth impurities inside thin film cubi alloys *Phys. Rev. Lett.* **109** 156602

[139] Edelstein V M 1990 Spin polarization of conduction electrons induced by electric current in two-dimensional asymmetric electron systems *Solid State Commun.* **73** 233-235

[140] Chernyshov A, Overby M, Liu X, Furdyna J K, Lyanda-Geller Y and Rokhinson L P 2009 Evidence for reversible control of magnetization in a ferromagnetic material by means of spin–orbit magnetic field *Nat. Phys.* **5** 656-659

[141] Bychkov Y A and Rashba E I 1984 Oscillatory effects and the magnetic susceptibility of carriers in inversion layers *J. Phys. C: Solid State Phys.* **17** 6039-6045

[142] Bandyopadhyay S and Cahay M 2015 Introduction to Spintronics 2nd edn (Boca Raton, FL: CRC Press)

[143] Liu L, Moriyama T, Ralph D C and Buhrman R A 2011 Spin-torque ferromagnetic resonance induced by the spin Hall effect *Phys. Rev. Lett.* **106** 036601

[430] Valenzuela S O and Tinkham M 2006 Direct electronic measurement of the spin Hall effect *Nature* **442** 176-179
[431] Kato Y K, Myers R C, Gossard A C and Awschalom D D 2004 Observation of the spin Hall effect in semiconductors *Science* **306** 1910
[432] Saitoh E, Ueda M, Miyajima H and Tatara G 2006 Conversion of spin current into charge current at room temperature: Inverse spin-Hall effect *Appl. Phys. Lett.* **88** 182509
[433] Sinova J, Culcer D, Niu Q, Sinitsyn N A, Jungwirth T and MacDonald A H 2004 Universal intrinsic spin Hall effect *Phys. Rev. Lett.* **92** 126603
[434] Demasius K-U, Phung T, Zhang W, Hughes B P, Yang S-H, Kellock A, Han W, Pushp A and Parkin S S P 2016 Enhanced spin–orbit torques by oxygen incorporation in tungsten films *Nat. Commun.* **7** 10644
[435] Khang N H D, Ueda Y and Hai P N 2018 A conductive topological insulator with large spin Hall effect for ultralow power spin–orbit torque switching *Nat. Mater.* **17** 808-813
[436] Tserkovnyak Y, Brataas A and Bauer G E W 2002 Enhanced Gilbert damping in thin ferromagnetic films *Phys. Rev. Lett.* **88** 117601
[437] Tserkovnyak Y, Brataas A, Bauer G E W and Halperin B I 2005 Nonlocal magnetization dynamics in ferromagnetic heterostructures *Rev. Mod. Phys.* **77** 1375-1421
[438] Tserkovnyak Y, Brataas A and Bauer G E W 2002 Spin pumping and magnetization dynamics in metallic multilayers *Phys. Rev. B* **66** 224403
[439] Shaw J M, Nembach H T and Silva T J 2012 Determination of spin pumping as a source of linewidth in sputtered $Co_{90}Fe_{10}$/Pd multilayers by use of broadband ferromagnetic resonance spectroscopy *Phys. Rev. B* **85** 054412
[440] Zhang W, Han W, Jiang X, Yang S-H and S. P. Parkin S 2015 Role of transparency of platinum–ferromagnet interfaces in determining the intrinsic magnitude of the spin Hall effect *Nat. Phys.* **11** 496-502
[441] Panda S N, Mondal S, Sinha J, Choudhury S and Barman A 2019 All-optical detection of interfacial spin transparency from spin pumping in β-Ta/CoFeB thin films *Sci. Adv.* **5** eaav7200
[442] Dzyaloshinsky I 1958 A thermodynamic theory of "weak" ferromagnetism of antiferromagnetics *J. Phys. Chem. Solids* **4** 241-255
[443] Moriya T 1960 New mechanism of anisotropic superexchange interaction *Phys. Rev. Lett.* **4** 228-230
[444] Moriya T 1960 Anisotropic superexchange interaction and weak ferromagnetism *Phys. Rev.* **120** 91-98
[445] Fert A and Levy P M 1980 Role of anisotropic exchange interactions in determining the properties of spin-glasses *Phys. Rev. Lett.* **44** 1538-1541
[446] Ryu K-S, Yang S-H, Thomas L and Parkin S S P 2014 Chiral spin torque arising from proximity-induced magnetization *Nat. Commun.* **5** 3910
[447] Yang H, Thiaville A, Rohart S, Fert A and Chshiev M 2015 Anatomy of Dzyaloshinskii-Moriya interaction at Co/Pt interfaces *Phys. Rev. Lett.* **115** 267210
[448] Belabbes A, Bihlmayer G, Bechstedt F, Blügel S and Manchon A 2016 Hund's Rule-Driven Dzyaloshinskii-Moriya Interaction at 3d-5d Interfaces *Phys. Rev. Lett.* **117** 247202
[449] Torrejon J, Kim J, Sinha J, Mitani S, Hayashi M, Yamanouchi M and Ohno H 2014 Interface control of the magnetic chirality in CoFeB/MgO heterostructures with heavy-metal underlayers *Nat. Commun.* **5** 4655
109